\documentclass[traditabstract]{aa} % for the abstract without structuration (traditional abstract) 

\usepackage{graphicx}
\usepackage{lscape}
\usepackage{array}
\usepackage{amssymb}
\usepackage{savesym}
\usepackage{amsmath}
\savesymbol{iint} %otherwise get LaTeX Error: Command \iint already defined.
\usepackage{txfonts}
\restoresymbol{TXF}{iint}
\usepackage{bm}
\usepackage{natbib}
\bibpunct{(}{)}{;}{a}{}{,} % to follow the A&A style

\newcommand{\var}{\text{Var}}
\newcommand{\cov}{\text{Cov}}
\def\kms{\hbox{$~$km$~$s$^{-1}$}}
\def\l{\ifmmode\lambda\else$\lambda$\fi}
\def\snia{SN~Ia}
\def\sneia{SN~Ia}

\def\catwo{Ca\,{\sc II}}
\def\fetwo{Fe\,{\sc II}}
\def\fethree{Fe\,{\sc III}}
\def\mgtwo{Mg\,{\sc II}}

\def\naoned{Na\,{\sc I}\,D}

\def\stwo{S\,{\sc II}}
\def\sitwo{Si\,{\sc II}}
\def\titwo{Ti\,{\sc II}}

\begin{document}
\title{Do spectra improve distance measurements of Type Ia supernovae?}

\author{S. Blondin\inst{1}
        \and
        K.~S.~Mandel\inst{2}
        \and
	R.~P.~Kirshner\inst{2}
        }

\institute{
Centre de Physique des Particules de Marseille (CPPM), CNRS/IN2P3, 163
avenue de Luminy, 13288 Marseille Cedex 9, France\\ 
\email{blondin@cppm.in2p3.fr}
\and
Harvard-Smithsonian Center for Astrophysics (CfA), 60 Garden Street,
Cambridge, MA 02138, USA
}

\date{Received 20 September 2010 / Accepted 24 November 2010}

%%%%%%%%%%%%%%%%%%%%%%%%%%%%%%%%%%%%%%%%%%%%%%%%%%%%%%%%%%%%%%%%%%
%%
%%   Abstract
%%
%%%%%%%%%%%%%%%%%%%%%%%%%%%%%%%%%%%%%%%%%%%%%%%%%%%%%%%%%%%%%%%%%%

\abstract{
We investigate the use of a wide variety of spectroscopic measurements
to determine distances to low-redshift Type Ia supernovae (\sneia) in
the Hubble flow observed through the CfA Supernova Program. We
consider linear models for predicting distances to SN Ia using
light-curve width and color parameters (determined using the SALT2
light-curve fitter) and a spectroscopic indicator, and evaluate the
resulting Hubble diagram scatter using a cross-validation procedure. 
We confirm the ability of spectral flux ratios alone at maximum light
to reduce the scatter of Hubble residuals by $\sim10$\% [weighted rms,
  or ${\rm WRMS}=0.189\pm0.026$\,mag for the flux ratio
  $\mathcal{R}(6630/4400)$] with respect to the standard combination
of light-curve width and color, for which ${\rm
  WRMS}=0.204\pm0.029$\,mag. When used in combination with the SALT2
color parameter, the color-corrected flux ratio
$\mathcal{R}^c(6420/5290)$ at maximum light leads to an even lower
scatter (${\rm WRMS}=0.175\pm0.025$\,mag), although the improvement
has low statistical significance ($<2\sigma$) given the size of our
sample (26 \sneia). We highlight the importance of an accurate
relative flux calibration and the failure of this method for
highly-reddened objects. Comparison with synthetic spectra from 2D
delayed-detonation explosion models shows that the correlation of
$\mathcal{R}(6630/4400)$ with \snia\ absolute magnitudes can be
largely attributed to intrinsic color variations and not to reddening
by dust in the host galaxy. We consider flux ratios at other ages, as
well as the use of pairs of flux ratios, revealing the presence of
small-scale intrinsic spectroscopic variations in the iron-group-dominated
absorption features around $\sim4300$\,\AA\ and
$\sim4800$\,\AA. The best flux ratio overall is the color-corrected 
$\mathcal{R}^c(4610/4260)$ at $t=-2.5$\,d from maximum light, which
leads to $\sim30$\% lower scatter (${\rm WRMS}=0.143\pm0.020$\,mag)
with respect to the standard combination of light-curve width and
color, at $\sim2\sigma$ significance. We examine other spectroscopic
indicators related to line-profile morphology (absorption velocity,
pseudo-equivalent width etc.), but none appear to lead to a
significant improvement over the standard light-curve width and color
parameters. We discuss the use of spectra in measuring more precise
distances to \sneia\ and the implications for future surveys which
seek to determine the properties of dark energy.
}


\maketitle

%%%%%%%%%%%%%%%%%%%%%%%%%%%%%%%%%%%%%%%%%%%%%%%%%%%%%%%%%%%%%%%%%%
%%
%%   Introduction
%%
%%%%%%%%%%%%%%%%%%%%%%%%%%%%%%%%%%%%%%%%%%%%%%%%%%%%%%%%%%%%%%%%%%

\section{Introduction}\label{sect:intro}

Precise distances to Type Ia supernovae (\snia) formed the cornerstone 
of the discovery of cosmic acceleration \citep{R98,P99}.  These
measurements use the shape of supernova light curves and their colors
to tell which supernovae are bright and which are intrinsically dim
\citep{Phillips:1993,MLCS,Prieto/Rest/Suntzeff:2006,MLCS2k2,SALT2,Conley/etal:2008,Mandel/etal:2009}.
In this paper we explore the suggestion of \cite{Bailey/etal:2009}
that spectra can contribute to improved distance measurements.  We
apply statistical tests to a subset of the $\sim250$ \snia\ for
which we have good light curves and spectra based on the ongoing
program of supernova observations at the Harvard-Smithsonian Center
for Astrophysics (CfA;
\citealt{Matheson/etal:2008,Hicken/etal:2009a}).

It is important to construct the best possible distance indicators to
extract the maximum cosmological information from supernova
surveys. The present state-of-the-art gives distances to well-observed
individual objects with uncertainties of order 10\%, so that samples
of nearby \citep{Hicken/etal:2009b} and distant \sneia\
(ESSENCE, \citealt{Miknaitis/etal:2007}; SNLS,
\citealt{Astier/etal:2006}) can be combined to constrain the
equation-of-state for dark energy, noted $w$.  The first results
show that for a flat universe with constant $w$, the dark 
energy is compatible with a cosmological constant (for which $w=-1$)
within about 10\% \citep{Astier/etal:2006,Wood-Vasey/etal:2007}.
Constraints on the variation of $w$ with redshift come from from high-redshift 
observations with the Hubble Space Telescope
\citep{Riess/etal:2004,Riess/etal:2007}. Present-day limits are weak,
but future work with large, carefully calibrated samples from the
ground (Pan-STARRS, Dark Energy Survey, LSST) and from space (Euclid,
WFIRST) will contribute to distinguishing the nature of dark energy
\citep{Albrecht/etal:2009}.  In designing the follow-up observations
for these enterprises, it is worth knowing whether spectra will be
useful only for classification and precise redshifts, or whether
the spectra of the 
supernovae themselves can be used to improve the precision of the
distances.  The way we explore this is to analyze the CfA sample,
using the difference between the distance derived from Hubble
expansion with the distance predicted from our various
models.  This difference is the Hubble residual, which we use as a
measure of the power of a particular model to predict the supernova
distance.  As described below, we explore models that combine
quantitative information from the spectrum with information on light
curve shape and color.

Spectroscopic information is fundamental to the success of employing
\sneia\ as distance indicators in large surveys. Cleanly
separating Type Ia supernovae from core-collapse events like SN~Ib
and SN~Ic improves the purity of the sample.
More directly, \cite{Nugent/etal:1995} showed that some easily-measured
line ratios in \snia\ spectra are correlated with the luminosity. 
Measurements of line velocities (and gradients thereof), strengths,
and widths and their relation to supernova luminosity have been
explored recently by several authors
\citep{Benetti/etal:2005,Blondin/etal:2006,Bongard/etal:2006,Hachinger/Mazzali/Benetti:2006,Bronder/etal:2008}.
Likewise, \cite{Matheson/etal:2008} revealed spectroscopic
variability amongst \sneia\ of similar luminosity.
But the first application of spectroscopic clues to improve 
distance estimates has come from \cite{Bailey/etal:2009}. Using
spectra of 58 \sneia\ from the Nearby Supernova Factory,
they showed that the ratio of fluxes in selected wavelength bins (flux
ratios) could reduce the scatter of Hubble residuals by $\sim20$\% compared
to the usual combination of light-curve width and color parameters
($\sigma = 0.128\pm0.012$\,mag cf. $0.161\pm0.015$\,mag). By
using a flux ratio measured on a de-reddened spectrum in combination
with a color parameter they found a further $\sim5$\% improvement
($\sigma = 0.119\pm0.011$\,mag). We have sought first to see if we can
reproduce their results using the CfA data set, and then to test
additional ideas about ways to use spectra to improve the estimates of
supernova distances.

In practice, the standardization of \snia\ magnitudes involves a term
related to the width of the light curve and a correction due to
color. While some methods attempt to separate intrinsic color
variations from reddening by dust in the host galaxy (e.g., MLCS2k2;
\citealt{MLCS2k2}) others use a single parameter for both effects
(e.g. SALT2; \citealt{SALT2}), exploiting the degeneracy between the
two: underluminous \sneia\ are also intrinsically redder than
overluminous \sneia\ (e.g. \citealt{Tripp:1998}). 
We adopt the latter approach in this paper, to match the method
used by \cite{Bailey/etal:2009}. 
An active area of
research involves the use of \snia\ spectra to provide independent or
complementary information on \snia\ luminosities that would help improve
their use as distance indicators.

We consider models for predicting distances to \sneia\ of the form:
\begin{equation}\label{eqn:intro}
\mu = m_B - M + (\alpha \times {\rm width}) - (\beta \times
{\rm color}) + (\gamma \times {\rm spec}),
\end{equation}
where $m_B$ is the apparent rest-frame $B$-band magnitude at peak, $M$
is a reference absolute magnitude, ``width'' and ``color'' are the usual
light-curve parameters, and ``spec'' is some spectroscopic indicator;
$(\alpha,\beta,\gamma)$ are fitting constants. We study the following
five models:
\begin{itemize}
\item[1.]{only a spectroscopic indicator is used [i.e. $(\alpha,\beta)=(0,0)$],}
\item[2.]{both a spectroscopic indicator and a light-curve width 
  parameter are used, but no color parameter (i.e. $\beta=0$),}
\item[3.]{both a spectroscopic indicator and a color parameter are used,
  but no light-curve width parameter (i.e. $\alpha=0$),}
\item[4.]{a spectroscopic indicator is used in addition to the
  light-curve width and color parameters.}
\item[5.]{both light-curve width and color parameters are used, but no
  spectroscopic indicator (i.e. $\gamma=0$). We refer to this as the
  ``standard'' model.}
\end{itemize}
We refer to the set of light-curve parameters and spectroscopic
indicators in a given model as the ``predictors'' for that model, as
is common practice in the field of statistics.
We can evaluate the use of including a spectroscopic indicator
(models 1-4) by comparing the resulting scatter of Hubble diagram
residuals with that from the standard model (No. 5).

The paper is organized as follows: in \S~\ref{sect:method} we present
our light-curve fitting  and training method, as well as a
cross-validation procedure to evaluate the impact of each spectroscopic
indicator. We present the CfA data set in \S~\ref{sect:data}. In
\S~\ref{sect:fluxratio} we study the flux ratios of
\cite{Bailey/etal:2009}, while in \S~\ref{sect:specindic} we consider
other spectroscopic indicators. We discuss the use of \snia\
spectra for distance measurements in \S~\ref{sect:disc} and conclude
in \S~\ref{sect:ccl}.

%%%%%%%%%%%%%%%%%%%%%%%%%%%%%%%%%%%%%%%%%%%%%%%%%%%%%%%%%%%%%%%%%%
%%
%%   Methodology
%%
%%%%%%%%%%%%%%%%%%%%%%%%%%%%%%%%%%%%%%%%%%%%%%%%%%%%%%%%%%%%%%%%%%

\section{Methodology}\label{sect:method}

\subsection{Light-curve fitting}

We use the SALT2 light-curve fitter of \cite{SALT2} to
determine the width and color parameters for each \snia\
in our sample.  A model relating distance, apparent magnitudes, and
linear dependencies of the absolute magnitude is: 
\begin{equation}\label{eqn:salt2}
\mu = m_B - M + \alpha x_1 - \beta c + \gamma \mathcal{S},
\end{equation}
where $(x_1,c)$ are the SALT2 light curve width and color parameters, and
$\mathcal{S}$ is some spectroscopic indicator.  The rest-frame peak
apparent $B$-band magnitude is $m_B$, also obtained from 
the SALT2 fit to a supernova's light curve.
The distance modulus predicted from the light curve and spectral
indicators is $\mu$, and the constant $M$ is a reference absolute
magnitude.  The distance modulus estimated from the redshift is
$\mu(z) = 25 + 5 \log_{10}[ D_L(z) \text{Mpc}^{-1}]$ under a fixed
cosmology, where $D_L$ is the luminosity distance.

We use the exact same SALT2 options as \cite{SALT2} to fit the
\snia\ light curves in our sample, and only trust the result when the
following conditions are met: reduced $\chi^2_\nu \le 2$; at least one
$B$-band point before +5\,d from $B$-band maximum, and one after
+10\,d; at least 5 $B$- and $V$-band points in the age range $-15 \le t \le
+60$\,d; finally, we impose a cut on the SALT2 $x_1$ parameter, namely
$-3\le x_1 \le 2$. This last condition is equivalent to considering
\sneia\ in the range $0.8 \lesssim \Delta m_{15}(B) \lesssim 1.7$
(i.e. subluminous 1991bg-like \sneia\ are excluded). We examined all
the light-curve fits by eye to ensure they were satisfactory given
this set of conditions. Approximately 170 of the $\sim250$ \sneia\
with light curves from the CfA SN program pass these requirements.

\subsection{Training}\label{sect:training}

For estimating the coefficients of the model (training), we use a
custom version of the luminosity distance fitter {\tt
  simple\_cosfitter}\footnote{http://qold.astro.utoronto.ca/conley/simple\_cosfitter} 
(A. Conley 2009, private communication) based on the {\tt Minuit}
function minimization package \citep{James/Roos:1975}.
This code minimizes the following expression with respect to the
parameters $(\alpha,\beta,\gamma,\mathcal{M})$:
\begin{equation}\label{eqn:chisqr}
\begin{split}
\chi^2 &= \sum_{s=1}^N \frac{[m_{B,s} - m_{{\rm pred},s}(z_s;  x_1^s, c_s, \mathcal{S}_s; \alpha,\beta,\gamma,\mathcal{M})]^2}{\sigma_s^2} \\
&= \sum_{s=1}^N \frac{ [\mu(m_B^s, x_1^s, c_s, \mathcal{S}_s; \alpha,\beta,\gamma, M) - \mu(z_s)]^2}{\sigma_s^2}
\end{split}
\end{equation}
where $m_{B,s}$ is the rest-frame peak apparent $B$-band magnitude of
the $s^{\rm th}$ \snia, and $m_{{\rm pred},s}$ is the predicted peak
apparent $B$-band magnitude, given by: 
\begin{equation}\label{eqn:mpred}
m_{{\rm pred},s} = 5 \log_{10} D_L(z_s) - \alpha
x_{1,s} + \beta c_s - \gamma \mathcal{S}_s + \mathcal{M},
\end{equation}
where $D_L$ is the luminosity distance at redshift $z_s$ for a given
cosmological model described by the standard parameters
$(w,\Omega_m,\Omega_\Lambda)$.
Since our analysis only includes objects at low redshifts ($z <
0.06$), we do not solve for these parameters and simply assume a flat,
cosmological constant-dominated model with
$(w,\Omega_m,\Omega_\Lambda)=(-1,0.27,0.73)$. The $\mathcal{M}$ term
is a collection of constants including the reference $M$.

The variance $\sigma_s^2$ that appears in the denominator of
Eq.~\ref{eqn:chisqr} includes an error on the corrected magnitude
(Eq.~\ref{eqn:salt2}), using the estimation error covariance of the
light-curve parameters and spectroscopic indicators, 
a variance due to peculiar velocities [$\sigma_{{\rm pec},s}=(v_{\rm
  pec}/cz_s)(5/\ln 10)$, where we take the rms peculiar velocity
$v_{\rm pec}=300$\,\kms], and an intrinsic dispersion of
\snia\ magnitudes: 
\begin{equation}\label{eqn:magerr}
\sigma_s^2 = \sigma_{m,s}^2 + \sigma_{{\rm pec},s}^2 + \sigma_{\rm int}^2,
\end{equation}
where $\sigma_{\rm int}$ is 
adjusted iteratively until $\chi^2_\nu\approx1$ [typically
$\sigma_{\rm int}\lesssim 0.2$\,mag for the standard $(x_1,c)$
model].   For any particular model, the intrinsic variance
$\sigma_{\rm int}^2$ accounts for deviations in magnitude in the
Hubble diagram beyond that explained by measurement error or random
peculiar velocities, and hence represents a floor to how
accurately the model can predict distances. To limit 
the impact of the peculiar velocity error we restrict our analysis to
\sneia\ at redshifts $z > 0.015$ ($\sigma_{\rm pec}<0.15$\,mag).
Of the 170 \sneia\ with satisfactory SALT2 fits, 114 are at
redshifts greater than 0.015.

\subsection{Cross-validation}\label{sect:cv}

We consider several models described by Eq. \ref{eqn:salt2} that use
different subsets of the predictors
$(x_1, c, \mathcal{S}$).   If we train a model on the data
of all the SN in the sample to estimate the coefficients
$(\hat{\alpha}, \hat{\beta}, \hat{\gamma}, \hat{\mathcal{M}})$, we can
evaluate the fit of the model by computing the training error,
e.g. the mean squared distance modulus residual, $\mu(m_B^s, x_1^s,
c_s, \mathcal{S}_s; \hat{\alpha},\hat{\beta},\hat{\gamma}, \hat{M}) -
\mu(z_s)$, over all SN $s$ in the training set. 

For finite samples, the average Hubble diagram residual of the
training set SN is an optimistic estimate of the ability of the
statistical model, Eq. \ref{eqn:salt2}, to make accurate predictions
given the supernova observables.   This is because it uses the
supernova data twice: first for estimating the model parameters
(training), and second in evaluating the residual error. 
Hence, the training set residuals underestimate the prediction error,
which is the expected error in estimating the distance of a SN that
was not originally in the finite training set.
We refer to these data as ``out-of-sample''.
Furthermore, with a fixed, finite, and noisy training data set, it is
always possible to reduce the residual, or training, error of the fit
by introducing more predictors to the model.  However,
this may lead to over-fitting, in which apparently significant
predictors are found in noisy data, even though in reality there was
no trend.  These relationships are sensitive to the finite training
set and would not generalize to out-of-sample cases.   To evaluate
predictive performance and guard against over-fitting with a
statistical model based on finite data, we should estimate the
prediction error for out-of-sample cases. To do so,
we use a cross-validation (CV) procedure to evaluate the impact of
using a spectroscopic indicator $\mathcal{S}$, alone and in conjunction 
with standard light curve parameters, on the accuracy of distance
predictions in the Hubble diagram.   

Cross-validation seeks to
estimate prediction error and to test the sensitivity of the trained
statistical model to the data set by partitioning the full data set
into smaller subsets. One subset is held out for testing
predictions of the model, while its complement is used to train the
model.  This process is repeated over partitions of the full data set.
This method avoids using the same data simultaneously for training the
model and for estimating its prediction error. Cross-validation was
used before for statistical modeling of \sneia\ by \cite{Mandel/etal:2009},
who applied the .632 bootstrap method to evaluate distance predictions
for \sneia\ using near infrared light curves. A careful
implementation of a cross-validation method is particularly important
for small samples, as is the case in this paper (e.g. 26 \sneia\ at
maximum light; see \S~\ref{sect:tmaxres}).

In this paper, the cross-validation method we use is known as $K$-fold
CV. The idea is to divide our \snia\ sample into $K$ subsets, train a
given model on $K-1$ subsets, and validate it on
the remaining subset. This procedure is repeated $K$ times, at which
point all \sneia\ have been part of a validation set once. Typical
choices of $K$ are 5 or 10 (e.g.,
\citealt{Hastie/Tibshirani/Friedman:2009}). The case $K=N$, where $N$
is the number of \sneia\ in our sample, is known as ``leave-one-out''
CV. In this case, each \snia\ in turn is used as a validation set, and
the training is repeated $N$ times on $N-1$ \sneia.

In practice, we run $K$-fold CV as follows:

\begin{itemize}

\item[1.]{the sample of $N$ \sneia\ is randomly divided into $K$
  subsets of equal size (when $N$ is not a multiple of $K$, the
  number of \sneia\ between any two subsets differs by at most
  one).}

\item[2.]{Looping over each $K$ fold:
\begin{itemize}
\item[2a.]{all the \sneia\ in the $K^{\rm th}$ subset are removed
  from the sample: they form the validation set. The
  remaining \sneia\ define the training set.}
\item[2b.]{the objects in the training set are then used to
  determine the best-fit values for the parameters
  $(\hat{\alpha}_K, \hat{\beta}_K, \hat{\gamma}_K,
  \hat{\mathcal{M}}_K)$ in Eq.~\ref{eqn:chisqr}, as well as the
  intrinsic dispersion $\sigma_{\rm int}$ in Eq.~\ref{eqn:magerr}}.
\item[2c.]{using this set of parameters we predict the magnitudes
  of the \sneia\ in the validation set (indexed $j$):
\begin{equation}
m_{{\rm pred},j} = 5 \log_{10} D_L(z_j) - \alpha_K x_{1,j} + \beta_K
c_j - \gamma_K \mathcal{S}_j + \mathcal{M}_K. 
\end{equation}
The Hubble residual, or error, of the predicted distance modulus is then
\begin{equation}
\Delta \mu_j = m_{B,j} - m_{{\rm pred},j} = \mu(m_B^j, x_1^j,
c_j, \mathcal{S}_j; \alpha,\beta,\gamma, M) - \mu(z_j).
\end{equation}
}
\end{itemize}
}

\item[3.]{When the magnitude or distance of each \snia\ has been
  predicted once using the above scheme, we analyze the prediction
  errors (\S\ref{sect:stats}). 
  When doing so, we check that the set of best-fit
  $(\hat{\alpha}_K,\hat{\beta}_K,\hat{\gamma}_K,\hat{\mathcal{M}}_K)$ 
  are consistent amongst all training sets.
}

\end{itemize}

For all the spectroscopic indicators we consider in this paper, we run
$K$-fold CV with $K=2,5,10$, and $N$ to make sure our results are not
sensitive to the exact choice of $K$ (the impact on the weighted rms
of prediction Hubble residuals is $\lesssim0.002$\,mag). Moreover, we
run each $K$-fold CV 10 times to check the outcome is insensitive to
how the starting \snia\ sample is divided into $K$ subsets (the impact
on the weighted rms of prediction Hubble residuals is
$\lesssim0.003$\,mag). In what follows we report our results based on
$K=10$.

\subsection{Comparing model predictions}\label{sect:stats}

For each model, which we label by its predictors, e.g. $(x_1, c,
\mathcal{S})$, cross-validation gives us a set of prediction errors
$\{ \Delta \mu_s\}$ for each SN $s$.  To summarize the total
dispersion of predictions, we computed the weighted mean squared
error, 
\begin{equation}
\text{WRMS}^2 = \left(\sum_{s=1}^N w_s\right)^{-1} \sum_{s=1}^N w_s ~ \Delta \mu_s^2,
\end{equation}
the square root of which is the weighted rms.  We weight the
contribution from each SN by the inverse of its expected total
variance (the precision) $w_s = \sigma_s^{-2}$.  We prefer to use the
rms of the prediction residuals rather than the sample standard
deviation, since the former measures the average squared deviation of
the distance prediction from the Hubble distance $\mu(z)$, whereas the
latter measures the average squared deviation of prediction errors
from the mean prediction error.  Note that the mean squared error is
equal to the sample variance plus the square of the mean error.
Thus, the mean squared error will be larger than the sample variance
if the mean error, or bias, is significant, but the two statistics
will be the same if it is not.   Since the mean prediction error is
not guaranteed to be zero, we use the WRMS statistic to assess the
total dispersion of distance prediction errors.  We also estimate the
sampling variance of this statistic (see Appendix~\ref{app:wrms}). 

The WRMS measures the \emph{total} dispersion in the Hubble diagram.
However, we expect that some of that scatter is due to random peculiar
velocities [influencing $\mu(z)$ with variance $\sigma^2_{\rm pec}$],
and some due to measurement error ($\sigma_m^2$).
Using the cross-validated distance errors, we also estimate how
precisely we can expect a particular model to predict the distance to
a \snia\ when these other sources of error are negligible.  We call this
variance estimate the \emph{rms intrinsic prediction error}, a property of
the model itself, and label it $\sigma_{\rm pred}^2$.  Intuitively,
this is the result of subtracting from the total dispersion the
expected contributions of peculiar velocities and measurement
uncertainties. It is similar to the intrinsic variance
$\sigma_{\rm int}$ discussed in \S~\ref{sect:training}, in that it
represents a floor to how accurately the model can predict
distances. It is not strictly equivalent, however, since $\sigma_{\rm
  int}$ is adjusted during the training process so that
$\chi^2_\nu\approx1$, while $\sigma_{\rm pred}$ is estimated using the
cross-validated distance modulus prediction errors.
In Appendix~\ref{app:mle}, we describe a maximum likelihood
estimate for $\sigma_{\rm pred}$ and its standard error from the
set of distance predictions.

We are also interested in the \emph{intrinsic covariance} of the
distance prediction errors generated by two different models.  Imagine
that peculiar velocities and 
measurement error were negligible, and model $P$ and model $Q$ predict
distances to the same set of \sneia.  We calculate the prediction errors,
$\{ \Delta \mu_s^P, \Delta \mu_s^Q\}$ from each model.   There is a
positive intrinsic  covariance if $\Delta \mu_s^P$ tends to be
positive when $\Delta \mu_s^Q$ is positive, and a negative intrinsic
covariance if they tend to make errors in opposite directions.  The
intrinsic correlation is important because it suggests how useful it
would be to combine the distance predictions of two models.  If two
models tend to make prediction errors in the same direction (positive
correlation), then the combined model is not likely to do much better
than the most accurate of the two original models.  However, if two
models tend to make prediction errors that are wrong in different ways
(zero or negative correlation), then we expect to see a gain from
averaging the two models. 

Even if two models make prediction errors that are intrinsically
uncorrelated, random peculiar velocities will tend to induce a
positive correlation in the realized errors $\{ \Delta \mu_s^P, \Delta
\mu_s^Q\}$ if the methods are used on the same set of SN.  This is
because the unknown peculiar velocity for a given SN is the same
regardless of the model we use to generate its distance prediction.
Hence, the expected contribution of random peculiar velocities to the
sample covariance of predictions must be removed to estimate the
intrinsic covariance between two models.  In Appendix~\ref{app:mle}, we
describe a maximum likelihood estimator for the intrinsic covariance
and its standard error using the set of distance predictions.

We use the maximum likelihood estimation method to estimate the
intrinsic prediction error and intrinsic covariance of each model
compared to the reference model $(x_1, c)$ that uses only light curve
information.

%%%%%%%%%%%%%%%%%%%%%%%%%%%%%%%%%%%%%%%%%%%%%%%%%%%%%%%%%%%%%%%%%%
%%
%%   Data
%%
%%%%%%%%%%%%%%%%%%%%%%%%%%%%%%%%%%%%%%%%%%%%%%%%%%%%%%%%%%%%%%%%%%

\section{Spectroscopic data}\label{sect:data}

We have used a large spectroscopic data set obtained through the CfA
Supernova Program. Since 1994, we have obtained $\sim 2400$ optical
spectra of $\sim 450$ low-redshift ($z \lesssim 0.05$) \sneia\ with
the 1.5\,m Tillinghast telescope at FLWO using the FAST spectrograph
\citep{Fabricant/etal:1998}. Several spectra were published in studies
of specific supernovae (e.g., SN~1998bu; \citealt{Jha/etal:1999}),
while 432 spectra of 32 \sneia\ have recently been published by 
\cite{Matheson/etal:2008}. We also have complementary multi-band
optical photometry for a subset of $\sim 250$ \snia\
\citep{Riess/etal:1999a,Jha/etal:2006,Hicken/etal:2009a}, as well as
NIR $JHK_s$ photometry for the brighter ones
\citep{Wood-Vasey/etal:2008}. All published data are available via the
CfA Supernova
Archive\footnote{http://www.cfa.harvard.edu/supernova/SNarchive.html}.

All the spectra were obtained with the same telescope and instrument,
and reduced in a consistent manner (see \citealt{Matheson/etal:2008}
for details). The uniformity of this data set is unique and
enables an accurate estimate of our measurement errors.

%%%%%%%%%%%%%%%%%%%%%%%%%%%%%%%%%%%%%%%%%%%%%%%%%%%%%%%%%%%%%%%%%%
%%
%%   Flux ratios
%%
%%%%%%%%%%%%%%%%%%%%%%%%%%%%%%%%%%%%%%%%%%%%%%%%%%%%%%%%%%%%%%%%%%

\section{Spectral flux ratios}\label{sect:fluxratio}

\subsection{Measurements}

\begin{figure}
\centering
\resizebox{0.475\textwidth}{!}{\includegraphics{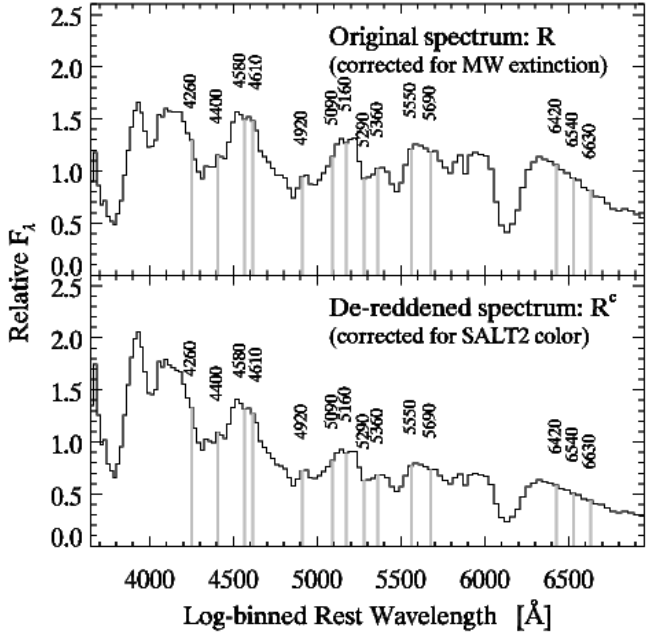}}
\caption{\label{fig:frfig}
Illustration of the flux ratio measurement. The upper panel shows the
input spectrum (de-redshifted and corrected for Galactic reddening;
here SN~1999gd around maximum light), binned on a log-wavelength
scale. The gray vertical lines represent the fluxes in characteristic
wavelength bins mentioned throughout the text. The lower panel shows
the same spectrum corrected for SALT2 color, which is used to measure
the color-corrected flux ratios $\mathcal{R}^c$.
}
\end{figure}

\cite{Bailey/etal:2009} introduced a new spectroscopic indicator,
calculated as the ratio of fluxes in two wavelength regions of a
\snia\ spectrum binned on a logarithmic wavelength scale. This ratio,
noted $\mathcal{R}(\l_X/\l_Y)=F(\l_X)/F(\l_Y)$ [$\l_X$ and $\l_Y$
being the rest-frame wavelength coordinates in \AA\ of a given bin
center], is measured on a de-redshifted spectrum corrected for
Galactic reddening using the \cite{CCM89} extinction law with
$R_V=3.1$ in combination with the dust maps of \cite{SFD98}. A
color-corrected version of this flux ratio, noted
$\mathcal{R}^c(\l_X/\l_Y)$, is measured on a spectrum additionally
corrected for the SALT2 color parameter using the color law of
\cite{SALT2}. Figure~\ref{fig:frfig} illustrates both measurements.

%%% Fig. frfig

We use the same binning as \cite{Bailey/etal:2009}, namely 134 bins
equally spaced in $\ln \l$ between 3500\,\AA\ and 8500\,\AA\ (rest
frame), although most of the CfA spectra used here do not extend
beyond $\sim7100$\,\AA\ (see \S~\ref{sect:tmaxres}). The resulting
$\sim2000$\,\kms\ bin size is significantly less than the typical
width of a \snia\ feature ($\sim10000$\,\kms). The error on 
$\mathcal{R}$ includes a flux error (from the corresponding variance
spectrum), an error due to the relative flux calibration accuracy (see
\S~\ref{sect:frel}), and an error due to the SALT2 color
precision. When there are several spectra of a given \snia\ within
$\Delta t=2.5$\,d of the age we consider (see \S~\ref{sect:tmaxres}
for spectra at maximum light; \S~\ref{sect:othertres} for spectra at
other ages), we use the error-weighted mean and standard deviation of
all flux ratios as our measurement and error,
respectively. \cite{Bailey/etal:2009} also chose $\Delta t=2.5$\,d in
their analysis, and we find that increasing $\Delta t$ worsens the
results while decreasing it leads to too small a sample.

\cite{Bailey/etal:2009} cross-checked the results
for their best single flux ratio $\mathcal{R}(6420/4430)$ using the
the sample of \snia\ spectra published by \cite{Matheson/etal:2008}
[and available through the CfA SN Archive].
We checked the validity of our flux ratio measurements by comparing
the values of $\mathcal{R}(6420/4430)$ in the
\cite{Matheson/etal:2008} sample with those reported in Table 2 of
\cite{Bailey/etal:2009}. In all cases, our measurements agree well
within the $1\sigma$ errors. This also holds for SN~1998bu,
accidentally removed from the \cite{Matheson/etal:2008} sample by
\cite{Bailey/etal:2009} [H. Fakhouri 2010, private communication].
We note that we were unable to cross-check the flux ratio
measurements of \cite{Bailey/etal:2009} in a similar fashion, since
none of their 58 \snia\ spectra are publicly available.

\subsection{Impact of relative flux calibration and SALT2 color}\label{sect:frel}

\begin{figure}
\centering
\resizebox{0.475\textwidth}{!}{\includegraphics{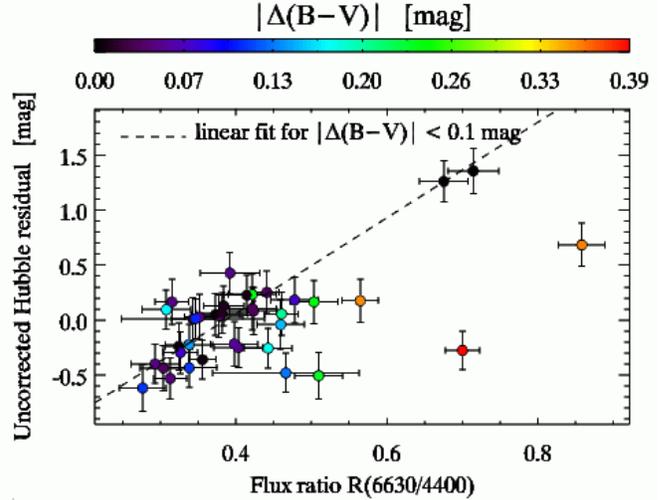}}
\caption{\label{fig:frdbmv}
Uncorrected Hubble residual {\it vs.} flux ratio
$\mathcal{R}(6630/4400)$ at maximum light, color-coded according to
the absolute difference in $B-V$ color derived from the spectrum and
that derived from the photometry, noted $|\Delta(B-V)|$. 
The dashed line is a linear fit to the \sneia\ with $|\Delta(B-V)| <
0.1$\,mag. The highly-reddened SN~2006br is not shown here.
}
\end{figure}

When the $\l_X$ and $\l_Y$ wavelength bins have a large separation
($\gtrsim1000$\,\AA), $\mathcal{R}(\l_X/\l_Y)$ is essentially a color
measurement. We therefore expect flux ratios to be sensitive to the
relative flux calibration accuracy of the
spectra. Fig.~\ref{fig:frdbmv} shows the relation between uncorrected
Hubble residuals [i.e. $m_B-M-\mu(z)$] and
our most highly-ranked flux ratio $\mathcal{R}(6630/4400)$ at maximum
light (see \S~\ref{sect:tmaxres}). There is one data point per \snia,
color-coded according to the absolute difference in $B-V$ color at
maximum light derived from the spectrum and that derived from the
photometry, noted $|\Delta(B-V)|$, which we use as a proxy for
relative flux calibration accuracy. The bulk of the sample defines a
highly correlated relation ({\it dashed line}), with several outliers
all having $|\Delta(B-V)| \ge 0.1$\,mag. We therefore restrict our
analysis to \sneia\ with spectra that have a relative flux
calibration better than 0.1\,mag. 

%%% Fig. Uncorrected hubres vs. R(6630/4400) colorcode=dBmV 

\begin{figure*}
\centering
\resizebox{\textwidth}{!}{\includegraphics{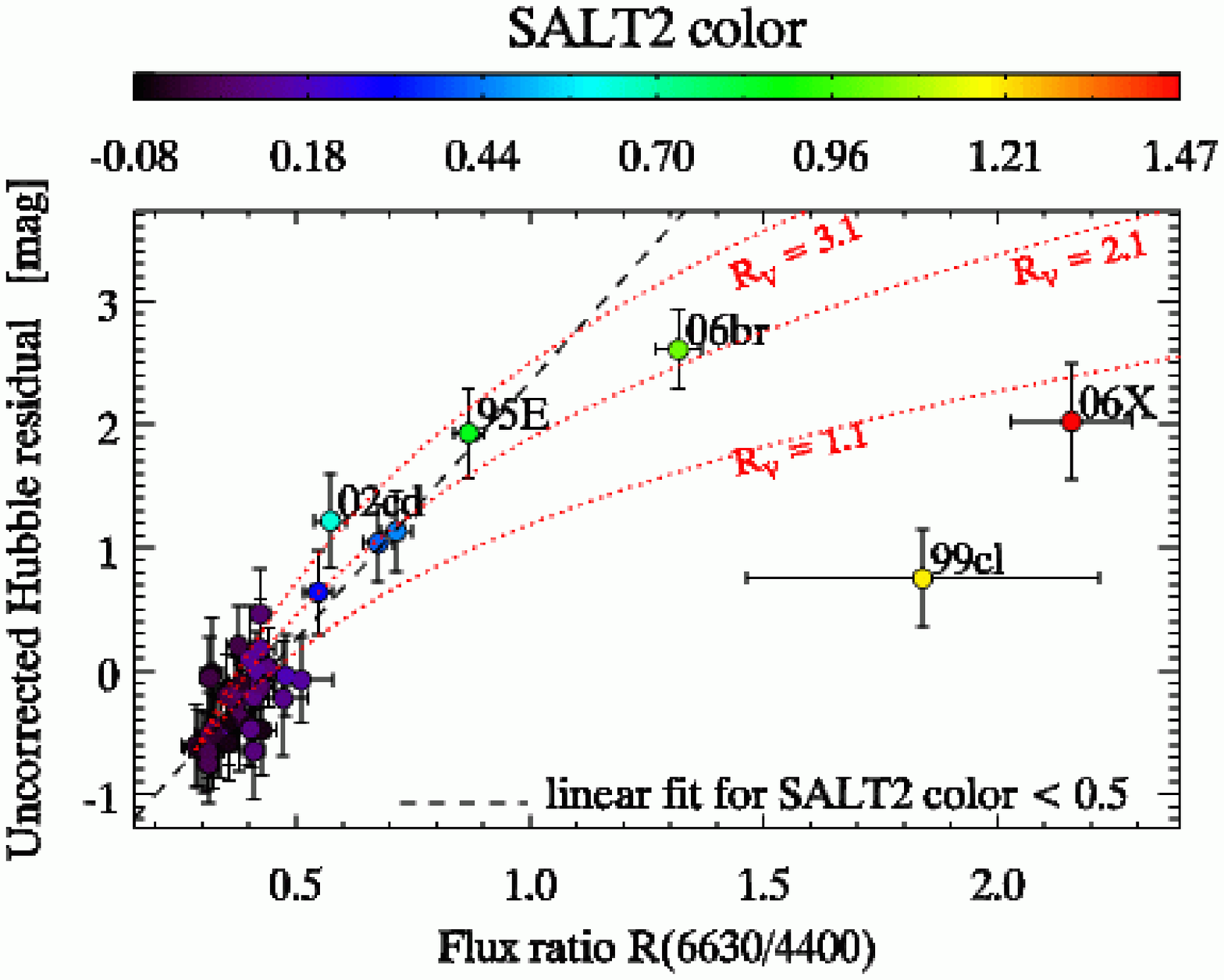}\hspace{1.5cm}\includegraphics{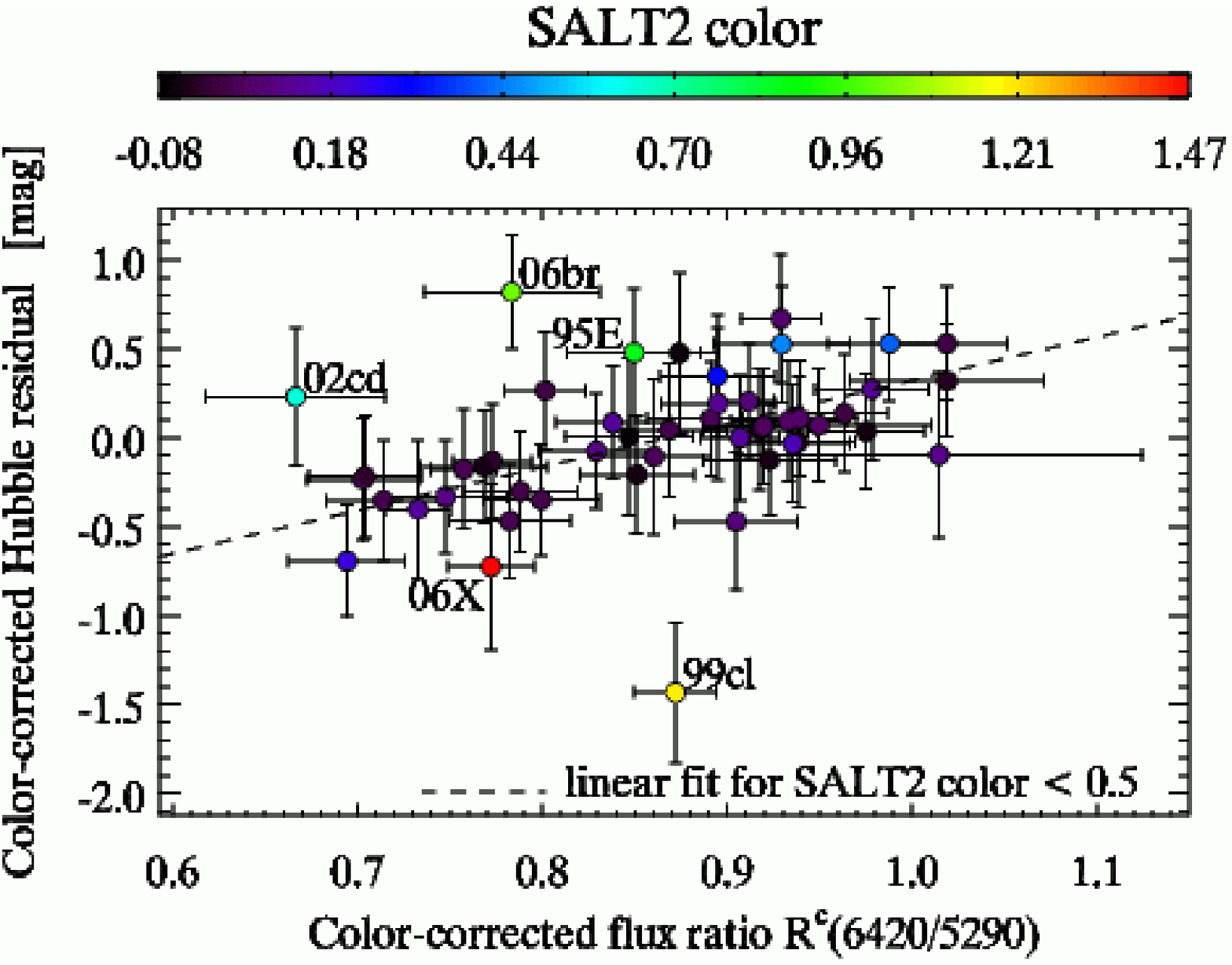}}
\caption{\label{fig:frredcol}
{\it Left:} Uncorrected Hubble residual {\it vs.} flux ratio
$\mathcal{R}(6630/4400)$ at maximum light for \sneia\ at $z>0.005$
with $|\Delta(B-V)| < 0.1$\,mag, color-coded according to the SALT2
color parameter, $c$. Points corresponding to \sneia\ with $c>0.5$ are
labeled. The dashed line is a linear fit to the \sneia\ with
$c<0.5$. The dotted lines are reddening curves for different values of
$R_V$, normalized to the smallest $\mathcal{R}(6630/4400)$ value. 
{\it Right:} Color-corrected Hubble residual  {\it vs.}
color-corrected flux ratio $\mathcal{R}^c(6420/5290)$ at maximum light.
}
\end{figure*}

\begin{figure}
\centering
\resizebox{0.475\textwidth}{!}{\includegraphics{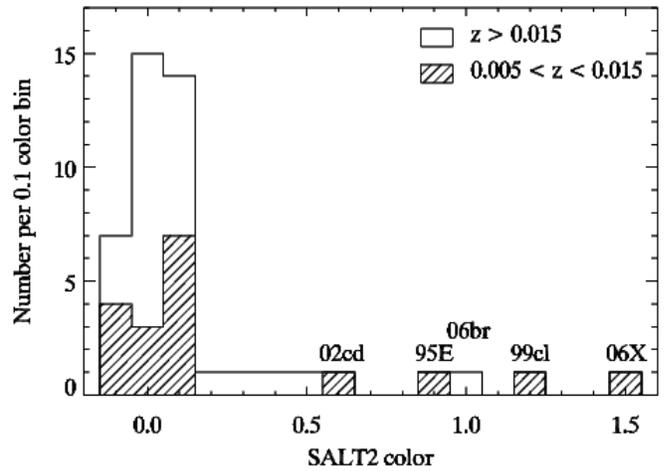}}
\caption{\label{fig:colorhist}
Histogram of the SALT2 color parameter ($c$) for \sneia\ at $z >
0.015$ ({\it open}) and $0.005<z<0.015$ ({\it hatched}). Bins that
include \sneia\ with $c>0.5$ are labeled.
}
\end{figure}

\cite{Bailey/etal:2009} noted that the highly-reddened SN~1999cl was a
large outlier in their analysis, and attributed this to the
non-standard nature of the extinction towards this SN ($R_V\approx1.5$;
\citealt{Krisciunas/etal:2006}). To explore the effects of reddening, in
Fig.~\ref{fig:frredcol} ({\it left}), we show the relation between
uncorrected Hubble residual and $\mathcal{R}(6630/4400)$ at maximum
light, for \sneia\ at redshifts $z>0.005$ that satisfy our requirement
on the relative flux calibration accuracy. Using this lower redshift
bound has the effect of including several highly-reddened \sneia\ (including
SN~1999cl; see Fig.~\ref{fig:colorhist}), which are otherwise
excluded based on the redshift cut we use elsewhere this paper
($z>0.015$). For \sneia\ with $c<0.5$, $\mathcal{R}(6630/4400)$ is
highly correlated with uncorrected Hubble residuals, but those with
red colors ($c>0.5$) tend to deviate significantly from this relation
({\it dashed line}; this is not the case for SN~1995E, for which
  $c\approx0.9$), 
the two largest outliers corresponding to the
reddest \sneia\ (SN~1999cl and SN~2006X). Both are subject to high
extinction by non-standard dust in their respective host galaxies
($A_V\approx2$\,mag for $R_V\approx1.5$;
\citealt{Krisciunas/etal:2006,WangX/etal:2008a}) and display
time-variable \naoned\ absorption, whose circumstellar or interstellar
origin is still debated
\citep{Patat/etal:2007a,Blondin/etal:2009}. The reddening curves in
Fig.~\ref{fig:frredcol} ({\it dotted lines}) seem to corroborate the 
fact that the nonlinear increase of flux ratios at high values of the
SALT2 color parameter is mainly due to reddening by dust with low
$R_V$. Nonetheless, SN~1999cl still stands out in this respect as it
would require a value of $R_V\lesssim0.5$ inconsistent with that found
by \cite{Krisciunas/etal:2006}. Moreover, while we obtain
  consistent $R_V$ estimates for SN~2006X using other flux ratios, this
is not the case for SN~2006br, for which some flux ratios are
consistent with $R_V=3.1$.

%%% Fig. Uncorrected hubres vs. R(6630/4400) colorcode=SALT2 col 
%%% Fig. color-corrected hubres vs. Rc(6420/5290) colorcode=SALT2 col 

%%% Fig. SALT2 color distribution for z > 0.005

The right panel of Fig.~\ref{fig:frredcol} shows the relation between
color-corrected Hubble residual [i.e. $m_B-M-\beta c -\mu(z)$]
and our most highly-ranked color-corrected flux ratio
$\mathcal{R}^c(6420/5290)$ at maximum light 
(see \S~\ref{sect:tmaxres}) for the same sample. \sneia\ with a SALT2
color $c>0.5$ are again outliers. As noted by
\citealt{Bailey/etal:2009}, this shows that a single color parameter
cannot encompass the variety of \snia\ intrinsic colors and extinction
by non-standard dust. We therefore impose a cut on SALT2 color
in our analysis, only considering \sneia\ with $c<0.5$. Four of the
five SNe with $c>0.5$ in Fig.~\ref{fig:frredcol} are rejected anyway
based on our redshift cut. The remaining one, SN~2006br, is
then rejected based on our color cut.

\subsection{Results on Maximum-light Spectra}\label{sect:tmaxres}

\subsubsection{Selecting the best flux ratios}\label{sect:bestfr}

After selecting \sneia\ that satisfy both requirements on relative
flux calibration accuracy and SALT2 color parameter, we are left with
26 \sneia\ at $z>0.015$ with spectra within $\Delta t=2.5$\,d from maximum
light (see Table~\ref{tab:sninfo}, where we also present selected flux
ratio measurements). The spectra show no sign of
significant contamination by host-galaxy light, which can also bias
the flux ratio measurements. We make no cut based on the
signal-to-noise ratio (S/N) of our spectra, as they are generally well
in excess of 100 per log-wavelength bin. We only consider flux ratios
for wavelength bins represented in all the spectra. This leads to 98
bins between $\sim 3690$\,\AA\ and $\sim 7060$\,\AA, i.e. 9506
independent flux ratios.

\begin{table*}
\scriptsize
\caption{\snia\ sample for flux ratio measurements at maximum light}\label{tab:sninfo}
\begin{tabular}{lccrrccccc}
\hline\hline
SN & $z_{\rm CMB}$ & $m_B$ & \multicolumn{1}{c}{$x_1$} & \multicolumn{1}{c}{$c$} & $\mathcal{R}(6630/4400)$ & $\mathcal{R}^c(6420/5290)$ & $\mathcal{R}^c(5690/5360)$ & $\mathcal{R}^c(5160/5290)$ & $\mathcal{R}^c(5690/5550)$ \\
\hline
1998V  & 0.0170 & 15.085 (0.020) & $-0.435$ (0.161) & $ 0.031$ (0.015) & 0.330 (0.005) & 0.744 (0.006) & 1.045 (0.004) & 1.485 (0.006) & 0.933 (0.004) \\
1998dx & 0.0539 & 17.536 (0.037) & $-1.890$ (0.457) & $-0.051$ (0.027) & 0.365 (0.018) & 0.949 (0.027) & 1.086 (0.024) & 1.445 (0.033) & 0.946 (0.018) \\
1998eg & 0.0237 & 16.096 (0.016) & $-0.956$ (0.366) & $ 0.048$ (0.019) & 0.378 (0.023) & 0.920 (0.022) & 1.070 (0.011) & 1.478 (0.013) & 0.995 (0.007) \\
1999aa & 0.0152 & 14.698 (0.009) & $ 0.896$ (0.073) & $-0.019$ (0.009) & 0.286 (0.027) & 0.704 (0.025) & 1.054 (0.012) & 1.448 (0.011) & 0.994 (0.006) \\
1999cc & 0.0316 & 16.760 (0.010) & $-1.891$ (0.175) & $ 0.057$ (0.012) & 0.406 (0.013) & 0.994 (0.017) & 1.001 (0.015) & 1.513 (0.019) & 0.850 (0.011) \\
1999ek & 0.0176 & 15.587 (0.009) & $-1.075$ (0.127) & $ 0.163$ (0.010) & 0.478 (0.039) & 0.936 (0.036) & 1.091 (0.016) & 1.461 (0.014) & 0.952 (0.007) \\
1999gd & 0.0191 & 16.940 (0.022) & $-1.210$ (0.193) & $ 0.455$ (0.022) & 0.714 (0.013) & 0.929 (0.016) & 1.081 (0.024) & 1.455 (0.020) & 1.028 (0.015) \\
2000dk & 0.0165 & 15.347 (0.021) & $-2.658$ (0.301) & $ 0.055$ (0.022) & 0.423 (0.017) & 0.933 (0.017) & 1.038 (0.009) & 1.425 (0.010) & 1.006 (0.006) \\
2000fa & 0.0218 & 15.883 (0.023) & $ 0.311$ (0.127) & $ 0.100$ (0.018) & 0.410 (0.046) & 0.829 (0.042) & 1.082 (0.020) & 1.459 (0.017) & 0.902 (0.009) \\
2001eh & 0.0363 & 16.575 (0.018) & $ 1.457$ (0.222) & $ 0.020$ (0.017) & 0.312 (0.047) & 0.799 (0.043) & 1.040 (0.020) & 1.451 (0.019) & 0.931 (0.010) \\
2002ck & 0.0302 & 16.303 (0.048) & $-0.183$ (0.147) & $-0.017$ (0.023) & 0.348 (0.046) & 0.752 (0.042) & 1.030 (0.020) & 1.463 (0.017) & 0.945 (0.010) \\
2002hd & 0.0360 & 16.738 (0.038) & $-0.748$ (0.456) & $ 0.100$ (0.022) & 0.403 (0.027) & 0.748 (0.027) & 0.955 (0.018) & 1.378 (0.020) & 0.899 (0.013) \\
2002hu & 0.0359 & 16.587 (0.012) & $ 0.052$ (0.143) & $-0.052$ (0.012) & 0.293 (0.027) & 0.768 (0.029) & 1.048 (0.021) & 1.509 (0.029) & 0.984 (0.017) \\
2002jy & 0.0187 & 15.702 (0.019) & $ 0.660$ (0.212) & $ 0.013$ (0.015) & 0.305 (0.015) & 0.795 (0.015) & 1.073 (0.010) & 1.566 (0.012) & 0.963 (0.007) \\
2002kf & 0.0195 & 15.654 (0.033) & $-1.493$ (0.189) & $ 0.009$ (0.023) & 0.361 (0.035) & 0.953 (0.035) & 1.132 (0.024) & 1.550 (0.024) & 0.969 (0.014) \\
2003U  & 0.0279 & 16.471 (0.046) & $-2.536$ (0.558) & $ 0.033$ (0.035) & 0.373 (0.009) & 0.891 (0.016) & 1.048 (0.016) & 1.445 (0.025) & 0.965 (0.014) \\
2003ch & 0.0256 & 16.659 (0.022) & $-1.655$ (0.297) & $ 0.012$ (0.019) & 0.402 (0.025) & 1.024 (0.024) & 1.218 (0.010) & 1.570 (0.009) & 1.043 (0.005) \\
2003it & 0.0240 & 16.342 (0.028) & $-1.815$ (0.359) & $ 0.084$ (0.029) & 0.432 (0.011) & 0.918 (0.011) & 1.087 (0.007) & 1.456 (0.009) & 0.950 (0.005) \\
2003iv & 0.0335 & 16.961 (0.026) & $-2.473$ (0.486) & $-0.031$ (0.028) & 0.420 (0.005) & 1.092 (0.011) & 1.081 (0.011) & 1.467 (0.015) & 0.940 (0.009) \\
2004as & 0.0321 & 16.956 (0.018) & $-0.017$ (0.206) & $ 0.128$ (0.016) & 0.415 (0.006) & 0.838 (0.008) & 1.065 (0.009) & 1.573 (0.011) & 0.975 (0.007) \\
2005ki & 0.0208 & 15.551 (0.029) & $-2.123$ (0.153) & $-0.059$ (0.026) & 0.398 (0.032) & 0.975 (0.030) & 1.091 (0.015) & 1.441 (0.015) & 0.957 (0.009) \\
2006ax & 0.0180 & 15.010 (0.010) & $-0.062$ (0.062) & $-0.049$ (0.009) & 0.304 (0.019) & 0.851 (0.018) & 1.082 (0.008) & 1.558 (0.009) & 1.009 (0.005) \\
2006gj & 0.0277 & 17.668 (0.033) & $-2.073$ (0.280) & $ 0.409$ (0.023) & 0.675 (0.008) & 0.988 (0.012) & 1.011 (0.011) & 1.377 (0.014) & 0.930 (0.009) \\
2006sr & 0.0232 & 16.126 (0.017) & $-1.754$ (0.220) & $ 0.060$ (0.015) & 0.422 (0.019) & 0.939 (0.018) & 1.031 (0.010) & 1.474 (0.010) & 0.913 (0.006) \\
2007ca & 0.0151 & 15.933 (0.013) & $ 0.289$ (0.122) & $ 0.305$ (0.012) & 0.547 (0.028) & 0.895 (0.026) & 1.097 (0.012) & 1.569 (0.012) & 0.986 (0.006) \\
2008bf & 0.0257 & 15.703 (0.010) & $ 0.097$ (0.095) & $ 0.031$ (0.010) & 0.315 (0.038) & 0.806 (0.035) & 1.047 (0.016) & 1.526 (0.013) & 0.924 (0.007) \\
\hline
\end{tabular}
\tablefoot{Spectra within 2.5\,d from maximum light for these \sneia\
 are available via the CfA Supernova Archive.}
\end{table*}

We run the $K$-fold cross-validation procedure outlined in
\S~\ref{sect:cv}, and consider the five models for
estimating distances to \sneia\ described in \S~\ref{sect:intro}:
\begin{eqnarray}
\mu &=& m_B - M + \gamma \mathcal{R} \label{eqn:frarr1}\\
\mu &=& m_B - M + \alpha x_1 + \gamma \mathcal{R} \label{eqn:frarr2}\\
\mu &=& m_B - M - \beta c + \gamma \mathcal{R}^c \label{eqn:frarr3}\\
\mu &=& m_B - M + \alpha x_1 - \beta c + \gamma \mathcal{R}^c \label{eqn:frarr4}\\
\mu &=& m_B - M + \alpha x_1 - \beta c.\label{eqn:frarr5}
\end{eqnarray}
When no color correction is involved
(Eqs.~\ref{eqn:frarr1}-\ref{eqn:frarr2}), we use the uncorrected flux
ratio $\mathcal{R}$. When a color correction is involved (the $-\beta
c$ term in Eqs.~\ref{eqn:frarr3}-\ref{eqn:frarr4}), we use the
color-corrected version of the flux ratio $\mathcal{R}^c$. Using
$\mathcal{R}$ in combination with color, or $\mathcal{R}^c$ alone or
in combination with $x_1$, severely degrades the predictive power of
the model, so we do not report results using $(c,\mathcal{R})$;
$\mathcal{R}^c$ alone; or $(x_1,\mathcal{R}^c)$.

\begin{figure}
\centering
\resizebox{0.475\textwidth}{!}{\includegraphics{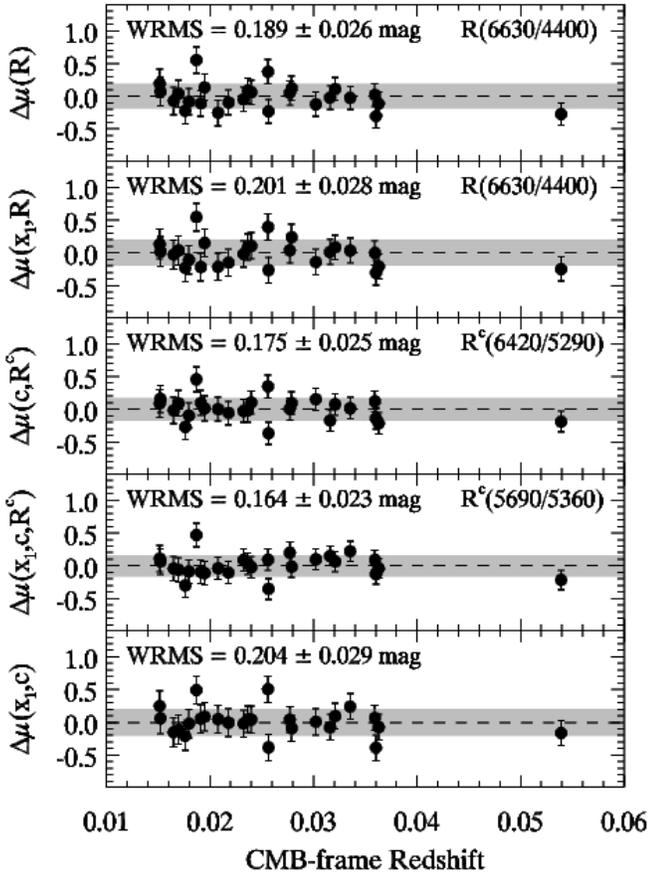}}
\caption{\label{fig:frtmaxhubdiag}
Hubble diagram residuals for the highest-ranked flux ratios at maximum
light. 
{\it From top to bottom:} prediction residuals using $\mathcal{R}$
only; $(\mathcal{R},x_1)$; $(\mathcal{R}^c,c)$;
$(\mathcal{R}^c,x_1,c)$; and using the standard SALT2 fit
parameters $(x_1,c)$. In each case we indicate the weighted rms of prediction
Hubble residuals ({\it gray highlighted region}).
}
\end{figure}

We rank the flux ratios in each case based on the intrinsic prediction
error ($\sigma_{\rm pred}$; see \S~\ref{sect:stats}), but note that
ranking based on the weighted rms of prediction Hubble residuals makes
almost no difference. The results for the top five
flux ratios are displayed in Table~\ref{tab:kfoldcv_10_p0}.
We also report the best-fit $\gamma$, the weighted rms of prediction
Hubble residuals (WRMS), the intrinsic correlation of residuals with
those found using the standard $(x_1,c)$ predictors (noted $\rho_{x_1,c}$;
see \S~\ref{sect:stats}), and the difference in intrinsic prediction
error with respect to the standard $(x_1,c)$ model, noted
$\Delta_{x_1,c}$. Since we 
compute the error on $\Delta_{x_1,c}$ (see Appendix~\ref{app:mle}), we
also report the significance of this difference with respect to the
standard $(x_1,c)$ predictors. This is a direct measure of whether a
particular model predicts more accurate distances to \sneia\ when
compared to the standard approach, and if so how significant is the
improvement. Fig.~\ref{fig:frtmaxhubdiag} shows the resulting Hubble
diagram residuals {\it vs.} redshift for the best flux ratio in each
of the four models given by Eqs.~\ref{eqn:frarr1}-\ref{eqn:frarr4},
and using the standard $(x_1,c)$ predictors.

\begin{table*}
\small
\caption{Top 5 flux ratios at maximum light from 10-fold CV on 26 \sneia}\label{tab:kfoldcv_10_p0}
\begin{tabular}{cccrccrr}
\hline\hline
Rank & $\lambda_X$ & $\lambda_Y$ & \multicolumn{1}{c}{$\gamma$} & WRMS & $\sigma_{\rm pred}$ & \multicolumn{1}{c}{$\rho_{x_1,c}$} & \multicolumn{1}{c}{$\Delta_{x_1,c}$} \\
\hline
\multicolumn{8}{l}{$\mathcal{R}$} \\
\hline
 1 & 6630 & 4400 & $-4.37 \pm 0.09$ & $0.189 \pm 0.026$ & $0.163 \pm 0.030$ & $ 0.80 \pm 0.09$ & $-0.018 \pm 0.025\ (0.7\sigma)$ \\
 2 & 6630 & 4430 & $-4.44 \pm 0.12$ & $0.191 \pm 0.027$ & $0.166 \pm 0.030$ & $ 0.60 \pm 0.15$ & $-0.015 \pm 0.033\ (0.5\sigma)$ \\
 3 & 6630 & 4670 & $-5.16 \pm 0.11$ & $0.197 \pm 0.027$ & $0.171 \pm 0.031$ & $ 0.57 \pm 0.17$ & $-0.007 \pm 0.032\ (0.2\sigma)$ \\
 4 & 6900 & 4460 & $-5.71 \pm 0.12$ & $0.196 \pm 0.027$ & $0.171 \pm 0.031$ & $ 0.54 \pm 0.17$ & $-0.009 \pm 0.036\ (0.2\sigma)$ \\
 5 & 6420 & 4430 & $-3.40 \pm 0.10$ & $0.196 \pm 0.028$ & $0.173 \pm 0.031$ & $ 0.56 \pm 0.16$ & $-0.010 \pm 0.033\ (0.3\sigma)$ \\
\hline
\multicolumn{8}{l}{$(x_1,\mathcal{R})$} \\
\hline
 1 & 6630 & 4400 & $-4.51 \pm 0.15$ & $0.201 \pm 0.028$ & $0.176 \pm 0.032$ & $ 0.74 \pm 0.10$ & $-0.006 \pm 0.028\ (0.2\sigma)$ \\
 2 & 6630 & 4430 & $-4.57 \pm 0.20$ & $0.204 \pm 0.029$ & $0.180 \pm 0.032$ & $ 0.55 \pm 0.17$ & $-0.001 \pm 0.033\ (0.0\sigma)$ \\
 3 & 6900 & 4460 & $-5.81 \pm 0.16$ & $0.206 \pm 0.029$ & $0.182 \pm 0.032$ & $ 0.52 \pm 0.17$ & $ 0.002 \pm 0.032\ (0.1\sigma)$ \\
 4 & 6900 & 4370 & $-4.46 \pm 0.04$ & $0.206 \pm 0.029$ & $0.182 \pm 0.032$ & $ 0.69 \pm 0.13$ & $ 0.001 \pm 0.030\ (0.0\sigma)$ \\
 5 & 6990 & 4370 & $-4.95 \pm 0.05$ & $0.207 \pm 0.029$ & $0.183 \pm 0.032$ & $ 0.61 \pm 0.15$ & $ 0.003 \pm 0.035\ (0.1\sigma)$ \\
\hline
\multicolumn{8}{l}{$(c,\mathcal{R}^c)$} \\
\hline
 1 & 6420 & 5290 & $-1.75 \pm 0.10$ & $0.175 \pm 0.025$ & $0.148 \pm 0.029$ & $ 0.80 \pm 0.09$ & $-0.032 \pm 0.023\ (1.4\sigma)$ \\
 2 & 6630 & 4890 & $-2.19 \pm 0.14$ & $0.181 \pm 0.024$ & $0.148 \pm 0.031$ & $ 0.81 \pm 0.08$ & $-0.023 \pm 0.022\ (1.0\sigma)$ \\
 3 & 4890 & 6630 & $ 0.73 \pm 0.06$ & $0.182 \pm 0.024$ & $0.148 \pm 0.031$ & $ 0.83 \pm 0.08$ & $-0.023 \pm 0.023\ (1.0\sigma)$ \\
 4 & 4890 & 6810 & $ 0.60 \pm 0.05$ & $0.180 \pm 0.024$ & $0.148 \pm 0.030$ & $ 0.73 \pm 0.11$ & $-0.026 \pm 0.026\ (1.0\sigma)$ \\
 5 & 6540 & 4890 & $-1.84 \pm 0.12$ & $0.179 \pm 0.024$ & $0.149 \pm 0.030$ & $ 0.88 \pm 0.05$ & $-0.026 \pm 0.020\ (1.3\sigma)$ \\
\hline
\multicolumn{8}{l}{$(x_1,c,\mathcal{R}^c)$} \\
\hline
 1 & 5690 & 5360 & $-2.78 \pm 0.20$ & $0.164 \pm 0.023$ & $0.134 \pm 0.028$ & $ 0.69 \pm 0.13$ & $-0.044 \pm 0.028\ (1.6\sigma)$ \\
 2 & 5360 & 5690 & $ 3.23 \pm 0.25$ & $0.167 \pm 0.023$ & $0.137 \pm 0.028$ & $ 0.69 \pm 0.13$ & $-0.042 \pm 0.028\ (1.5\sigma)$ \\
 3 & 5660 & 5290 & $-1.95 \pm 0.24$ & $0.173 \pm 0.024$ & $0.142 \pm 0.029$ & $ 0.72 \pm 0.11$ & $-0.038 \pm 0.029\ (1.3\sigma)$ \\
 4 & 5690 & 5290 & $-1.67 \pm 0.21$ & $0.171 \pm 0.024$ & $0.142 \pm 0.028$ & $ 0.75 \pm 0.11$ & $-0.037 \pm 0.022\ (1.7\sigma)$ \\
 5 & 5290 & 5660 & $ 2.75 \pm 0.33$ & $0.174 \pm 0.024$ & $0.144 \pm 0.029$ & $ 0.73 \pm 0.11$ & $-0.036 \pm 0.028\ (1.3\sigma)$ \\
\hline
\multicolumn{8}{l}{$(x_1,c)$} \\
\hline
$\cdots$ & $\cdots$ & $\cdots$ & \multicolumn{1}{c}{$\cdots$}& $0.204 \pm 0.029$ & $0.181 \pm 0.032$ & \multicolumn{1}{c}{$\cdots$} & \multicolumn{1}{c}{$\cdots$} \\
\hline
\end{tabular}
\tablefoot{
WRMS is the error-weighted rms of prediction Hubble 
residuals (in magnitudes; see Appendix~\ref{app:wrms}); $\sigma_{\rm
  pred}$ is the intrinsic prediction error (in magnitudes);
$\rho_{x_1,c}$ is the intrinsic correlation in prediction error with
the standard $(x_1,c)$ predictors; last, $\Delta_{x_1,c}$ is the
difference in intrinsic prediction error with respect to $(x_1,c)$
[see Appendix~\ref{app:mle}].
}
\end{table*}

%%% Fig. Hubble diagram for Tmax kfoldcv_10 results

All the flux ratios listed in Table~\ref{tab:kfoldcv_10_p0} lead to an
improvement over the standard $(x_1,c)$ correction
(i.e. $\Delta_{x_1,c}<0$), as found by
\cite{Bailey/etal:2009}, but the significance is low: $<1\sigma$ for
$\mathcal{R}$ only; $\ll 1\sigma$ for $(x_1,\mathcal{R})$;
$\sim1.5\sigma$ for $(c,\mathcal{R}^c)$ and  $(x_1,c,\mathcal{R}^c)$.
This is in part due to the small number of \sneia\ in our sample.
Note that $\rho_{x_1,c}>0.5$ in all cases, i.e. the models that
include a flux ratio tend to make prediction errors in the same
direction as $(x_1,c)$, and we do not expect to gain much by
combining these models.

Using best single flux ratio $\mathcal{R}(6630/4400)$ by itself
reduces the weighted rms of prediction residuals (as well as the
intrinsic prediction error, $\sigma_{\rm pred}$) by $\lesssim
10$\% when compared with $(x_1,c)$ [${\rm
 WRMS}=0.189\pm0.026$\,mag cf. $0.204\pm0.029$\,mag], although
as noted above the significance of the difference in intrinsic
prediction error is negligible ($\Delta_{x_1,c}=-0.018\pm0.025$\,mag,
or $0.7\sigma$).

Using $\mathcal{R}$ in combination with $x_1$ leads to no improvement
over using $\mathcal{R}$ alone (although this is not reported by
\citealt{Bailey/etal:2009}, it is consistent with their findings;
S. Bailey 2009, private communication), and even leads to
systematically worse results. Our best single flux ratio
$\mathcal{R}(6630/4400)$ yields a difference in intrinsic prediction
error with respect to $(x_1,c)$ of $\Delta_{x_1,c}=-0.018\pm0.025$\,mag,
when used on its own, while it yields
$\Delta_{x_1,c}=-0.006\pm0.028$\,mag when combined with 
$x_1$. These differences are statistically indistinguishable from one
another given the size of the error on $\Delta_{x_1,c}$, but they are
systematic regardless of the flux ratio we consider.

This seems counter-intuitive, as one might expect that
including an additional predictor would result in more accurate
distance predictions.  However, this is not necessarily the case under
cross-validation.  The reason is that $x_1$ by itself is a poor
predictor of Hubble residuals, and one does not gain anything by
combining it with $\mathcal{R}(6630/4400)$. This is not surprising, as
the relation between light-curve width and luminosity is only valid if
the \sneia\ are corrected for color or extinction by dust beforehand.
In fact, $\mathcal{R}(6630/4400)$ by 
itself accounts for most of the variation in Hubble residuals. When we
cross-validate, the extra coefficient $\alpha$ will tend to fit some
noise in a given training set, and this relation will not generalize
to the validation set. This results in an increase in prediction error
because the added information is not useful. We see from
Table~\ref{tab:kfoldcv_10_p0} that adding $x_1$ affects the best-fit
value for $\gamma$ [$\gamma=-4.51\pm0.15$ cf. $-4.37\pm0.09$
for $\mathcal{R}(6630/4400)$ only]; moreover, we obtain
$\alpha\lesssim0$ when using $[x_1,\mathcal{R}(6630/4400)]$ where
$\alpha\approx0.15$ when using $(x_1,c)$, which again shows that
$\alpha$ is fitting noise when $\mathcal{R}(6630/4400)$ is combined
with $x_1$. This illustrates the advantage of using cross-validation
in guarding against over-fitting noise as more parameters and 
potential predictors are added.

\begin{figure}
\centering
\resizebox{0.475\textwidth}{!}{\includegraphics{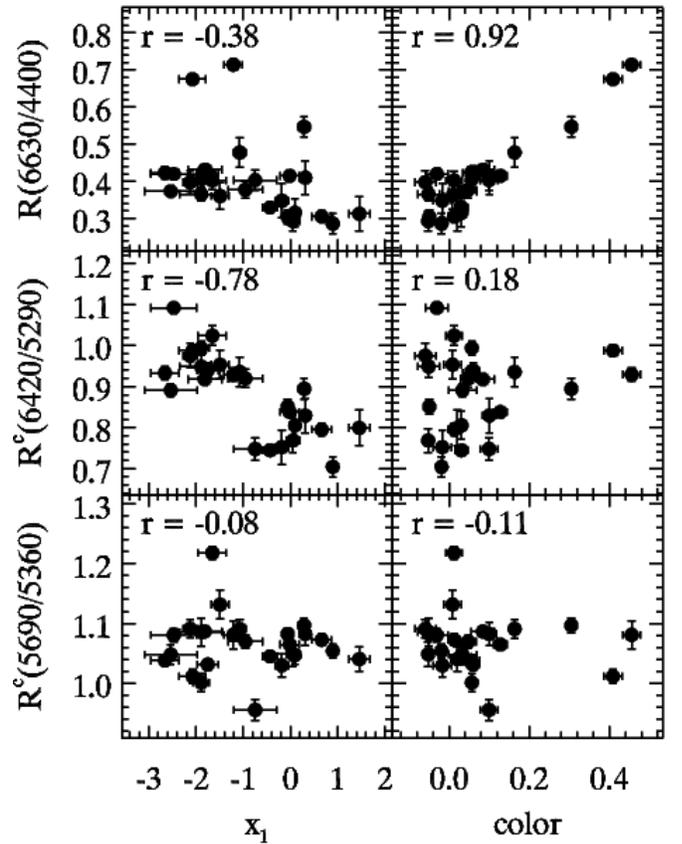}}
\caption{\label{fig:rrcx1c}
Correlation between the highest-ranked $(\mathcal{R},\mathcal{R}^c)$
at maximum light and the SALT2 fit parameters $(x_1,c)$.
}
\end{figure}

Figure~\ref{fig:rrcx1c} ({\it upper panel}) shows why
$\mathcal{R}(6630/4400)$ alone is a good predictor of Hubble
residuals. Its strong correlation with SALT2 color (Pearson
correlation coefficient $r=0.92$) shows that this ratio is essentially
a color measurement. The correlation with $x_1$ is less pronounced
($r=-0.38$), but this is largely due to a small number of outliers:
removing the three largest outliers results in a Pearson correlation
coefficient $r=-0.65$. The flux ratio $\mathcal{R}(6630/4400)$ by
itself is thus as useful a predictor as $x_1$ and $c$ combined.

The relation between $\mathcal{R}(6630/4400)$ and $x_1$ is not linear,
but it is certainly true that \sneia\ with higher $x_1$ (i.e. broader
light curves) tend to have lower $\mathcal{R}(6630/4400)$ [the same is
true for $\mathcal{R}(6420/4430)$, the highest-ranked flux ratio by
\citealt{Bailey/etal:2009}]. Since the width of the lightcurve is a
parameter intrinsic to each \snia\ (although its measurement can be subtly
affected by host-galaxy reddening; see \citealt{Phillips/etal:1999}),
the correlation between $x_1$ and $\mathcal{R}(6630/4400)$ shows that
the color variation measured by $\mathcal{R}(6630/4400)$ is intrinsic 
in part. This is consistent with the so-called ``brighter-bluer''
relation of \cite{Tripp:1998}: overluminous \sneia\ are  intrinsically
bluer than underluminous \sneia\ (see also \citealt{MLCS}).

Using a color-corrected flux ratio $\mathcal{R}^c$ in combination with
color results in even lower Hubble residual scatter when compared with
the single flux ratio case. Our best flux ratio in this case,
$\mathcal{R}^c(6420/5290)$, reduces the weighted rms of prediction
residuals by $\sim15$\% with respect to $(x_1,c)$ [${\rm
      WRMS}=0.175\pm0.025$\,mag cf. $0.204\pm0.029$\,mag],
and the intrinsic prediction error by $\sim20$\% [$\sigma_{\rm
      pred}=0.148\pm0.029$\,mag cf. $0.181\pm0.032$\,mag].
Again, the significance of this
difference is only $1.4\sigma$ ($\Delta_{x_1,c}=-0.032\pm0.023$\,mag). We
see from Fig.~\ref{fig:rrcx1c} ({\it middle panel}) that
$\mathcal{R}^c(6420/5290)$ is strongly anti-correlated with $x_1$
($r=-0.78$), and that dereddening the spectra using the SALT2 color
law is effective in removing any dependence of
$\mathcal{R}^c(6420/5290)$ on color, as expected. 

%%% Fig. Best R/Rc vs. x1,col

One would naively think that combining our best color-corrected ratio
$\mathcal{R}^c(6420/5290)$ with $(x_1,c)$ would lead to an even
further improvement, but this is not the case. In fact,
$\mathcal{R}^c(6420/5290)$ ranks 298$^{\rm th}$ when we consider the
set of predictors $(x_1,c,\mathcal{R}^c)$. This is due to the strong
anti-correlation of $\mathcal{R}^c(6420/5290)$ with $x_1$. Adding
$x_1$ as an extra predictor when $\mathcal{R}^c(6420/5290)$ already
includes this information means $\alpha$ will tend to fit noise in a
given training set, as was the case for the set of $(x_1,\mathcal{R})$
predictors when compared with $\mathcal{R}$-only. Indeed, the best-fit
value for $\alpha$ for $[x_1,c,\mathcal{R}^c(6420/5290)]$ is again
consistent with 0.

Nonetheless, several color-corrected flux ratios do result in a
further reduced scatter when combined with $(x_1,c)$, although the
wavelength baseline for these ratios is much smaller
($\lesssim400$\,\AA) and the wavelength bins forming the ratios are
all concentrated in the region of the \stwo\,\l\l5454,5640
doublet. Our highest-ranked flux ratio in this case,
$\mathcal{R}^c(5690/5360)$, reduces the weighted rms of prediction 
residuals by $\sim20$\% with respect to $(x_1,c)$ [${\rm
      WRMS}=0.164\pm0.023$\,mag cf. $0.204\pm0.029$\,mag],
and the intrinsic prediction error by $\sim25$\% [$\sigma_{\rm
    pred}=0.134\pm0.028$ cf. $0.181\pm0.032$\,mag]. Again, the
significance of this difference is only $1.6\sigma$
($\Delta_{x_1,c}=-0.044\pm0.028$\,mag). We see from
Fig.~\ref{fig:rrcx1c} ({\it lower panel}) that this ratio is not
correlated with $x_1$ ($r=-0.08$) or $c$ ($r=0.11$), and thus
constitutes a useful additional predictor of distances to \snia.

\subsubsection{Two-dimensional maps of all flux ratios}\label{sect:2dmaps}

\begin{figure*}
\begin{center}
\begin{tabular}{m{.12\linewidth} m{.38\linewidth} m{.38\linewidth}}
 & \multicolumn{1}{c}{WRMS residuals\ \ \ [mag]} & \multicolumn{1}{c}{Absolute Pearson correlation} \\
$\mathcal{R}$ & 
\includegraphics[bb=18 14 350 281,width=6cm]{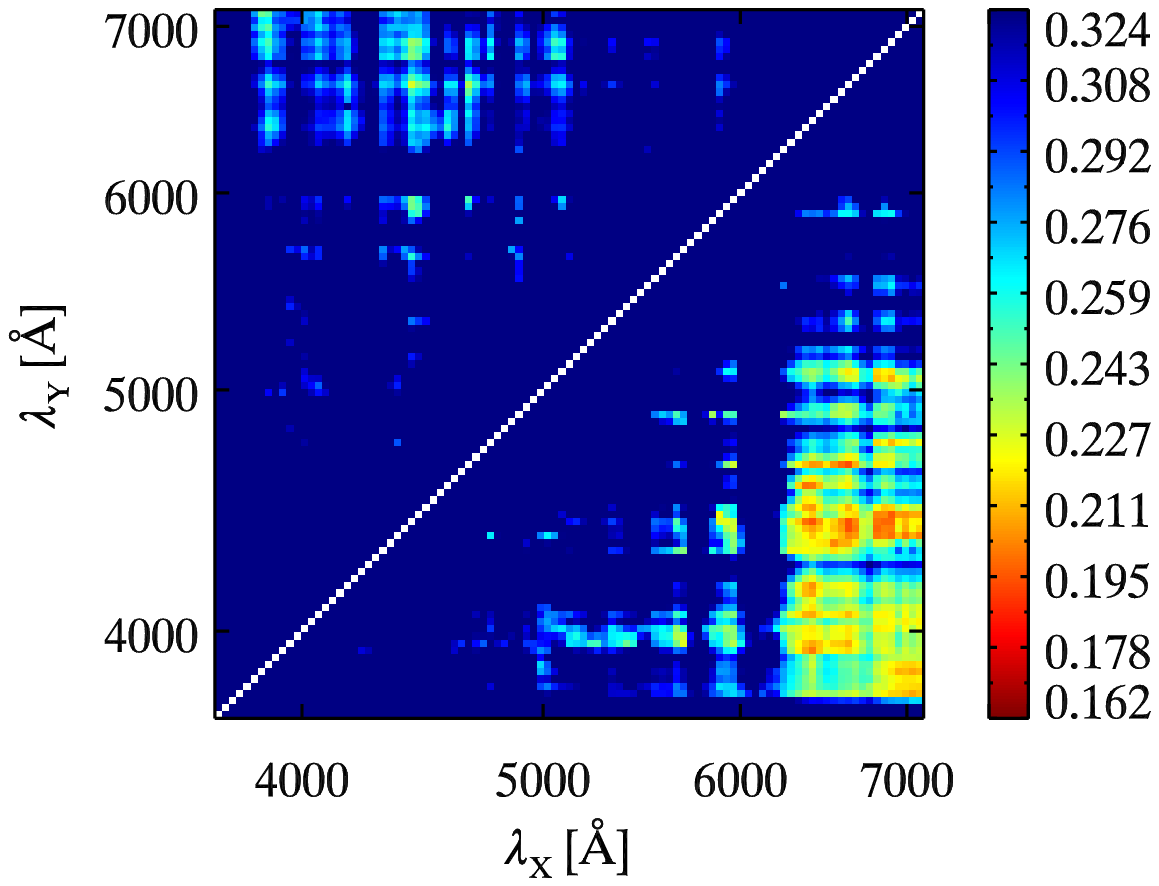} &
\includegraphics[bb=18 14 350 281,width=6cm]{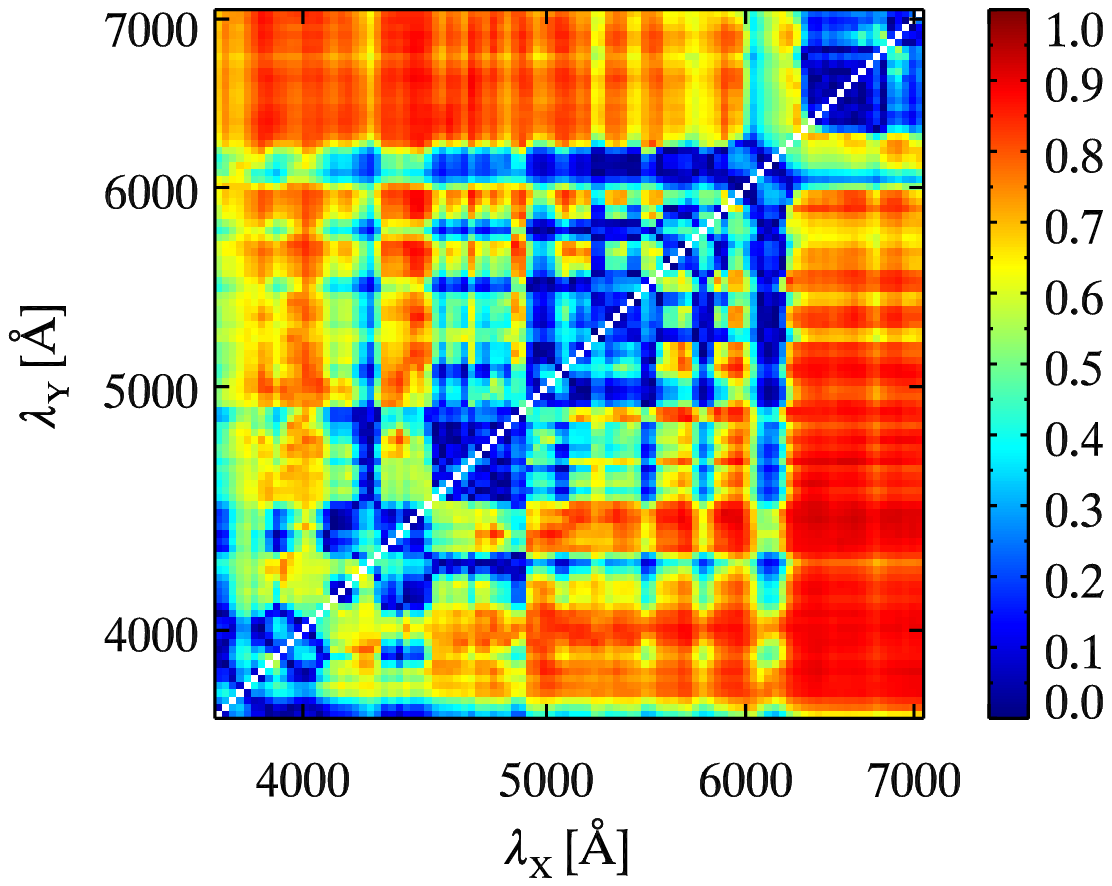} \\
$(x_1,\mathcal{R})$ & 
\includegraphics[bb=18 14 350 281,width=6cm]{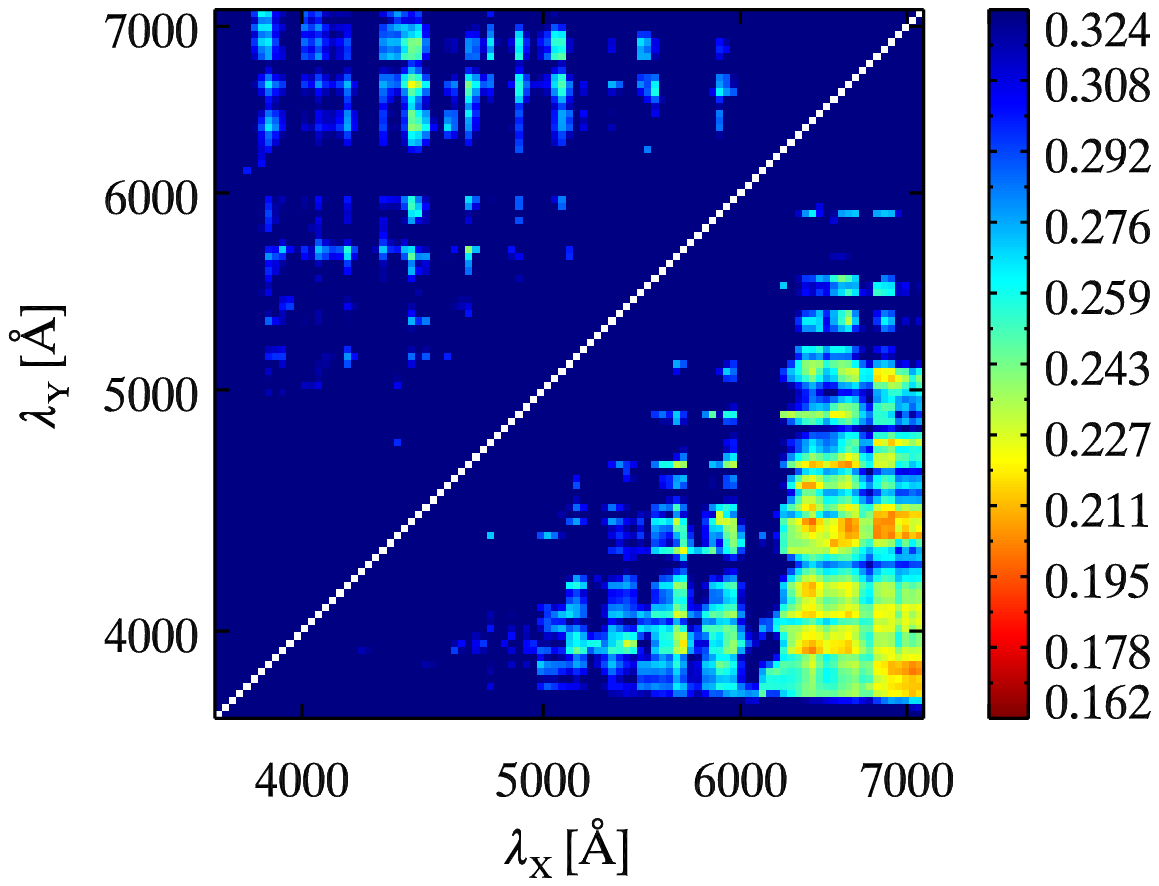} & 
\includegraphics[bb=18 14 350 281,width=6cm]{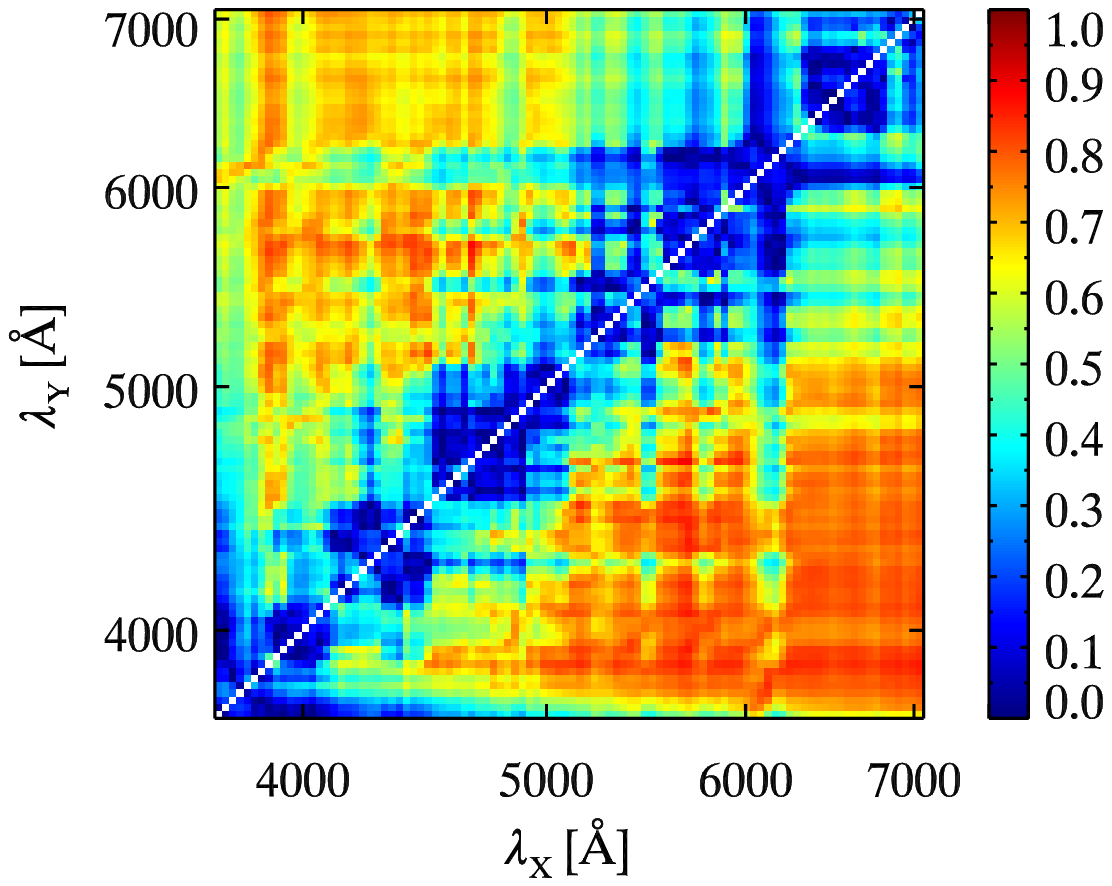} \\
$(c,\mathcal{R}^c)$ & 
\includegraphics[bb=18 14 350 281,width=6cm]{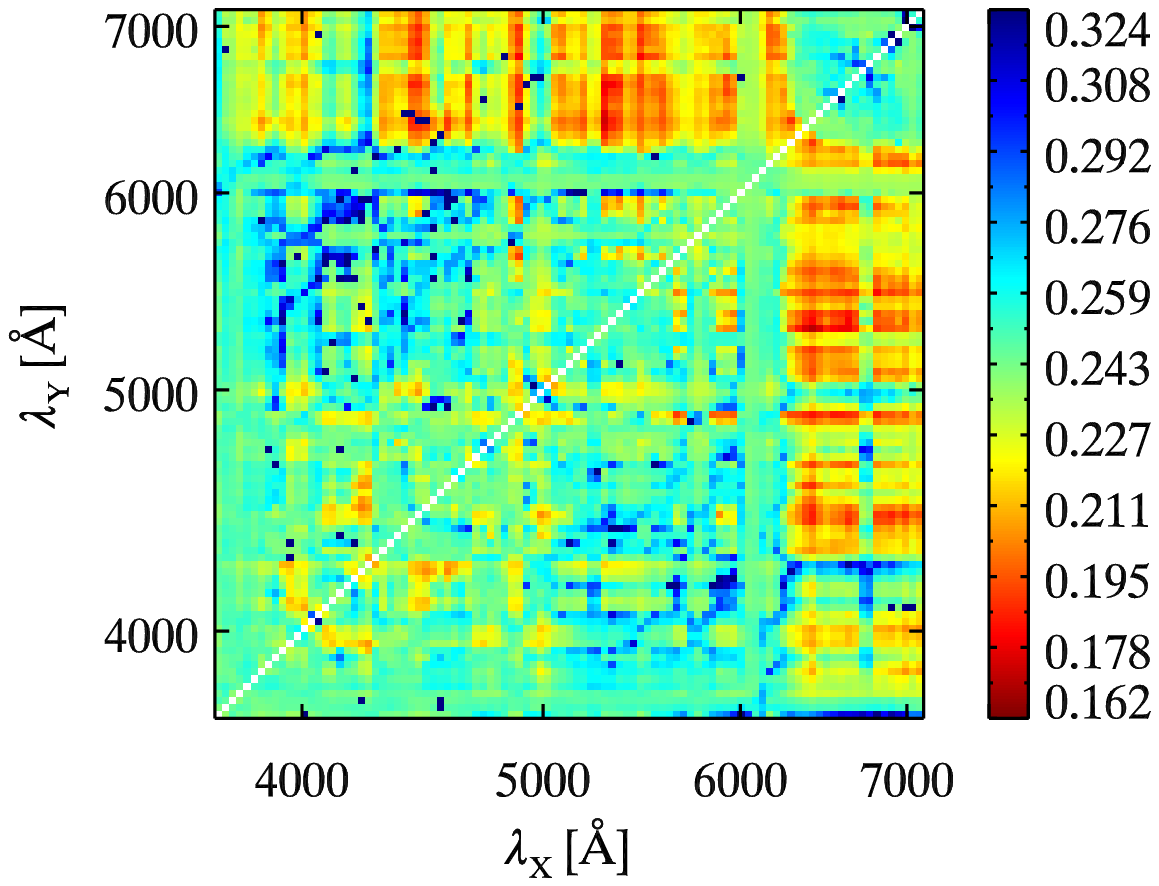} &
\includegraphics[bb=18 14 350 281,width=6cm]{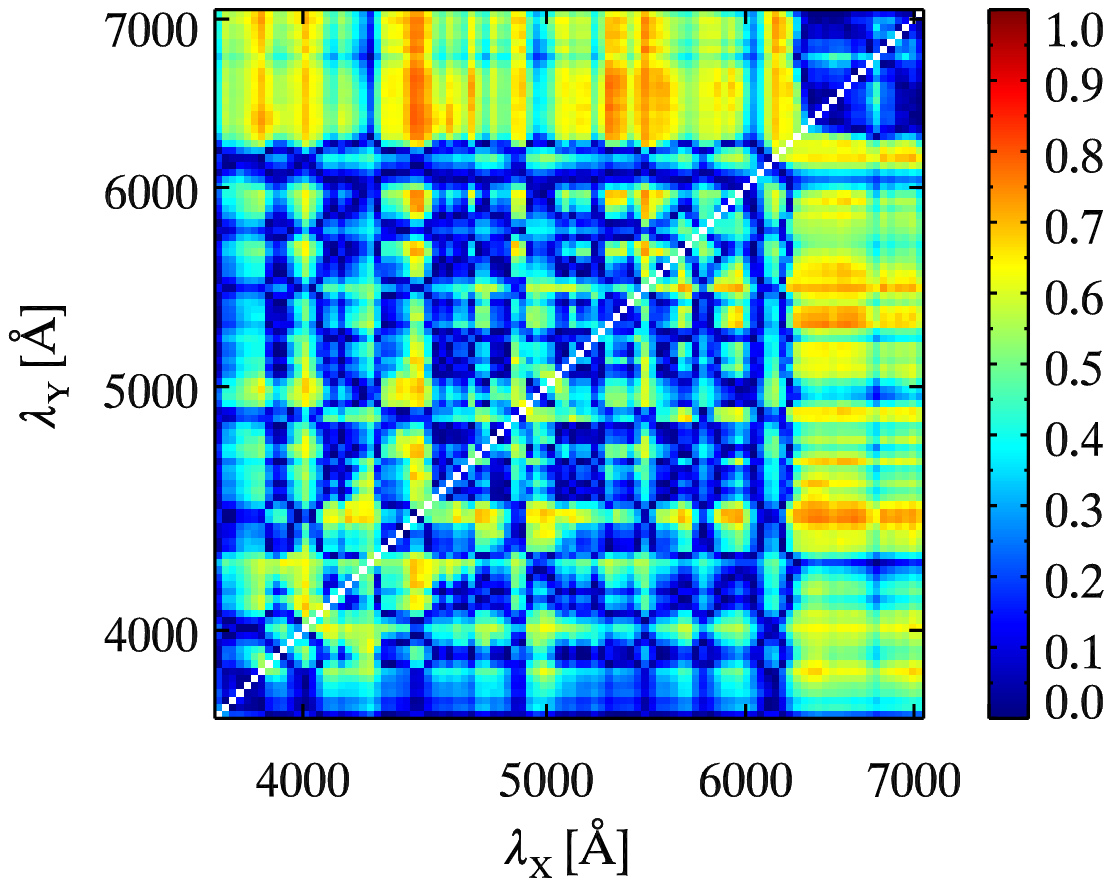} \\
$(x_1,c,\mathcal{R}^c)$ & 
\includegraphics[bb=18 14 350 281,width=6cm]{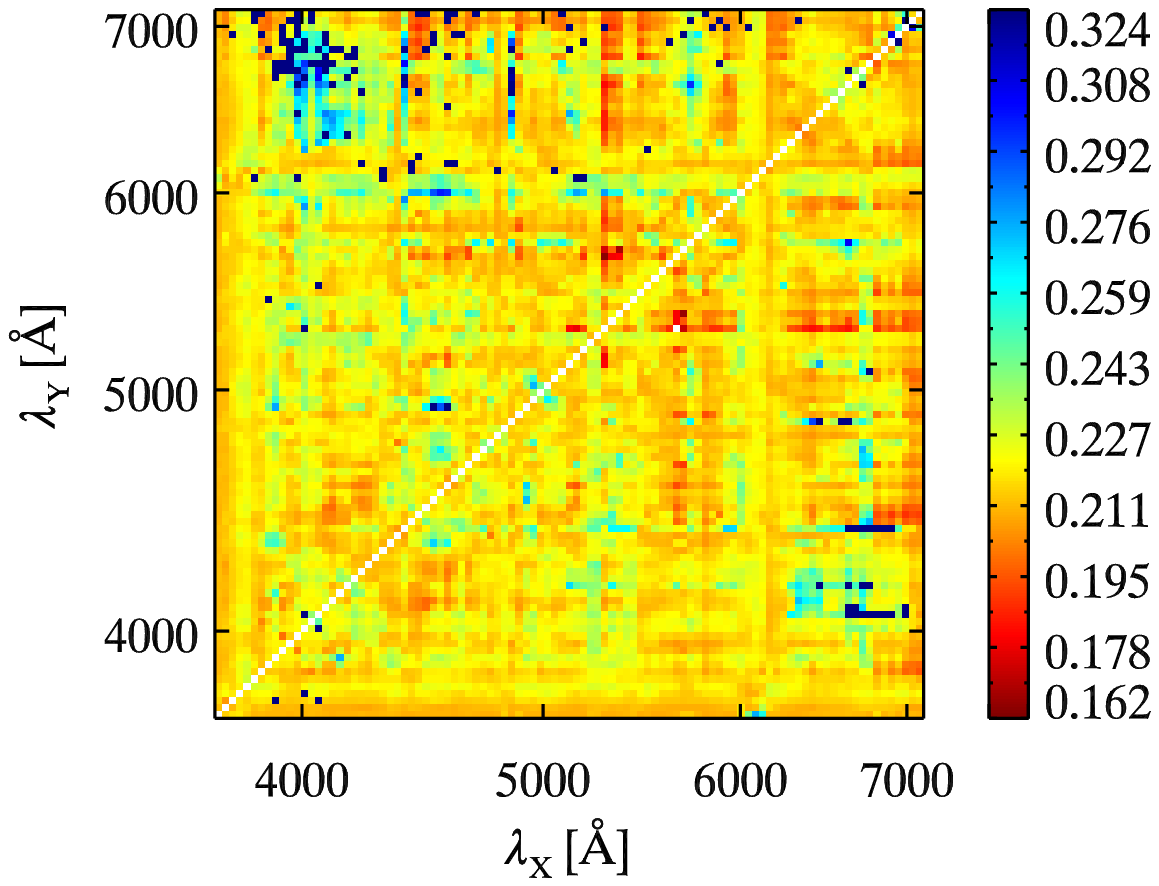} &
\includegraphics[bb=18 14 350 281,width=6cm]{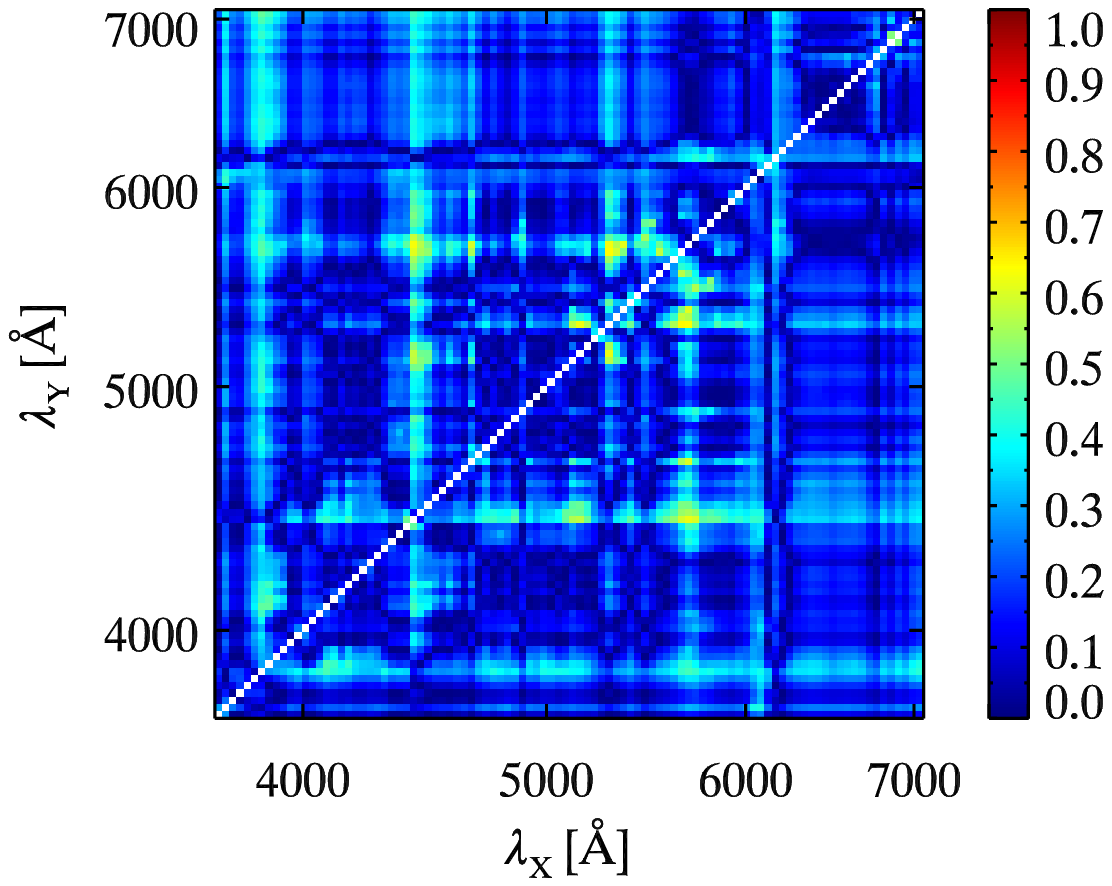}
\end{tabular}
\caption{\label{fig:kfoldtmaxres}
Results from 10-fold cross-validation on maximum-light spectra. 
{\it From top to bottom:} $\mathcal{R}$ only; ($x_1,\mathcal{R}$);
($c,\mathcal{R}^c$); ($x_1,c,\mathcal{R}^c$). The left column is
color-coded according to the weighted rms of prediction Hubble
residuals, while the right column corresponds to the absolute Pearson
cross-correlation coefficient of the correction terms with uncorrected
Hubble residuals.
}
\end{center}
\end{figure*}

The results for all 9506 flux ratios are displayed in
Fig.~\ref{fig:kfoldtmaxres}. The four rows correspond to the four
models for estimating \snia\ distances that include a flux ratio
(Eqs.~\ref{eqn:frarr1}-\ref{eqn:frarr4}). The 
left column is color-coded according to the weighted rms of prediction
Hubble residuals (flux ratios that result in ${\rm WRMS}>0.324$\,mag are
given the color corresponding to ${\rm WRMS}=0.324$\,mag), while the right
column is color-coded according to the absolute Pearson correlation
coefficient of the correction terms with uncorrected Hubble residuals
[e.g. for the set of predictors $(c,\mathcal{R}^c)$, the correlation
  of $(-\beta c + \gamma \mathcal{R}^c)$ with uncorrected residuals].

%%% Fig. Results from 10-fold cross-validation on maximum-light spectra

\begin{figure}
\centering
\resizebox{0.475\textwidth}{!}{\includegraphics{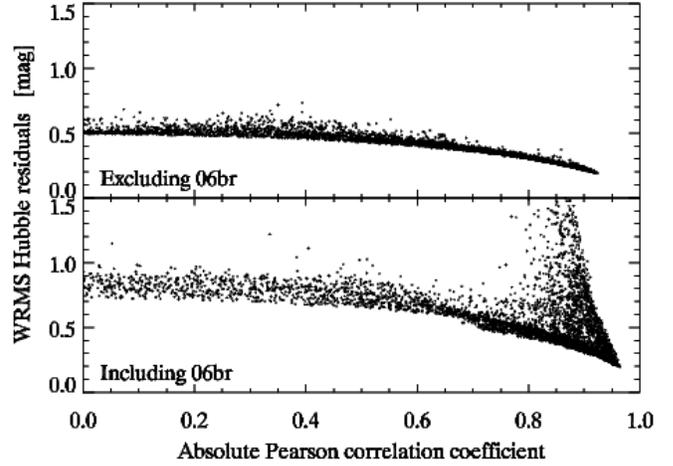}}
\caption{\label{fig:frxcorhubres}
Weighted rms of prediction Hubble residuals {\it vs.} absolute Pearson
cross-correlation coefficient for all flux ratios at maximum light,
excluding ({\it top}) and including ({\it bottom}) the highly-reddened
SN~2006br.
}
\end{figure}

Only a very restricted number of wavelength bins lead to a low
WRMS of prediction Hubble residuals when a flux ratio
$\mathcal{R}$ is used by itself (Fig.~\ref{fig:kfoldtmaxres}; {\it
upper left}), namely $\lambda_X \gtrsim 6300$\,\AA\ and
$\lambda_Y\approx4400$\,\AA\ (4 of the 5 best flux ratios in
Table~\ref{tab:kfoldcv_10_p0} for $\mathcal{R}$-only 
have $\lambda_Y\approx4400$\,\AA). This is in stark contrast with
the large number of flux ratios with absolute Pearson correlation
coefficients $|r|>0.8$ (Fig.~\ref{fig:kfoldtmaxres}; {\it upper
  right}). In general, a flux ratio with a higher correlation
coefficient will result in a Hubble diagram with less scatter, but
this is not systematically the case, and the relation between the two
is certainly not linear. For Pearson correlation coefficients
$|r|>0.8$, the standard deviation of Hubble residuals can vary by up to
0.1\,mag at any given $|r|$ (Fig.~\ref{fig:frxcorhubres}, {\it top
  panel}). This is because the cross-correlation coefficient does not
take into account errors on $\mathcal{R}$ or on the Hubble residual, and
is biased by outliers and reddened \sneia. The lower panel of
Fig.~\ref{fig:frxcorhubres} shows the impact of including the
highly-reddened SN~2006br: at any given $|r|$, the resulting weighted rms
of prediction Hubble residuals is 30-60\% higher. Moreover, many flux
ratios with high correlation coefficients ($|r|>0.8$) result in Hubble
diagrams with excessively large scatter (${\rm WRMS}>1$\,mag).
This is counter-intuitive, since the resulting scatter in these
cases appears to be larger than when no predictors at all are used
to determine distances to \sneia\ (in which case ${\rm WRMS}\approx
0.5$\,mag). The reason is that we consider the scatter under
cross-validation, as opposed to fitting all the \sneia\ at the same
time.  In these aberrant cases, the trained model is sensitive to the
inclusion or exclusion of some outlier in the training set, and this
leads to large errors when the outlier is in the validation set.
Last, including this SN leads to correlations with $|r|>0.95$, where
there are none otherwise. Fig.~\ref{fig:frxcorhubres} thus justifies
our excluding SN~2006br from the sample (already excluded based
  on our cut on SALT2 color; see \S~\ref{sect:frel}), and 
illustrates the advantage of selecting flux ratios based directly on
the weighted rms of prediction Hubble diagram residuals,
rather than on cross-correlation coefficients. As already mentioned in
\S~\ref{sect:bestfr}, using cross-validated prediction errors to
select the best flux ratios guards us against overfitting a small
sample: in the naive approach that consists in fitting the entire
\snia\ sample at once, adding more predictors always leads to a
lower scatter in Hubble residuals (this is known as
``resubstitution''; see, e.g., \citealt{Mandel/etal:2009}).

%%% Fig. WRMS hubres vs. xcor coefficient, with/without 06br

When the SALT2 color parameter is used in combination with a color-corrected
flux ratio $\mathcal{R}^c$, there are again restricted wavelength
regions that lead to a low weighted rms of prediction Hubble residuals
(Fig.~\ref{fig:kfoldtmaxres}; {\it third row left}) [4 of the 5 best
flux ratios in Table~\ref{tab:kfoldcv_10_p0} for $(c,\mathcal{R}^c)$
predictors involve wavelength bins at $\sim4900$\,\AA\ and
$\sim6500$\,\AA]. The SALT2 color parameter $c$ does not attempt to
distinguish between reddening by dust and intrinsic color
variations. Dereddening the spectra using this parameter corrects for
both effects regardless of their relative importance. However, since
the SALT2 color law is very similar to the \cite{CCM89} extinction law
with $R_V=3.1$ and $E(B-V)=0.1$\,mag \citep[their Fig.~3]{SALT2}, one
generally assumes that the color correction removes the bulk of
reddening by dust, and the remaining variations in the SED are
primarily intrinsic to the supernova. If this is so, it is intriguing
that the best flux ratios for the $\mathcal{R}$-only and
$(c,\mathcal{R}^c)$ models share similar wavelength bins. The recent survey
of 2D \snia\ models from \cite{Kasen/Roepke/Woosley:2009} suggests
that a significant part of the color variation measured by the
$\mathcal{R}(6630/4400)$ is indeed intrinsic (see
\S~\ref{sect:modelcomp}).

The second row of Fig.~\ref{fig:kfoldtmaxres} confirms that using the
$x_1$ parameter in combination with a flux ratio results in a slight
degradation in the weighted rms of prediction residuals, while the 
correlations with uncorrected Hubble residuals are degraded with
respect to cases where $\mathcal{R}$ is used by itself. Last, the
bottom row of Fig.~\ref{fig:kfoldtmaxres} is a visual demonstration
that $(x_1,c,\mathcal{R}^c)$ fares better than $(c,\mathcal{R}^c)$
overall, although the best color-corrected flux ratios do not perform
significantly better. We see from the right panel that the
correlations of $(\alpha x_1 - \beta c +\gamma \mathcal{R}^c)$ with
uncorrected residuals all have absolute Pearson correlation
coefficients $|r|\lesssim 0.5$. The two regions at
$\l_{X,Y}\approx5300$\,\AA\ stand out in the 2D plot of WRMS
residuals, and all the top color-corrected ratios for
$(x_1,c,\mathcal{R}^c)$ include a wavelength bin in that region (which
corresponds to the absorption trough of the \stwo\,\l5454 line).

\subsubsection{Comparison with \cite{Bailey/etal:2009}}

We confirm the basic result of \cite{Bailey/etal:2009} using an
independent sample and a different cross-validation method: the use of
a flux ratio alone or in combination with a color parameter results in
a Hubble diagram with lower scatter when compared to
the standard $(x_1,c)$ model. Using a flux ratio alone,
\cite{Bailey/etal:2009} find $\mathcal{R}(6420/4430)$ as their most
highly-ranked ratio, while we find $\mathcal{R}(6630/4400)$ [see
Table~\ref{tab:kfoldcv_10_p0}]. The wavelength bins are almost
identical, and in any case $\mathcal{R}(6420/4430)$ is amongst our
top 5 ratios. For this ratio we find $\gamma=-3.40\pm0.10$, in
agreement with $\gamma=-3.5\pm0.2$ found by
\cite{Bailey/etal:2009}\footnote{In fact \cite{Bailey/etal:2009} find
$\gamma=+3.5\pm0.2$, but this is due to a typo in 
their equation for the distance modulus: $\gamma\mathcal{R}$
really appears as a negative term in their paper (S.~Bailey 2010,
private communication).}.

The other four flux ratios given by \cite{Bailey/etal:2009} [their
  Table~1] are not part of our top-5 
$\mathcal{R}$. For two of these ratios the reason is trivial: they
include wavelength bins redder than 7100\,\AA, not covered by most of
our spectra. The other two flux ratios [$\mathcal{R}(6420/4170)$ and
$\mathcal{R}(6420/5120)$] lead to differences $<5$\% on the Hubble
diagram residual scatter with respect to the standard $(x_1,c)$
model according to \cite{Bailey/etal:2009}
[$\sigma=0.166\pm0.016$\,mag for $\mathcal{R}(6420/4170)$ and
$\sigma=0.154\pm0.015$\,mag for $\mathcal{R}(6420/5120)$,
cf. $0.161\pm0.015$\,mag for $(x_1,c)$], 
and they rank 
29$^{\rm th}$ and 658$^{\rm th}$ in our study, respectively. This
discrepancy is in part due to the selection method:
\cite{Bailey/etal:2009} select their best ratios based on
cross-correlation coefficients with uncorrected magnitudes, while we
select them based on the intrinsic prediction error from
cross-validated Hubble diagram residuals. However, ranking our ratios
using the same method as \cite{Bailey/etal:2009} does not resolve the
discrepancy. It is possible that \cite{Bailey/etal:2009} are sensitive
to their exact choice of training and validation samples, where we
have randomized the approach. We note however that the impact on the
weighted rms of prediction residuals is statistically indistinguishable
for many flux ratios given our sample size (e.g. error on WRMS
$\sim0.03$\,mag cf. differences of $\lesssim0.01$\,mag in WRMS for
the top 5 flux ratios; see Table~\ref{tab:kfoldcv_10_p0}), so that the 
exact ranking of flux ratios is not well determined and subject
to revisions from small changes in the input data.

Using both a color-corrected flux ratio $\mathcal{R}^c$ and the
SALT2 color parameter decreases the residual scatter further, as found by
\cite{Bailey/etal:2009}. Using the set of predictors
$[c,\mathcal{R}^c(6420/5290)]$ leads to $\sim15$\% lower WRMS with
respect to $(x_1,c)$ [${\rm WRMS}=0.175\pm0.025$\,mag
  cf. $0.204\pm0.029$\,mag], and to $\sim20$\% lower $\sigma_{\rm
  pred}$ [$\sigma_{\rm pred}=0.148\pm0.029$\,mag
  cf. $0.181\pm0.032$\,mag] at $1.4\sigma$ significance based on the
difference in intrinsic prediction error, $\Delta_{x_1,c}$. None of
the color-corrected flux ratios listed by \cite{Bailey/etal:2009}
[their Table~1] are part of our five highest-ranked $\mathcal{R}^c$,
although our top ratios are formed with almost the same wavelength bins
[$\mathcal{R}^c(6420/5290)$ in this paper; $\mathcal{R}^c(6420/5190)$
in \cite{Bailey/etal:2009}]. The other color-corrected ratios in
\cite{Bailey/etal:2009} rank well below in our study, whether we
select the best $\mathcal{R}^c$ according to the resulting Hubble
residual scatter or the cross-correlation of $(-\beta c + \gamma\mathcal{R}^c)$ with
uncorrected residuals. The same caveats apply here as when selecting
the best uncorrected flux ratios (see previous paragraph), although the
$\mathcal{R}^c$ measurement is probably even more sensitive to the relative flux
calibration accuracy of the spectra.

We also cross-checked the results of \cite{Bailey/etal:2009} by
simply validating their best flux ratios on our entire \snia\
sample. The results are displayed in Table~\ref{tab:validb09}, where
we give the weighted rms of Hubble residuals from a
simultaneous fit to the entire \snia\ sample (as done by
\citealt{Bailey/etal:2009}), as opposed to prediction residuals under
cross-validation.
For all flux ratios (both $\mathcal{R}$ and
$\mathcal{R}^c$) in Table~\ref{tab:validb09}, our own best-fit
$\gamma$ agrees within the $1\sigma$ errors with that found by
\cite{Bailey/etal:2009} [noted $\gamma$(B09) in
  Table~\ref{tab:validb09}] , although we have systematically larger 
errors. We note that most of the top ratios reported by
\cite{Bailey/etal:2009} lead to no significant improvement over
$(x_1,c)$, and even leads to slightly worse results for some ratios
[e.g. $\mathcal{R}(6420/5120)$ results in ${\rm WRMS}=0.237\pm0.032$\,mag
  cf. $0.194\pm0.027$\,mag for $(x_1,c)$]. A closer look at Table~1 of
\cite{Bailey/etal:2009} shows that this is also the case in their
paper: for the $\mathcal{R}$-only model, only one ratio out of five,
namely $\mathcal{R}(6420/4430)$, results in a lower Hubble diagram
residual scatter. The other four are either consistent with no
improvement [$\mathcal{R}(7720/4370)$ and  $\mathcal{R}(6420/5120)$],
or yield slightly worse results [$\mathcal{R}(6420/4170)$ and
  $\mathcal{R}(7280/3980)$]. Again, this results from the way
\cite{Bailey/etal:2009} selected their best ratios, based on the
correlation with uncorrected Hubble residuals.

\begin{table}
\scriptsize
\caption{Validation of top 5 flux ratios at maximum light from
  \cite{Bailey/etal:2009} [noted B09]}\label{tab:validb09}
\begin{tabular}{ccrrrr}
\hline\hline
$\lambda_{\rm X}$ & $\lambda_{\rm Y}$ & \multicolumn{1}{c}{$\gamma$} & \multicolumn{1}{c}{$\gamma$(B09)} & 
\multicolumn{1}{c}{WRMS\tablefootmark{a}} & \multicolumn{1}{c}{$\Delta_{x_1,c}$} \\
\hline
\multicolumn{6}{l}{$\mathcal{R}$} \\
\hline
6420 & 4430 & $-3.40 \pm 0.32$ & $-3.5 \pm 0.2$ & $0.184 \pm 0.026$ & $-0.007 \pm 0.031\ (0.2\sigma)$ \\
6420 & 4170 & $-4.43 \pm 0.41$ & $-4.9 \pm 0.2$ & $0.197 \pm 0.028$ & $ 0.008 \pm 0.028\ (0.3\sigma)$ \\
7720\tablefootmark{b} & 4370 & \multicolumn{1}{c}{$\cdots$} & $ 7.3 \pm 0.3$ & \multicolumn{1}{c}{$\cdots$} & \multicolumn{1}{c}{$\cdots$} \\ 
6420 & 5120 & $-4.34 \pm 0.42$ & $-4.7 \pm 0.3$ & $0.237 \pm 0.032$ & $ 0.052 \pm 0.034\ (1.5\sigma)$ \\
7280\tablefootmark{b} & 3980 & \multicolumn{1}{c}{$\cdots$} & $ 7.9 \pm 0.3$ & \multicolumn{1}{c}{$\cdots$} & \multicolumn{1}{c}{$\cdots$} \\ 
\hline
\multicolumn{6}{l}{$(c,\mathcal{R}^c)$} \\
\hline
6420 & 5190 & $-2.20 \pm 0.57$ & $-3.5 \pm 0.3$ & $0.176 \pm 0.025$ & $-0.012 \pm 0.016\ (0.8\sigma)$ \\
5770 & 6420 & $ 0.67 \pm 0.22$ & $ 1.4 \pm 0.1$ & $0.193 \pm 0.027$ & $ 0.005 \pm 0.011\ (0.5\sigma)$ \\
6420 & 5360 & $-1.71 \pm 0.43$ & $-2.3 \pm 0.2$ & $0.169 \pm 0.024$ & $-0.019 \pm 0.016\ (1.2\sigma)$ \\
6760 & 6420 & $ 1.79 \pm 1.09$ & $ 4.2 \pm 0.5$ & $0.217 \pm 0.032$ & $ 0.030 \pm 0.025\ (1.2\sigma)$ \\
6420 & 4430 & $-2.56 \pm 0.63$ & $-3.2 \pm 0.3$ & $0.168 \pm 0.024$ & $-0.027 \pm 0.023\ (1.2\sigma)$ \\
\hline
\multicolumn{6}{l}{$(x_1,c)$} \\
\hline
$\cdots$ & $\cdots$ & \multicolumn{1}{c}{$\cdots$} & \multicolumn{1}{c}{$\cdots$} & $0.194 \pm 0.027$ & \multicolumn{1}{c}{$\cdots$} \\
\hline
\end{tabular}
\tablefoot{
\tablefoottext{a}{Weighted rms of Hubble residuals from a simultaneous fit to
the entire \snia\ sample (as done by \citealt{Bailey/etal:2009}), as
opposed to prediction residuals under cross-validation. As
explained in \S~\ref{sect:cv}, the weighted rms of prediction Hubble
residuals is a more realistic estimate of the accuracy of a given
model in measuring distances to \sneia.}
\tablefoottext{b}{Wavelength bins redder than 7100\,\AA, not covered
  by most of our spectra.}
}
\end{table}

We cannot directly compare the resulting scatter in
Hubble diagram residuals with those reported in Table~1 of
\cite{Bailey/etal:2009}. First, they use the sample standard
deviation ($\sigma$), whereas we use the weighted rms (see
\S~\ref{sect:stats}). Second, the scatter they find for the standard
$(x_1,c)$ model is significantly lower than ours. We have refit the
data presented in Table~1 of \cite{Bailey/etal:2009} to derive the
weighted rms of Hubble residuals for the $(x_1,c)$ model from their
sample, and find ${\rm WRMS}=0.148\pm0.014$\,mag, which is almost
0.05\,mag smaller when compared to our sample
($0.194\pm0.027$\,mag). This difference in the Hubble residual scatter
between the SN-Factory and CfA samples is consistent with the
difference found amongst other nearby \snia\ samples by
\cite{Hicken/etal:2009b}.

Interestingly, using the WRMS statistic as opposed to the sample
standard deviation results in a smaller difference in residual scatter
between the $\mathcal{R}$-only and $(x_1,c)$ models. Using our own
fits of the data presented in Table~1 of \cite{Bailey/etal:2009}, we
find ${\rm WRMS}=0.131\pm0.014$\,mag for $\mathcal{R}(6420/4430)$,
i.e. $\sim11$\% smaller scatter when compared to $(x_1,c)$, where the
difference between the two models is $\sim20$\% when considering the
sample standard deviation.

\subsubsection{Comparison with 2D models}\label{sect:modelcomp}

We use synthetic spectra based on a recent 2D survey of
delayed-detonation \snia\ models by \cite{Kasen/Roepke/Woosley:2009}
to investigate the physical origin of the high correlation between
several flux ratios and uncorrected \snia\ magnitudes. These models
were found to reproduce the empirical relation between peak $B$-band
magnitude and post-maximum decline rate. A more detailed comparison of
\snia\ data with these models will be presented elsewhere.

\begin{figure*}
\centering
\resizebox{\textwidth}{!}{\includegraphics{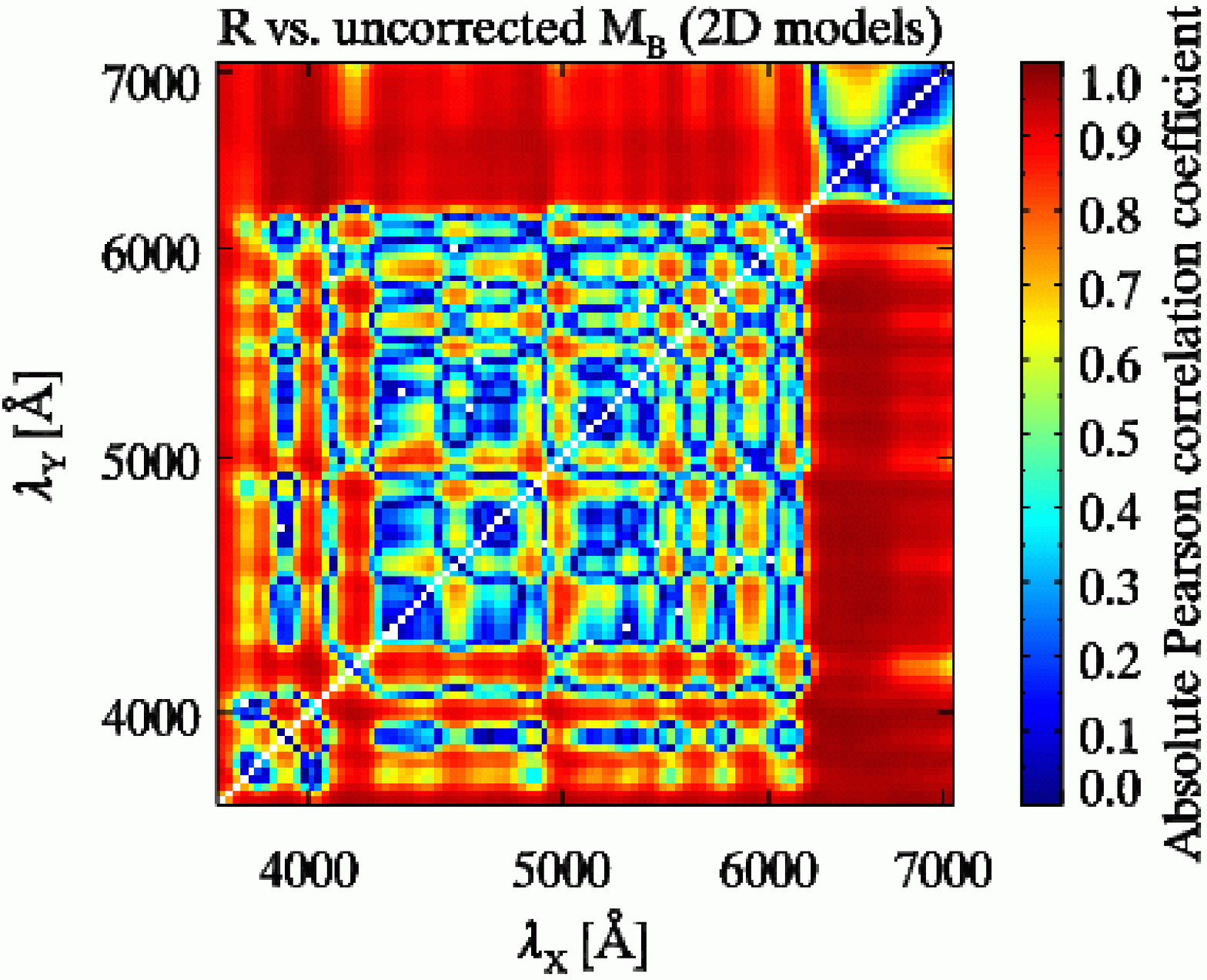}\hspace{1.5cm}\includegraphics{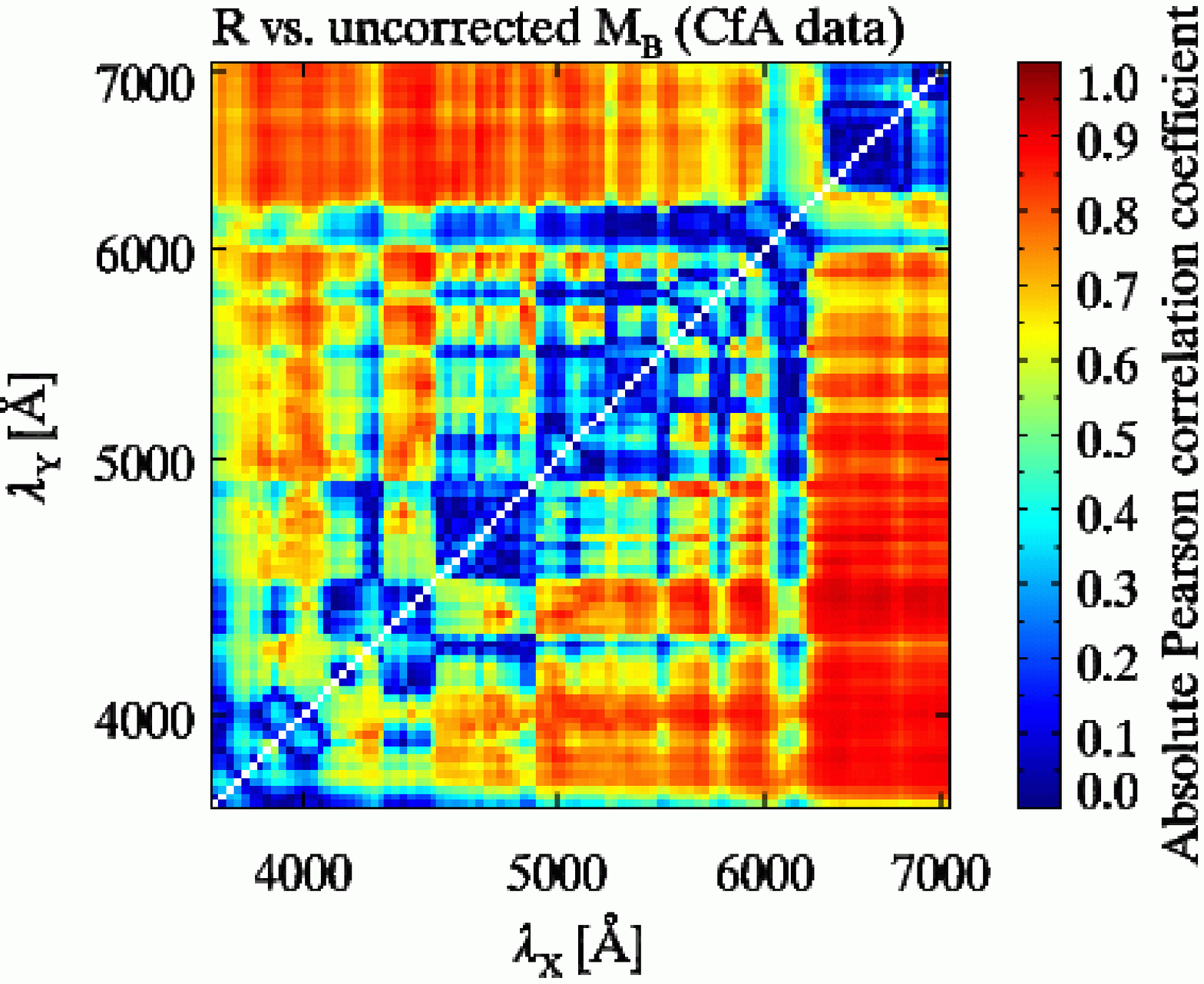}}
\caption{\label{fig:frdkmodel}
Absolute Pearson correlation coefficients of flux ratios at maximum
light with uncorrected absolute magnitudes $M_B$ in 2D
delayed-detonation \snia\ models of \cite{Kasen/Roepke/Woosley:2009}
({\it left}), and in data from the CfA \snia\ sample ({\it right}).
}
\end{figure*}

We measured flux ratios in the same manner as we did for our data, and
computed Pearson correlation coefficients with (uncorrected) absolute
magnitudes synthesized directly from the spectra. The 2D correlation
map is shown in Fig.~\ref{fig:frdkmodel} ({\it left panel}), alongside
the same map derived from the CfA \snia\ sample ({\it right
 panel}). At first glance, the two maps appear similar, with two large
$\sim1000$\,\AA-wide ``bands'' of flux ratios with strong correlations
with uncorrected magnitudes, for 
$\l_X (\l_Y) \gtrsim6200$\,\AA\ and $\l_Y (\l_X) \lesssim6000$\,\AA, 
although the correlations are even
stronger in the models (several flux ratios have absolute
Pearson correlation coefficients $|r|>0.95$, where there are
none in the data). A closer look reveals some important differences, 
the models having strong correlations for
$6000$\,\AA$\lesssim\l_X (\l_Y)\lesssim6200$\,\AA\ and
$\l_Y (\l_X) \gtrsim6200$\,\AA\ that are not present in the data. The same
applies to the regions with coordinates $\l_X (\l_Y)\approx4200$\,\AA. 
These differences are significant and illustrate the potential for
such comparisons to impose strong constraints on \snia\ models.

%%% Fig. R vs. uncorrected M_B from DK models and CfA data

\begin{figure}
\centering
\resizebox{0.475\textwidth}{!}{\includegraphics{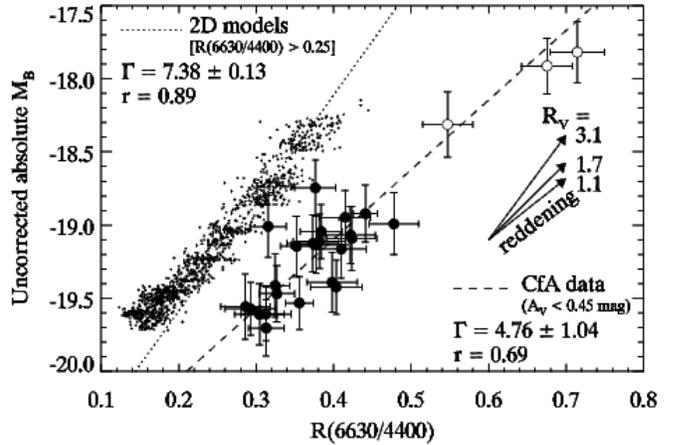}}
\caption{\label{fig:frmodeldata}
Absolute rest-frame $B$-band magnitude ($M_B$) {\it vs.} 
flux ratio $\mathcal{R}(6630/4400)$ at maximum light in 2D \snia\
models of \cite{Kasen/Roepke/Woosley:2009} [{\it small dots}], and in
data from the CfA \snia\ sample ($A_V<0.45$\,mag: {\it filled
circles}, $A_V>0.45$\,mag: {\it open circles}). The dotted and dashed
lines are linear fits to the models and data, respectively, where
models for which $\mathcal{R}(6630/4400)>0.25$ and data for which
$A_V>0.45$\,mag have been excluded from the fit. The slope ($\Gamma$)
and Pearson correlation coefficient ($r$) are indicated for both
cases. The data have been offset vertically for clarity. 
Including models for which $\mathcal{R}(6630/4400)<0.25$ results in
$\Gamma=5.71\pm0.03$ and $r=0.97$, while including data with
$A_V>0.45$\,mag results in $\Gamma=4.43\pm0.47$ and $r=0.92$.
The arrows indicate approximate reddening vectors for different values
of $R_V$.} 
\end{figure}

In Fig.~\ref{fig:frmodeldata} we show the correlation of uncorrected
absolute rest-frame $B$-band magnitudes ($M_B$) with our highest-ranked
flux ratio $\mathcal{R}(6630/4400)$, both from the 2D models and CfA data,
where we have used the redshift-based distance for the latter. The
vertical offset is arbitrary and solely depends on the normalization
adopted for the data, which we have chosen for sake of clarity. There
are 1320 model points, each corresponding to one of the 44 2D
delayed-detonation models of \cite{Kasen/Roepke/Woosley:2009} viewed
from one of 30 different viewing angles. The linear fits shown in
Fig.~\ref{fig:frmodeldata} are done over the range $0.25\lesssim
\mathcal{R}(6630/4400) \lesssim 0.50$, where the models and data
overlap. For the data this is equivalent to excluding the three most
highly-reddened \sneia\ ({\it open circles}), for which the host-galaxy
visual extinction $A_V$ was determined based on light-curve fits with
MLCS2k2 \citep{MLCS2k2}. This is justified since no reddening by dust
is applied to the models. The slope of the relation between $M_B$ and
$\mathcal{R}(6630/4400)$ is significantly steeper for the models
($\Gamma=7.38\pm0.13$) than for the data
($\Gamma=4.76\pm1.04$), and the correlation is much stronger ($r=0.89$
cf. 0.69 for the data). This is not surprising since the data are
subject to random measurement and peculiar velocity errors, which
degrade the correlation.
Including models for which
$\mathcal{R}(6630/4400)<0.25$ softens the slope to $\Gamma=5.71\pm0.03$
and results in a stronger correlation ($r=0.97$),
while including data with $A_V>0.45$\,mag results in
$\Gamma=4.43\pm0.47$ and a much stronger correlation $r=0.92$. This
last value for $\Gamma$ can be compared with the $\gamma$ fitting
parameter for this same flux ratio ($\gamma=-4.42\pm0.09$; see
Table~\ref{tab:kfoldcv_10_p0}), although the latter is based on a
formal cross-validation procedure and the opposite sign is a
consequence of the convention when using the flux ratio to predict
\snia\ distances. As noted in \S~\ref{sect:2dmaps}, the correlation
of $M_B$ with $\mathcal{R}(6630/4400)$ is largely biased by the
minority of highly-reddened \sneia.

%%% Fig. R(6630/4400) vs. uncorrected M_B/hubres from DK models and CfA data

The models yield values for $\mathcal{R}(6630/4400)$ ranging between
$\sim0.12$ and $\sim0.44$, all due to {\it intrinsic} color variations.
Since these models reproduce the relation between $M_B$ and post-maximum
decline rate of \cite{Phillips:1993}, they confirm the intrinsic
nature of the correlation between $\mathcal{R}(6630/4400)$ and $\{x_1,c\}$
shown in Fig.~\ref{fig:rrcx1c} ({\it upper left panel}).

The wavelength bins $\l_X=6630$\,\AA\ and $\l_Y=4400$\,\AA\ are close
to the central wavelengths of the standard $R$ and $B$ broadband
filters, hence $\mathcal{R}(6630/4400)$ is a rough measure of the
$B-R$ color at $B$-band maximum. The 2D models of Fig.~\ref{fig:frmodeldata}
indicate that a large part of the variation in
$\mathcal{R}(6630/4400)$ seen in the data is due to intrinsic
variations in $B-R$ color. Reddening in the host galaxy is then needed
to explain values of $\mathcal{R}(6630/4400) \gtrsim 0.4$, while at
lower values it is challenging at best to discriminate between the
effects of intrinsic color variations and extinction by dust, since
both affect $\mathcal{R}(6630/4400)$ in the same manner, as
illustrated by the reddening vectors in Fig.~\ref{fig:frmodeldata}
[they are really reddening {\it curves}, cf. Fig.~\ref{fig:frredcol},
but the behavior is almost linear over this small range in
$\mathcal{R}(6630/4400)$].

The models also give a physical explanation for the correlation of
$\mathcal{R}(6630/4400)$ with absolute magnitude. Indeed, the
variation of this ratio is largely caused by spectroscopic variations
around 4400\,\AA, a region dominated by lines of \fetwo\ and \fethree,
with contributions from \mgtwo\ (\titwo\ provides an important source
of opacity for the least luminous \sneia), while the region around
6630\,\AA\ has little intrinsic variation (this was noted by
\citealt{Bailey/etal:2009}). This translates to a standard deviation
of peak $B$-band magnitudes ($\sigma\approx0.40$\,mag) that is almost
twice as large as the $R$-band magnitude (at $B$ maximum;
$\sigma\approx0.26$\,mag) in the models. The relative contribution of
\fetwo\ and \fethree\ lines is related to the temperature of the
line-forming regions in the \snia\ ejecta, itself a function of peak
luminosity (dimmer \sneia\ are generally cooler; see,
e.g., \citealt{Kasen/Woosley:2007}). One thus expects a large
luminosity-dependent spectroscopic variation in this wavelength
region, although its exact shape and relation to temperature
  remains largely unknown.

While these models provide useful insights into the physical origin of
these correlations, a direct comparison with the data reveals some of
their shortcomings. In Fig.~\ref{fig:frmodeldata} we see that some
models predict values of the flux ratio $\mathcal{R}(6630/4400)
\lesssim 0.25$ for the most luminous \sneia, where the data are
limited to values greater than this. Our sample includes several
\sneia\ at the high luminosity end that show no sign of extinction in
their host galaxies ($A_V<0.05$\,mag based on light-curve fits with
MLCS2k2), so the differences are real and point to
discrepancies between the data and the models, some of the latter
having bluer $B-R$ colors at $B$-band maximum. 
This is not surprising, as the models explore a larger range of
parameter space than is realized in nature. Comparisons of this sort
can then help constrain the range of model input parameters.
A more detailed comparison of \snia\ data from the CfA SN program with
these models will be presented elsewhere.

\subsection{Results on spectra at other ages}\label{sect:othertres}

\cite{Bailey/etal:2009} restricted their analysis to spectra within
$\Delta t=2.5$\,d from $B$-band maximum. In this section we consider
flux ratios measured on spectra at other ages. We impose the same cuts
on relative flux calibration accuracy ($|\Delta(B-V)|<0.1$\,mag),
SALT2 color ($c<0.5$), redshift ($z>0.015$), and age range ($\Delta
t=2.5$\,d) as those used for the maximum-light spectra in the previous
section. We consider
all ages between $t=-2.5$\,d and $t=+7.5$\,d, in steps of 2.5\,d (for
ages earlier than $-2.5$\,d or later than +7.5\,d the number of
\sneia\ with spectra that satisfy our cuts falls below 20, and we do
not trust the results). We report the best ratio at each age in
Table~\ref{tab:kfoldcv_10_othert}, for both the $\mathcal{R}$-only
and $(c,\mathcal{R}^c)$ models.

\begin{table*}
\small
\caption{Top flux ratio at ages $-2.5\le t \le +7.5$\,d from 10-fold CV}\label{tab:kfoldcv_10_othert}
\begin{tabular}{cccrccrrc}
\hline\hline
Rank & $\lambda_X$ & $\lambda_Y$ & \multicolumn{1}{c}{$\gamma$} & WRMS
& $\sigma_{\rm pred}$ & \multicolumn{1}{c}{$\rho_{x_1,c}$} &
\multicolumn{1}{c}{$\Delta_{x_1,c}$} & $N_{\rm SNIa}$ \\
\hline
\multicolumn{9}{l}{$\mathcal{R}$} \\
\hline
 $-2.5$ & 6540 & 4580 & $-6.09 \pm 0.11$ & $0.182 \pm 0.025$ & $0.151 \pm 0.031$ & $ 0.70 \pm 0.13$ & $-0.032 \pm 0.028\ (1.1\sigma)$ & 24 \\
 $+0.0$ & 6630 & 4400 & $-4.37 \pm 0.09$ & $0.189 \pm 0.026$ & $0.163 \pm 0.030$ & $ 0.80 \pm 0.09$ & $-0.018 \pm 0.025\ (0.7\sigma)$ & 26 \\
 $+2.5$ & 6630 & 4040 & $-3.51 \pm 0.09$ & $0.203 \pm 0.027$ & $0.171 \pm 0.033$ & $ 0.63 \pm 0.13$ & $-0.012 \pm 0.033\ (0.4\sigma)$ & 26 \\
 $+5.0$ & 6590 & 4490 & $-4.69 \pm 0.12$ & $0.225 \pm 0.031$ & $0.203 \pm 0.034$ & $ 0.36 \pm 0.20$ & $ 0.022 \pm 0.041\ (0.5\sigma)$ & 26 \\
 $+7.5$ & 6590 & 4890 & $-3.40 \pm 0.20$ & $0.251 \pm 0.035$ & $0.229 \pm 0.039$ & $ 0.47 \pm 0.18$ & $ 0.044 \pm 0.043\ (1.0\sigma)$ & 25 \\
\hline
\multicolumn{9}{l}{$(c,\mathcal{R}^c)$} \\
\hline
 $-2.5$ & 4610 & 4260 & $ 2.19 \pm 0.14$ & $0.143 \pm 0.020$ & $0.106 \pm 0.028$ & $ 0.43 \pm 0.23$ & $-0.081 \pm 0.037\ (2.2\sigma)$ & 24 \\
 $+0.0$ & 6420 & 5290 & $-1.75 \pm 0.10$ & $0.175 \pm 0.025$ & $0.148 \pm 0.029$ & $ 0.80 \pm 0.09$ & $-0.032 \pm 0.023\ (1.4\sigma)$ & 26 \\
 $+2.5$ & 5550 & 6630 & $ 1.09 \pm 0.08$ & $0.169 \pm 0.022$ & $0.133 \pm 0.030$ & $ 0.79 \pm 0.09$ & $-0.049 \pm 0.027\ (1.8\sigma)$ & 26 \\
 $+5.0$ & 6540 & 5580 & $-5.18 \pm 0.48$ & $0.194 \pm 0.026$ & $0.166 \pm 0.031$ & $ 0.57 \pm 0.16$ & $-0.014 \pm 0.034\ (0.4\sigma)$ & 26 \\
 $+7.5$ & 6460 & 5510 & $-1.85 \pm 0.13$ & $0.200 \pm 0.028$ & $0.173 \pm 0.033$ & $ 0.79 \pm 0.09$ & $-0.007 \pm 0.027\ (0.3\sigma)$ & 25 \\
\hline
\end{tabular}
\end{table*}

\begin{figure*}
\begin{center}
\begin{tabular}{m{.12\linewidth} m{.38\linewidth} m{.38\linewidth}}
 & \multicolumn{1}{c}{WRMS$(\mathcal{R})$\ \ \ [mag]} & \multicolumn{1}{c}{WRMS$(c,\mathcal{R}^c)$\ \ \ [mag]} \\
$t=-2.5$\,d &
\includegraphics[bb=18 14 350 281,width=6cm]{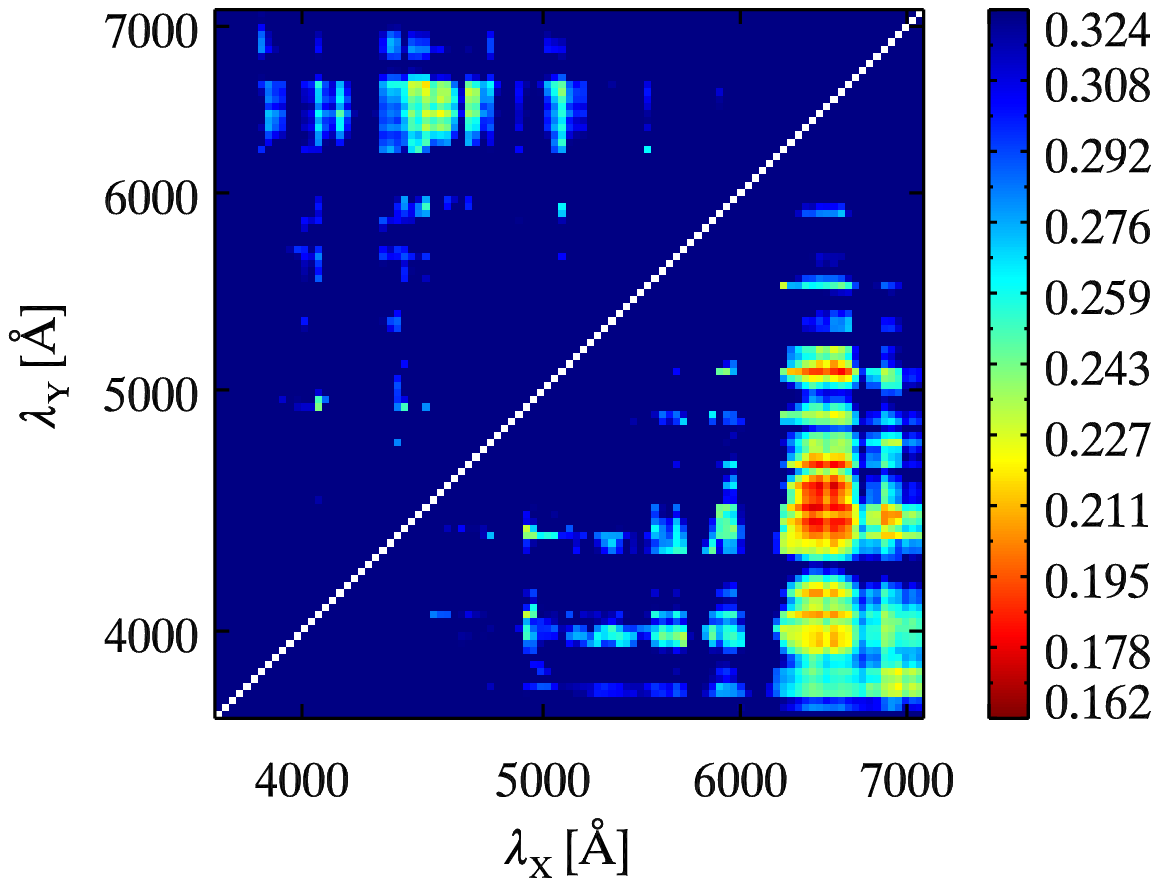} &
\includegraphics[bb=18 14 350 281,width=6cm]{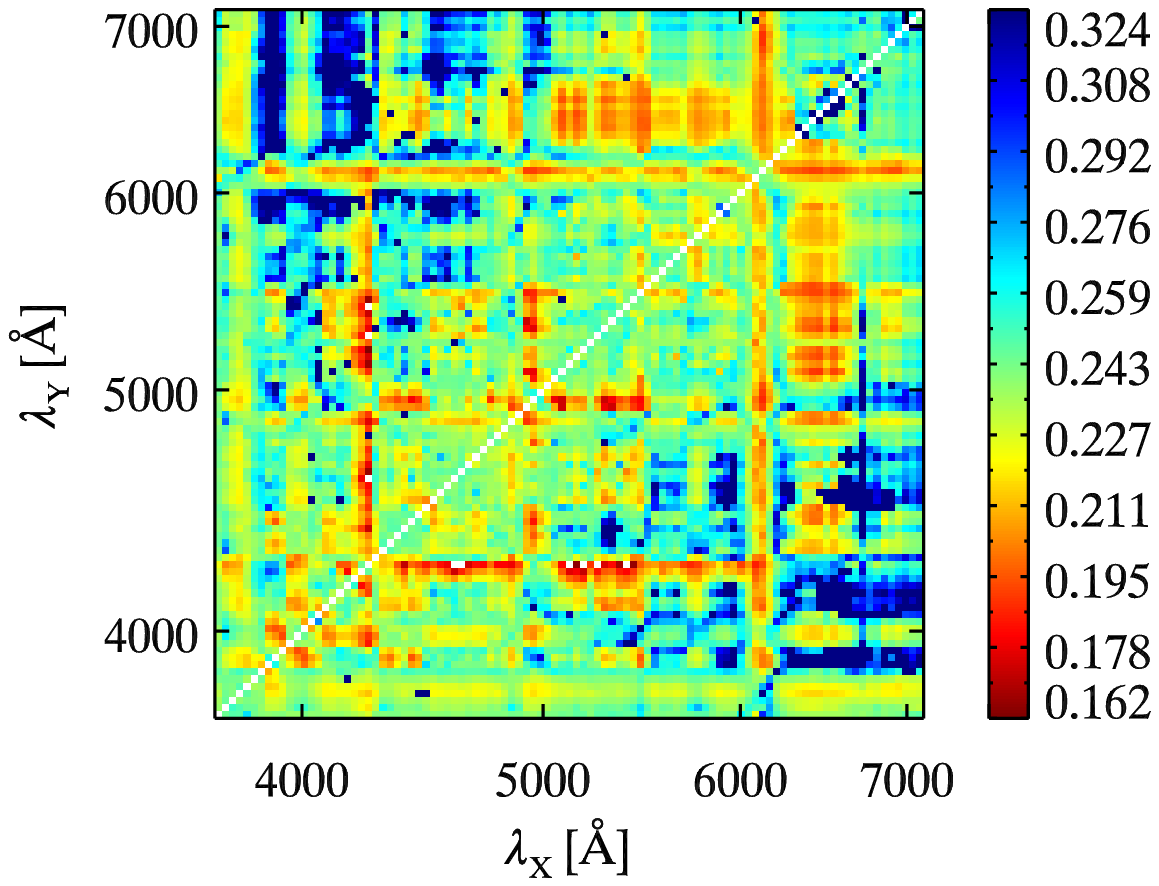} \\
$t=+0$\,d &
\includegraphics[bb=18 14 350 281,width=6cm]{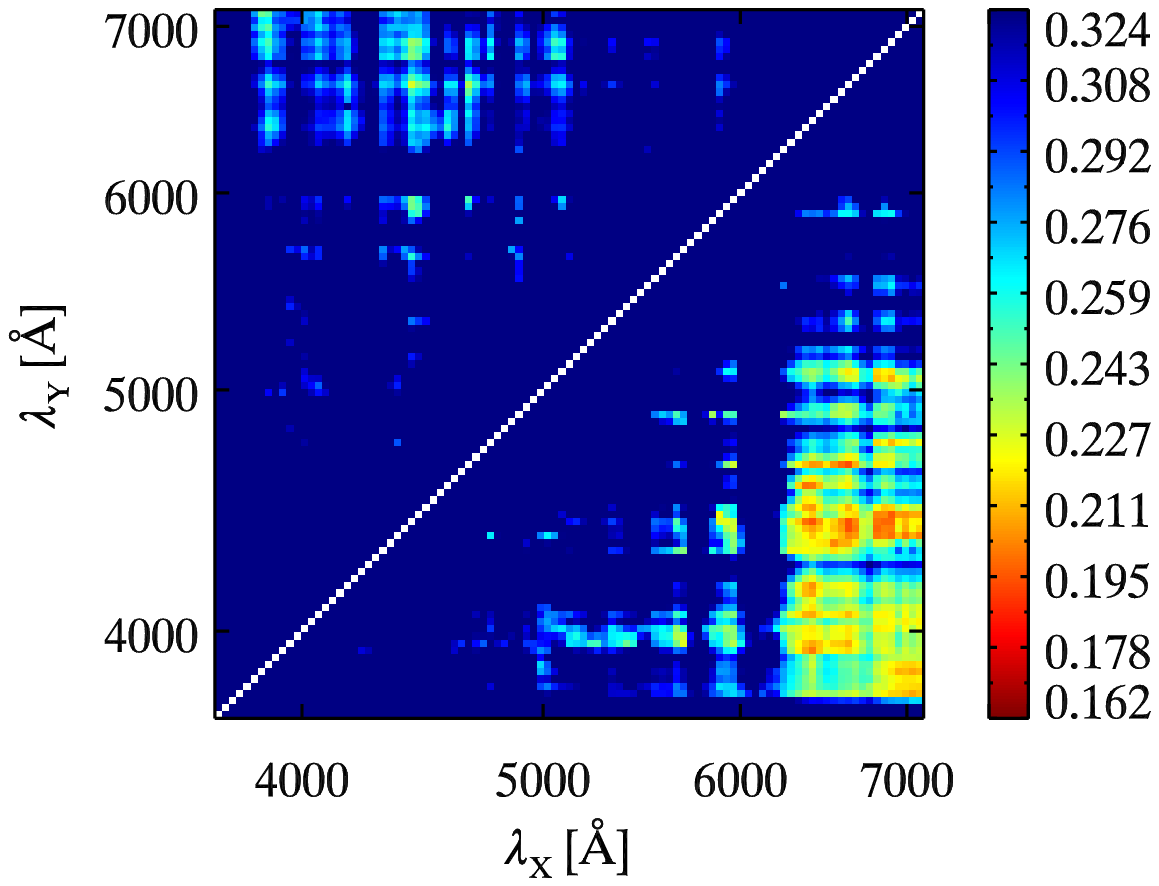} &
\includegraphics[bb=18 14 350 281,width=6cm]{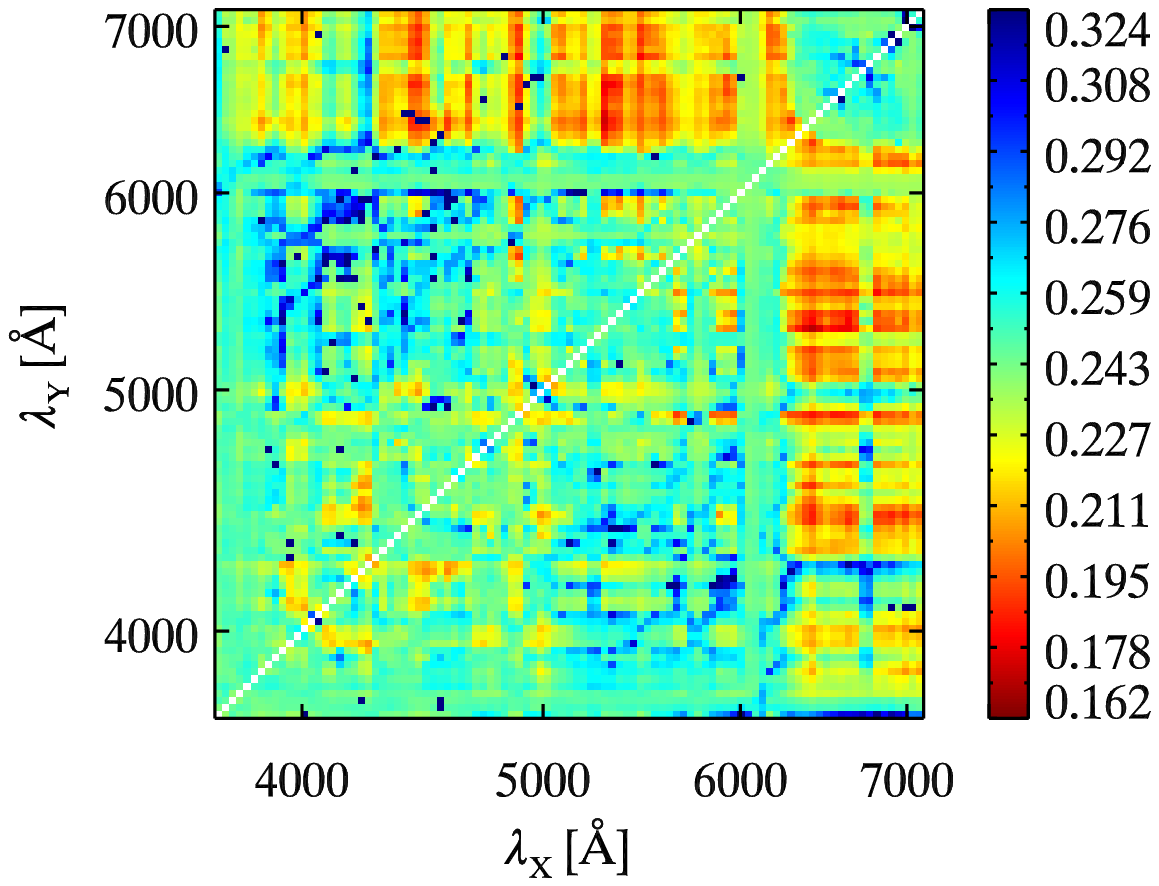} \\
$t=+5$\,d &
\includegraphics[bb=18 14 350 281,width=6cm]{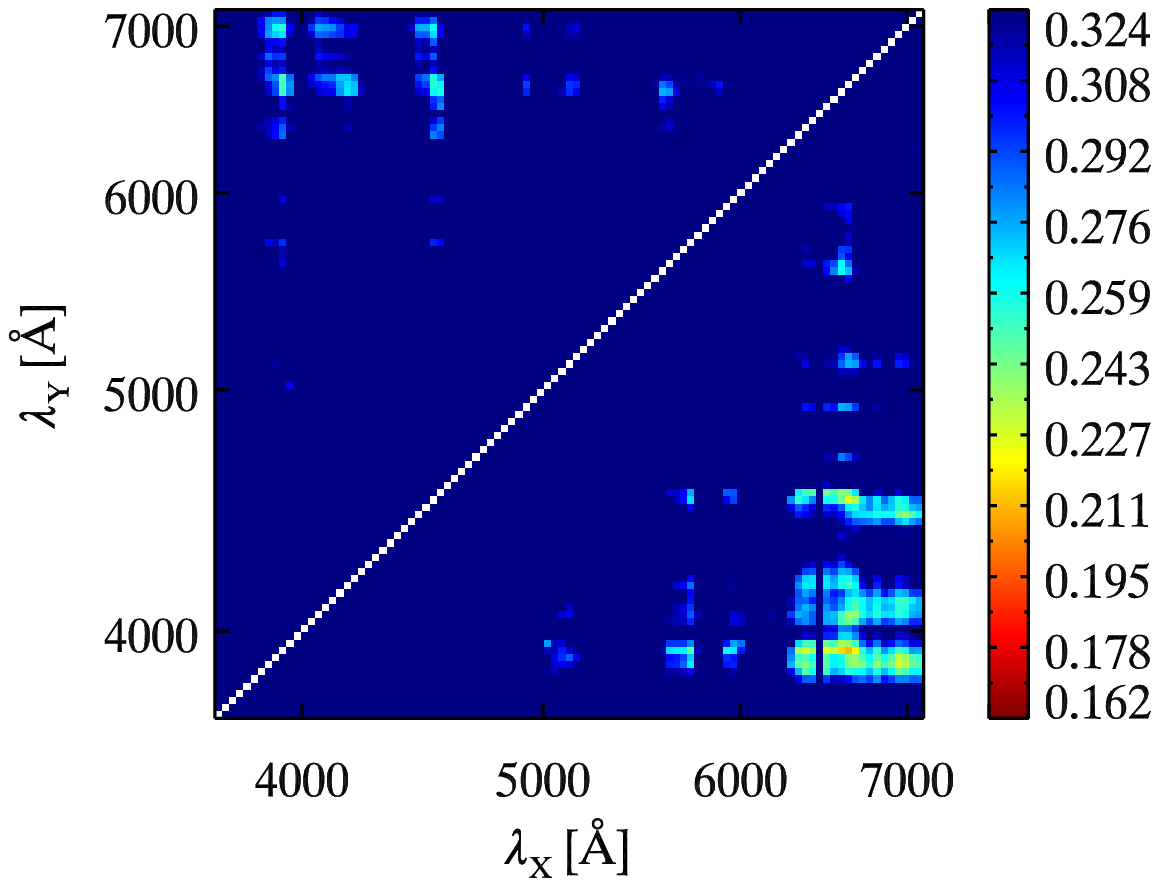} &
\includegraphics[bb=18 14 350 281,width=6cm]{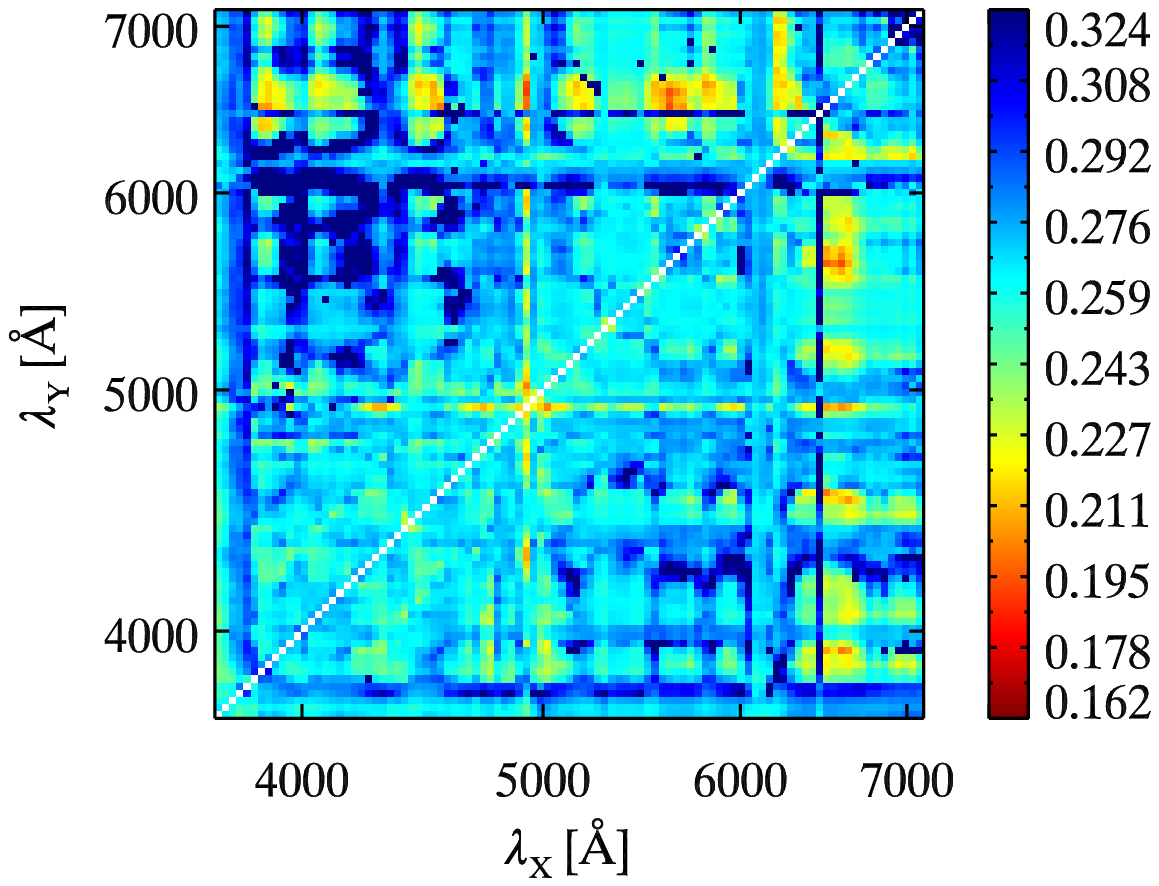} \\
$t=+7.5$\,d &
\includegraphics[bb=18 14 350 281,width=6cm]{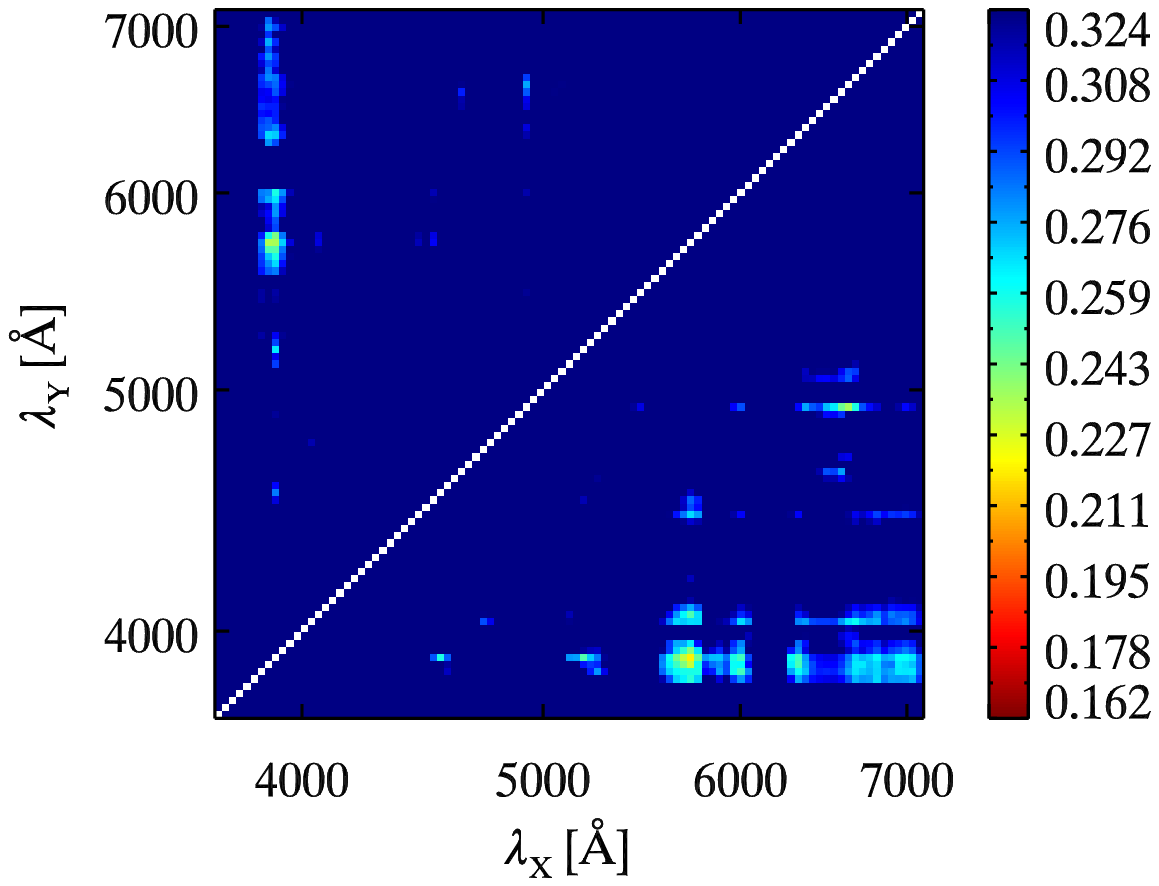} &
\includegraphics[bb=18 14 350 281,width=6cm]{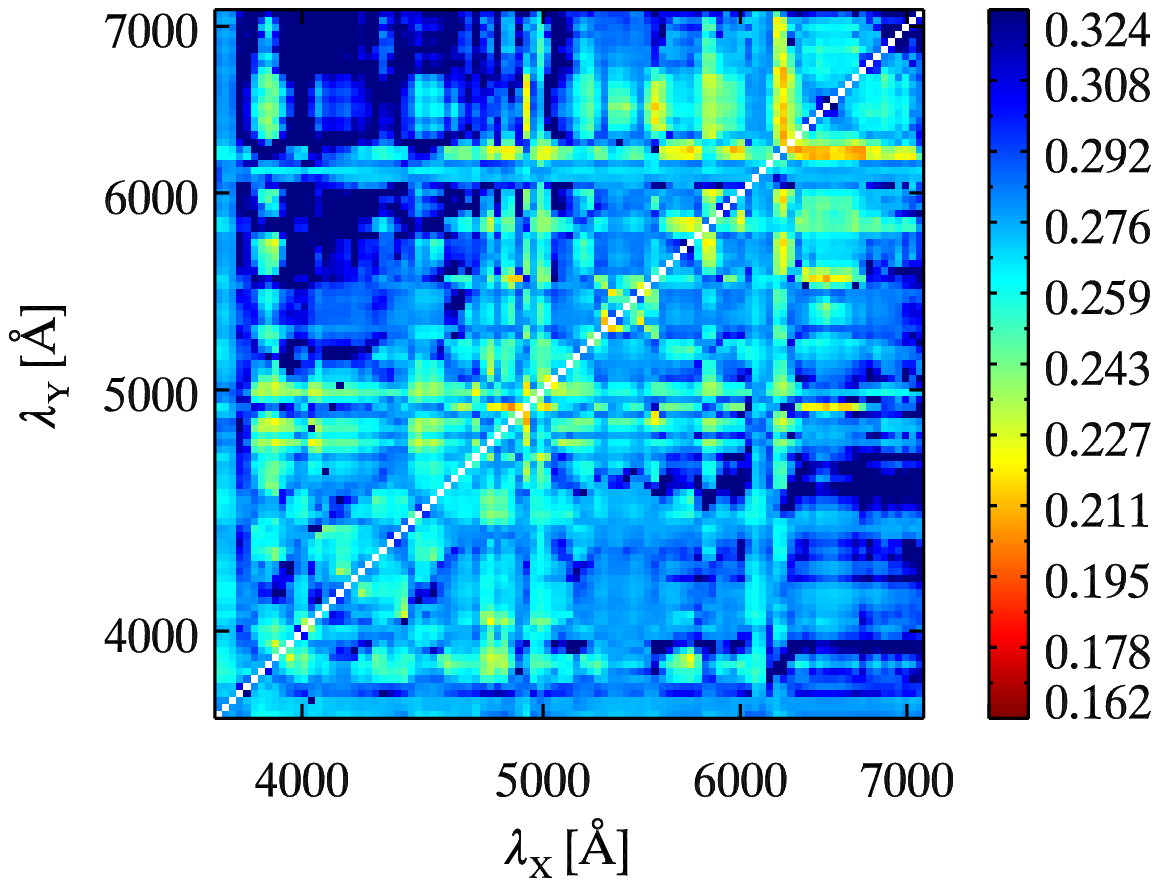} 
\end{tabular}
\caption{\label{fig:kfoldtres}
Results from 10-fold cross-validation on spectra at $t=-2.5,+0,+5,+7.5$\,d.
({\it from top to bottom}), color-coded according to the weighted rms 
of prediction Hubble residuals. The left column is corresponds to
$\mathcal{R}$ only, while the right column corresponds to the
$(c,\mathcal{R}^c)$ model.
}
\end{center}
\end{figure*}

The 2D maps of Hubble diagram residual scatter for spectra at ages
$t=-2.5$, +0, +5, and +7.5\,d are shown in Fig.~\ref{fig:kfoldtres}. As
was the case for maximum-light spectra, adding the SALT2 $x_1$ parameter
leads to slightly degraded results when compared with $\mathcal{R}$
alone, so we do not show plots for $(x_1,\mathcal{R})$ in
Fig.~\ref{fig:kfoldtres}. Moreover, we do not show results for
$(x_1,c,\mathcal{R}^c)$ since the best color-corrected flux ratios in
this case do not result in a significant improvement over
$(c,\mathcal{R}^c)$. At all the ages we consider here, the set of
predictors $(c,\mathcal{R}^c)$ results in lower weighted rms of
prediction residuals than $(x_1,c)$, although the significance of the
difference is $\lesssim2\sigma$ and is lower for $t\ge+5$\,d than
for $-2.5\le t \le +2.5$\,d. A flux ratio by itself only leads to an
improvement over $(x_1,c)$ near maximum light ($-2.5\le t \le
+2.5$\,d). As was the case at maximum light, there is a positive
intrinsic correlation in prediction error between all the distance
prediction models that include a flux ratio and $(x_1,c)$
[$0.4 \lesssim \rho_{x_1,c}\lesssim 0.8$].

%%% Fig. Results from 10-fold cross-validation on t=-5,+0,+5,+10 spectra

\begin{figure}
\centering
\resizebox{0.475\textwidth}{!}{\includegraphics{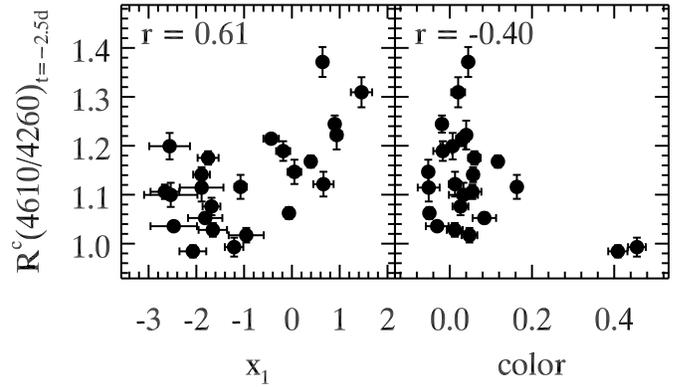}}
\caption{\label{fig:Rc_4610_4260_vs_x1c}
Correlation between $\mathcal{R}^c(4610/4260)$ at $t=-2.5$\,d and the
SALT2 fit parameters $(x_1,c)$. The Pearson coefficient of the correlation
with color drops to $r=0.03$ if we ignore the two points at $c>0.4$.
}
\end{figure}

The best set of predictors overall in the age range $-2.5\le t \le
+7.5$\,d is $[c,\mathcal{R}^c(4610/4260)]$ at $t=-2.5$\,d. The
weighted rms of prediction residuals is reduced by $\sim30$\% with
respect to $(x_1,c)$ [${\rm WRMS}=0.143\pm0.020$\,mag], and the
intrinsic prediction error by $\sim40$\% ($\sigma_{\rm
  pred}=0.106\pm0.028$\,mag), the significance of the difference being
$\sim2\sigma$ ($\Delta_{x_1,c}=-0.081\pm0.037$\,mag). We show the
correlation of $\mathcal{R}^c(4610/4260)$ at $t=-2.5$\,d and the SALT2
parameters $(x_1,c)$ in Fig.~\ref{fig:Rc_4610_4260_vs_x1c}. This
color-corrected ratio is mildly correlated with $x_1$ [$r=0.61$; this
  drops slightly to $r=0.47$ if we ignore the two points at
$\mathcal{R}^c(4610/4260)>1.3$] and uncorrelated with color 
($r=-0.40$; this drops to $r=0.03$ if we ignore the two points at
$c>0.4$). Interestingly, the wavelength bins that
constitute this ratio are part of the two prominent spectral
absorption features, predominantly due to iron-group elements,
that were found to vary intrinsically between \sneia\ based on the 2D
models discussed in \S~\ref{sect:modelcomp}.

%%% Fig. Rc(4610/4260) vs. x1,col

\subsection{Results using two flux ratios}\label{sect:twofr}

In this section we consider corrections using a linear combination of
two flux ratios as follows:
\begin{eqnarray}\label{eqn:fr2arr}
\mu &=& m_B - M + \gamma_1 \mathcal{R}_1 + \gamma_2 \mathcal{R}_2 \\
\mu &=& m_B - M - \beta c + \gamma_1 \mathcal{R}_1^c + \gamma_2 \mathcal{R}_2^c,
\end{eqnarray}
where the latter equation includes an additional correction due to
color, hence the use of color-corrected ratios $\mathcal{R}^c$. For
both cases, we fix $\mathcal{R}_1^{(c)}$ to the highest-ranked single
flux ratio [e.g., at maximum light: $\mathcal{R}_1(6630/4400)$ and
 $\mathcal{R}_1^c(6420/5290)$; see Table~\ref{tab:kfoldcv_10_p0}],
but leave both $(\gamma_1,\gamma_2)$ as free parameters (i.e. we do
not set $\gamma_1$ equal to $\gamma$ found in the single flux ratio
case).

\begin{figure}
\centering
\resizebox{0.475\textwidth}{!}{\includegraphics{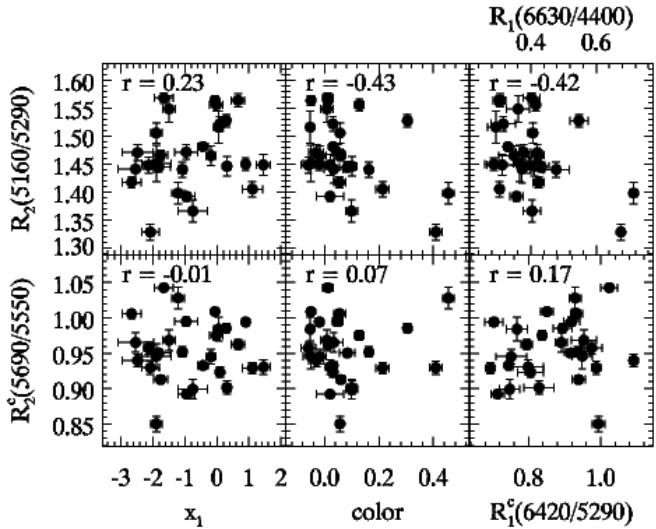}}
\caption{\label{fig:r2x1cr1}
Correlation between the highest-ranked
$(\mathcal{R}_2,\mathcal{R}^c_2)$ at maximum light and the SALT2
fit parameters $(x_1,c)$, and the highest-ranked
$(\mathcal{R}_1,\mathcal{R}^c_1)$.
}
\end{figure}

The results for the top five secondary flux ratios at maximum light
are displayed in Table~\ref{tab:kfoldcv_R2_p0}. In all cases,
including a second flux ratio further reduces the weighted rms of
prediction residuals by $\approx$15-20\% [${\rm
      WRMS}=0.162\pm0.022$\,mag for $\mathcal{R}_2(5160/5290)$; ${\rm
      WRMS}=0.151\pm0.021$\,mag for $\mathcal{R}_2^c(5690/5550)$]
with respect to the single flux ratio case [${\rm
      WRMS}=0.189\pm0.026$\,mag for $\mathcal{R}(6630/4400)$; ${\rm
      WRMS}=0.175\pm0.025$\,mag for
    $\mathcal{R}^c(6420/5290)$]. Again the significance of the
improvement ($\lesssim2\sigma$) is difficult to gauge given our sample
size, this despite the fact that $\gamma_2$ is significantly different
from zero in all cases. Our best secondary flux ratios,
$\mathcal{R}_2(5160/5290)$ and $\mathcal{R}^c_2(5690/5550)$, are
uncorrelated with the SALT2 fit parameters $(x_1,c)$ and with the
highest-ranked primary ratios $\mathcal{R}_1(6630/4400)$ and
$\mathcal{R}^c_1(6420/5290)$ [see Fig.~\ref{fig:r2x1cr1}], and hence
provide independent information on \snia\ luminosity. 

\begin{table*}
\small
\caption{Top 5 secondary flux ratios at maximum light from 10-fold CV on 26 \sneia}\label{tab:kfoldcv_R2_p0}
\begin{tabular}{cccrrccrr}
\hline\hline
Rank & $\lambda_X$ & $\lambda_Y$ & \multicolumn{1}{c}{$\gamma_1$} & \multicolumn{1}{c}{$\gamma_2$} & WRMS & $\sigma_{\rm pred}$ & \multicolumn{1}{c}{$\rho_{x_1,c}$} & \multicolumn{1}{c}{$\Delta_{x_1,c}$} \\
\hline
\multicolumn{9}{l}{$[\mathcal{R}_1(6630/4400),\mathcal{R}_2]$} \\
\hline
 1 & 5160 & 5290 & $-5.05 \pm 0.17$ & $-2.24 \pm 0.22$ & $0.162 \pm 0.022$ & $0.127 \pm 0.028$ & $ 0.32 \pm 0.23$ & $-0.052 \pm 0.038\ (1.4\sigma)$ \\
 2 & 5290 & 5160 & $-5.08 \pm 0.18$ & $ 4.78 \pm 0.50$ & $0.167 \pm 0.022$ & $0.133 \pm 0.029$ & $ 0.31 \pm 0.24$ & $-0.046 \pm 0.036\ (1.3\sigma)$ \\
 3 & 5120 & 5290 & $-5.19 \pm 0.11$ & $-1.75 \pm 0.20$ & $0.172 \pm 0.024$ & $0.141 \pm 0.029$ & $ 0.41 \pm 0.21$ & $-0.038 \pm 0.036\ (1.1\sigma)$ \\
 4 & 5690 & 5360 & $-3.93 \pm 0.12$ & $-2.06 \pm 0.21$ & $0.172 \pm 0.024$ & $0.141 \pm 0.029$ & $ 0.46 \pm 0.20$ & $-0.038 \pm 0.032\ (1.2\sigma)$ \\
 5 & 5690 & 5290 & $-4.14 \pm 0.11$ & $-1.26 \pm 0.13$ & $0.172 \pm 0.024$ & $0.142 \pm 0.029$ & $ 0.48 \pm 0.19$ & $-0.038 \pm 0.032\ (1.2\sigma)$ \\
\hline
\multicolumn{9}{l}{$[c,\mathcal{R}_1^c(6420/5290),\mathcal{R}_2^c]$} \\
\hline
 1 & 5690 & 5550 & $-1.69 \pm 0.09$ & $-1.93 \pm 0.21$ & $0.151 \pm 0.021$ & $0.115 \pm 0.027$ & $ 0.58 \pm 0.17$ & $-0.063 \pm 0.030\ (2.1\sigma)$ \\
 2 & 5690 & 5960 & $-2.34 \pm 0.11$ & $-1.90 \pm 0.23$ & $0.152 \pm 0.020$ & $0.116 \pm 0.028$ & $ 0.42 \pm 0.21$ & $-0.064 \pm 0.035\ (1.8\sigma)$ \\
 3 & 5960 & 5690 & $-2.35 \pm 0.11$ & $ 2.05 \pm 0.24$ & $0.153 \pm 0.021$ & $0.117 \pm 0.028$ & $ 0.43 \pm 0.21$ & $-0.063 \pm 0.036\ (1.8\sigma)$ \\
 4 & 5550 & 5690 & $-1.70 \pm 0.09$ & $ 1.72 \pm 0.20$ & $0.152 \pm 0.021$ & $0.118 \pm 0.027$ & $ 0.58 \pm 0.17$ & $-0.061 \pm 0.031\ (2.0\sigma)$ \\
 5 & 4890 & 5690 & $-1.25 \pm 0.10$ & $ 1.22 \pm 0.24$ & $0.158 \pm 0.021$ & $0.122 \pm 0.028$ & $ 0.62 \pm 0.15$ & $-0.052 \pm 0.026\ (2.0\sigma)$ \\
\hline
\end{tabular}
\end{table*}

%%% Fig. Best R2/R2c vs. x1,col

The secondary flux ratios $\mathcal{R}_2$ listed in
Table~\ref{tab:kfoldcv_R2_p0} have a much smaller wavelength baseline
than the primary ratios $\mathcal{R}_1$ and $\mathcal{R}^c_1$ (see
Table~\ref{tab:kfoldcv_10_p0}). The highest-ranked secondary ratios
have a 130\,\AA\ and 140\,\AA\ baseline, respectively. These ratios do
not measure \snia\ colors: they measure small-scale intrinsic
spectroscopic variations. Interestingly, all the wavelength bins that
form these secondary ratios are clustered around the
\stwo\,\l\l5454,5640 doublet and the iron-group-dominated absorption complex
\fetwo\,\l4800 mentioned in \S~\ref{sect:othertres} (see also
\S~\ref{sect:modelcomp}). As was the case for a single flux ratio, the
results using a secondary flux ratio are similar for $-2.5\le t \le
+2.5$\,d, and tend to be worse at later ages. We choose not 
to discuss them further.

In a recent paper\footnote{At the time of writing, the paper by
\cite{Yu/Yang/Lu:2009} has not been accepted for publication. Here we
refer to the 2$^{\rm nd}$ version of their paper, dated 30$^{\rm th}$
June 2010.}, \cite{Yu/Yang/Lu:2009} have searched for flux ratio pairs
that minimize Hubble diagram residuals (with no color correction), and
find several such pairs which achieve a standard deviation
$\sigma\lesssim0.10$\,mag at ages between $-3$\,d and +12\,d from
maximum light. We have validated the flux ratio pairs reported in their
Table~4 and find that none of them lead to an improvement compared
with the standard $(x_1,c)$ model. There could be several reasons
for this disagreement: \cite{Yu/Yang/Lu:2009} do not use a
cross-validation procedure and their sample size (anywhere
between 17 and 24 \sneia\ depending on the age and flux ratio pair
considered) suggests they may be overfitting a small sample. Moreover,
they use color-corrected flux ratios (actually corrected for the
host-galaxy extinction, $A_V$, as opposed to the SALT2 color
parameter) but do not include a color parameter in their equation to
correct for the \snia\ magnitudes. When we use the SALT2 color parameter in
addition to the same flux ratio pairs as \cite{Yu/Yang/Lu:2009},
several pairs indeed lead to an improvement over the standard
$(x_1,c)$ model, but not over the correction using a single flux
ratio (or a single flux ratio in combination with color) we find in
this paper. Last, \cite{Yu/Yang/Lu:2009} do not impose a redshift cut
on their sample: only 14 out of 38 \sneia\ are at redshifts $z>0.015$,
and 5 are at redshifts $z<0.005$, where the magnitude error due to
peculiar velocities is $\sigma_{\rm pec} > 0.4$\,mag. It is unclear
why this should lead to a lower scatter in Hubble residuals (on the
contrary one expects an increased scatter), but it certainly impacts
their analysis.

%%%%%%%%%%%%%%%%%%%%%%%%%%%%%%%%%%%%%%%%%%%%%%%%%%%%%%%%%%%%%%%%%%
%%
%%   Other spectroscopic indicators
%%
%%%%%%%%%%%%%%%%%%%%%%%%%%%%%%%%%%%%%%%%%%%%%%%%%%%%%%%%%%%%%%%%%%

\section{Other spectroscopic indicators}\label{sect:specindic}

In this section we consider other spectroscopic indicators, mostly
related to spectral line profile morphology. Some of these indicators
are also flux ratios, but the wavelengths correspond to precise
locations of absorption troughs or emission peaks in the \snia\
spectrum, as opposed to the ``blind'' approach of computing flux
ratios from all possible wavelength bins with no {\it a priori}
physical motivation. We use the same approach as for the flux ratios,
i.e. we consider models for predicting \snia\ distances which include a
spectroscopic indicator, possibly in combination with a light-curve 
parameter (cf. Eq.~\ref{eqn:intro}), and we validate each
model using $K$-fold cross-validation (and present results for
$K=10$ in this section).

\subsection{Measurements}

\begin{figure}
\centering
\resizebox{0.475\textwidth}{!}{\includegraphics{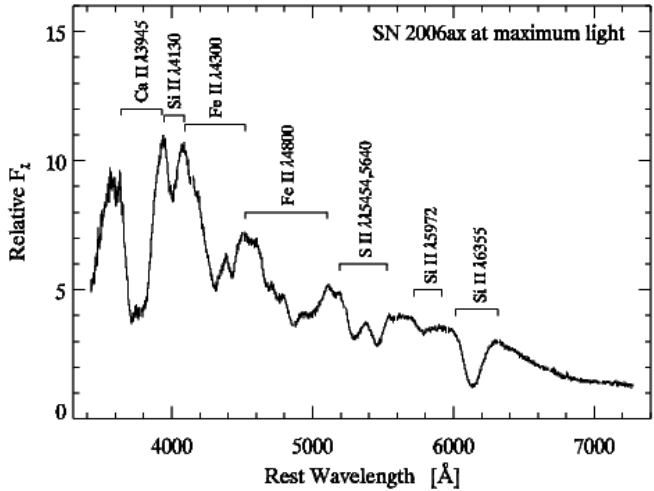}}
\caption{\label{fig:specline}
Wavelength bounds of spectroscopic features for which we measured the
various indicators shown in Fig.~\ref{fig:specindic}, illustrated
using the maximum-light spectrum of SN~2006ax.
}
\end{figure}

We divide the \snia\ spectrum into several ``features'', each labeled
according to the strongest line in that wavelength
range. Figure~\ref{fig:specline} shows the seven features we consider
in this paper, from \catwo\,\l3945 in the blue to \sitwo\,\l6355 in
the red. The wavelengths associated with each feature correspond to
the $gf$-weighted mean wavelength of the different atomic transitions
for the specified ion (e.g. 3945\,\AA\ for the \catwo\ H \& K lines),
except for the two large features dominated by \fetwo\ lines, where
the wavelengths denote the approximate location of the deepest
absorption ($\sim4300$\,\AA\ and $\sim4800$\,\AA). The \fetwo\,\l4300
feature also includes contributions from \mgtwo, and possibly
\fethree\ for the most luminous \sneia. For the faintest, 1991bg-like
\sneia, \titwo\ constitutes a dominating source of opacity in this
wavelength region. Since we do not include 1991bg-like \sneia\ in our
analysis, however, we do not present measurements for \titwo.

%%% Fig. specline

\begin{figure}
\centering
\resizebox{0.475\textwidth}{!}{\includegraphics{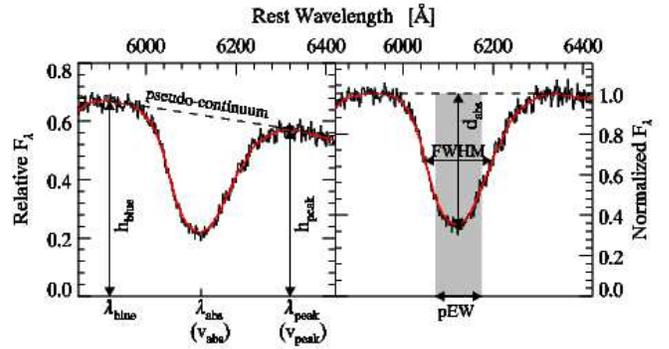}}
\caption{\label{fig:specindic}
Definition of the main spectroscopic indicators used in this paper,
here illustrated using the \sitwo\ \l6355 line profile in the
spectrum of SN~2005ki at $t=+1$\,d. The right panel shows the
pseudo-continuum ({\it dashed line}), as well as the wavelength
locations of the blue and red emission peaks ($\l_{\rm blue}$ and
$\l_{\rm peak}$) and their respective heights ($h_{\rm blue}$ and
$h_{\rm peak}$). The wavelength of maximum absorption ($\l_{\rm abs}$)
serves to define the absorption velocity, $v_{\rm abs}$. The peak
velocity $v_{\rm peak}$ is defined analogously.
The left panel shows the same line profile normalized to the
pseudo-continuum, and serves to define the (relative) absorption depth
($d_{\rm abs}$), FWHM, and pseudo-equivalent width (pEW; {\it shaded
  gray region}). In both panels, the thick line corresponds to the
smoothed flux, where we have used the inverse-variance weighted
Gaussian filter of \cite{Blondin/etal:2006} with a smoothing factor
$d\l/\l=0.005$.
}
\end{figure}

The various spectroscopic indicators we consider are illustrated in
Fig.~\ref{fig:specindic}, based on the \sitwo\,\l6355 line profile in
the spectrum of SN~2005ki at $t=+1$\,d from maximum light. We first
smooth the spectrum using the inverse-variance Gaussian filter of
\cite{Blondin/etal:2006} with a smoothing factor $0.001 < d\l/\l <
0.01$ determined based on a $\chi^2$ test using flux errors from the
variance spectra (Fig.~\ref{fig:specindic}; {\it thick line}). The
smoothed spectrum makes it easier to define wavelength locations of
local flux maxima on either side of the absorption component of the P
Cygni profile ($\l_{\rm blue}$ and $\l_{\rm peak}$), as well as the
location of maximum absorption ($\l_{\rm abs}$). The wavelengths $\l_{\rm abs}$ and
$\l_{\rm peak}$ are then used to define the absorption and peak
velocities, respectively ($v_{\rm abs}$ and $v_{\rm peak}$), using the
relativistic Doppler formula (see also
\citealt{Blondin/etal:2006}). We also measure the heights of the local
maximum ($h_{\rm blue}$ and $h_{\rm peak}$), and define a
pseudo-continuum between them. These latter quantities are measured on
the original, unsmoothed spectrum. Division by this pseudo-continuum
enables us to measure the relative absorption depth ($d_{\rm abs}$)
and full-width at half-maximum (FWHM) of the absorption component, as
well as its pseudo-equivalent width (pEW; defined analogously to the
equivalent width used by stellar spectroscopists for abundance
determinations, but without the physical basis, hence ``pseudo'' EW;
Fig.~\ref{fig:specindic}; {\it right panel}). 

%%% Fig. specindic

We measure these quantities for all the features presented in
Fig.~\ref{fig:specline}, except for the complex \fetwo\,\l4300 and 
\fetwo\,\l4800 features for which we only consider the
pseudo-equivalent width. The error on each measured quantity includes
errors due to redshift, relative flux calibration, host-galaxy
extinction and contamination, and of course the flux error. We only
consider measurements for which the mean S/N over the entire feature
is greater than 5 per \AA, and require a minimum of 20 \sneia\ with
valid measurements. Note that we do not impose cuts on relative flux
calibration accuracy or SALT2 color, as was the case for the flux
ratios, since these quantities are mostly local measurements which are
far less sensitive to the overall SED.

We also consider various spectroscopic ratios, which were found to
correlate with absolute magnitude, defined below:
\begin{eqnarray}\label{eqn:rcasi}
\mathcal{R}(\rm Ca)    &=& \frac{h_{\rm peak}(\rm \catwo\,\l3945)}{h_{\rm blue}(\rm \catwo\,\l3945)} \label{eqn:rcasi1} \\
\mathcal{R}(\rm CaS)   &=& \frac{\int_{3887}^{4012}F_\l d\l}{\int_{3620}^{3716}F_\l d\l} \label{eqn:rcasi2} \\
\mathcal{R}(\rm Si)    &=& \frac{d_{\rm abs}(\rm \sitwo\,\l5972)}{d_{\rm abs}(\rm \sitwo\,\l6355)} \label{eqn:rcasi3} \\
\mathcal{R}(\rm SiS)   &=& \frac{h_{\rm peak}(\rm \sitwo\,\l6355)}{h_{\rm peak}(\rm \stwo\,\l5640)} \label{eqn:rcasi4} \\
\mathcal{R}(\rm SiSS)  &=& \frac{\int_{5500}^{5700}F_\l d\l}{\int_{6200}^{6450}F_\l d\l} \label{eqn:rcasi5} \\
\mathcal{R}(\rm S,Si)  &=& \frac{\rm pEW(\stwo\,\l\l5454,5640)}{\rm pEW(\sitwo\,\l5972)} \label{eqn:rcasi6} \\
\mathcal{R}(\rm Si,Fe) &=& \frac{\rm pEW(\sitwo\,\l5972)}{\rm pEW(\fetwo\,\l4800)}. \label{eqn:rcasi7}
\end{eqnarray}
The ratios $\mathcal{R}(\rm Ca)$ and $\mathcal{R}(\rm Si)$ were both defined by
\cite{Nugent/etal:1995}, and found to correlate well with the
luminosity decline rate parameter $\Delta m_{15}(B)$. 
To increase the S/N of the $\mathcal{R}(\rm Ca)$ measurement,
\cite{Bongard/etal:2006} introduced the corresponding integral flux ratio
$\mathcal{R}(\rm CaS)$, also found to correlate with absolute
magnitude. Using a grid of LTE synthetic spectra to investigate the
$\mathcal{R}(\rm Si)$ wavelength region, \cite{Bongard/etal:2006} also 
defined a ratio of the red local maximum of \sitwo\,\l6355 to the red
local maximum of \stwo\,\l5640, noted $\mathcal{R}(\rm SiS)$. The
corresponding integral flux ratio is $\mathcal{R}(\rm SiSS)$, again
introduced by \cite{Bongard/etal:2006} to increase the S/N of the
$\mathcal{R}(\rm SiS)$ measurement. Last,
\cite{Hachinger/Mazzali/Benetti:2006} measured the absorption
velocities and pseudo-EW in 28 \snia\ spectra and found two additional
pEW ratios, $\mathcal{R}(\rm S,Si)$ and $\mathcal{R}(\rm Si,Fe)$, that
are good indicators of luminosity.
Note that $\mathcal{R}(\rm Ca)$ and $\mathcal{R}(\rm SiS)$ are in fact flux
ratios similar to those defined by \cite{Bailey/etal:2009}.

\subsection{Results}

We present our results using the absorption velocity ($v_{\rm abs}$;
units of $10^4$\,\kms), the full-width at half-maximum (FWHM; units of
$10^2$\,\AA), the relative absorption depth ($d_{\rm abs}$), the
pseudo-equivalent width (pEW; units of $10^2$\,\AA), and the various 
spectroscopic ratios $\mathcal{R}(X)$
[Eqs.~\ref{eqn:rcasi1}-\ref{eqn:rcasi7}] in
Tables~\ref{tab:vabs}-\ref{tab:rx} (Appendix~\ref{app:tabs}). We do not
present results for the peak velocity ($v_{\rm peak}$) as they are far
worse than for the other indicators. There were not enough valid
measurements for \catwo\,\l3945, hence the absence of this line in
Tables~\ref{tab:vabs}-\ref{tab:pew}. We only report results for the
bluer absorption of the \stwo\ doublet (\stwo\,\l5454) in
Tables~\ref{tab:vabs}-\ref{tab:dabs}, but the pseudo-equivalent width
is that of the entire doublet (see Table~\ref{tab:pew}).

Based on the difference in intrinsic prediction error 
with respect to the standard model which uses the SALT2 
fit parameters $(x_1,c)$, again noted $\Delta_{x_1,c}$, we see from
Tables~\ref{tab:vabs}-\ref{tab:rx} that none of these spectroscopic
indicators {\it alone} leads to a lower weighted rms of prediction
residuals (i.e. $\Delta_{x_1,c}>0$). At best they are consistent with
no improvement at all [e.g. pEW(\sitwo\,\l4130), for which
  $\Delta_{x_1,c}=0.041\pm0.033$\,mag)]. The same is true at ages other
than maximum light.

Nonetheless, several such indicators compete well with $(x_1,c)$, even
leading to small improvements (albeit statistically insignificant),
but only when combined with SALT2 color [pEW(\sitwo\,\l4130) and
$\mathcal{R}$(Si)] or in addition to $(x_1,c)$ [$v_{\rm
 abs}$(\sitwo\,\l6355 and $d_{\rm abs}$(\stwo\,\l5454)]. We discuss
these indicators in the two following sections.

\subsubsection{Spectroscopic indicators in combination with SALT2 color}

\begin{figure}
\centering
\resizebox{0.475\textwidth}{!}{\includegraphics{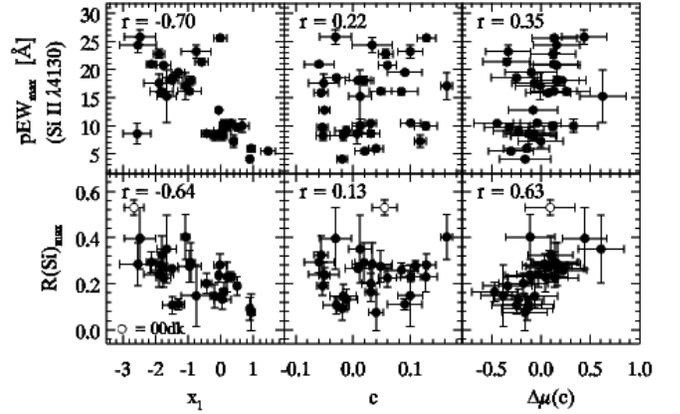}}
\caption{\label{fig:rsix1c}
Correlation between pEW(\sitwo\,\l4130) and $\mathcal{R}({\rm Si})$ at
maximum light and the SALT2 fit parameters $(x_1,c)$, and color-corrected
Hubble residual. The open circle in the lower panels corresponds to
SN~2000dk.
}
\end{figure}

When used in combination with the SALT2 color parameter, both the
pseudo-equivalent width of \sitwo\,\l4130 and the $\mathcal{R}$(Si)
spectroscopic ratio compete well with the standard $(x_1,c)$
predictors ($\Delta_{x_1,c}=0.006\pm0.014$\,mag and
$\Delta_{x_1,c}=-0.007\pm0.030$\,mag). Both indicators are strongly
anti-correlated with $x_1$ and uncorrelated with SALT2 color
(see Fig.~\ref{fig:rsix1c}, {\it left and middle panels}), while the
correlation with color-corrected Hubble residual is more pronounced
for $\mathcal{R}$(Si) [$r=0.63$] than for pEW(\sitwo\,\l4130)
[$r=0.35$]. In a sense, both indicators act to replace the light-curve
width parameter $x_1$. The anti-correlation of $\mathcal{R}$(Si) with
$x_1$ has been recovered by several authors since its publication by
\cite{Nugent/etal:1995}, while the relation between
pEW(\sitwo\,\l4130) and light-curve shape has more recently been
mentioned by \cite{Arsenijevic/etal:2008} and \cite{Walker/etal:2010}.

%%% Fig. pEW(SiII 4130) and R(Si) vs. x1,col,hubres

\begin{figure}
\centering
\resizebox{0.475\textwidth}{!}{\includegraphics{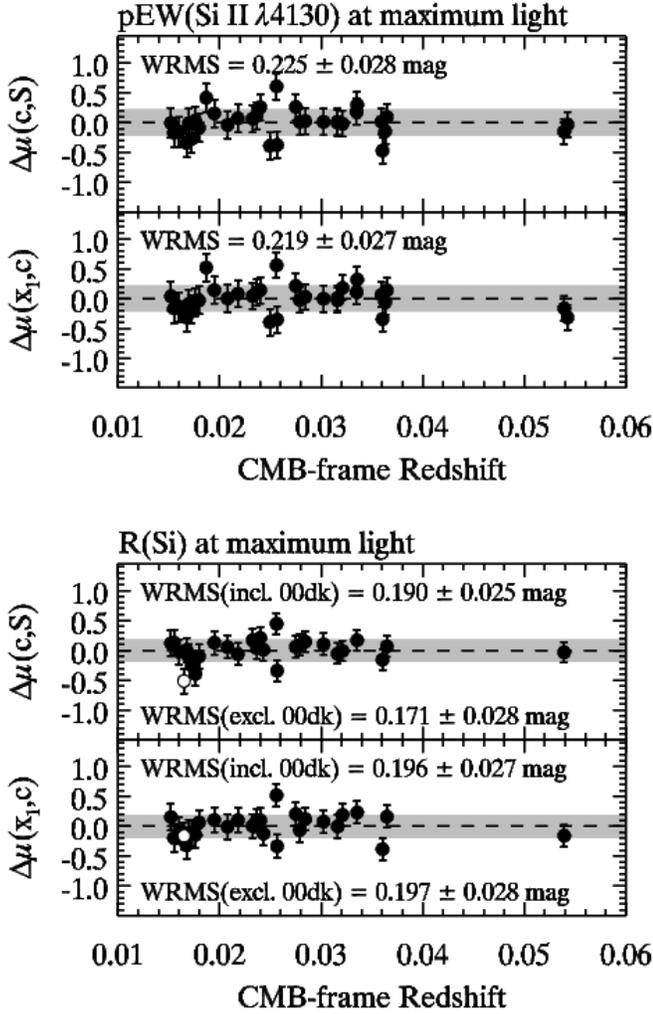}}
\caption{\label{fig:rsihubdiag}
Hubble diagram residuals for pEW(\sitwo\,\l4130) [{\it top}] and
$\mathcal{R}({\rm Si})$ [{\it bottom}] at maximum light. 
In each case we show the Hubble residuals obtained using SALT2 color
and the spectroscopic indicator ({\it upper panels}), and using the
standard SALT2 fit parameters $(x_1,c)$ ({\it lower panels}). 
We also indicate the weighted rms of Hubble
residuals ({\it gray highlighted region}). For the $\mathcal{R}({\rm
  Si})$ spectroscopic indicator, we report the weighted rms both
including and excluding SN~2000dk ({\it open circle}).
}
\end{figure}

We show the Hubble residuals obtained when using pEW(\sitwo\,\l4130)
and $\mathcal{R}$(Si) in combination with SALT2 color in
Fig.~\ref{fig:rsihubdiag}, where we also show the residuals from the
standard $(x_1,c)$ model. The subluminous (but not 1991bg-like)
SN~2000dk stands out as a $\lesssim2\sigma$ outlier for
$[c,\mathcal{R}$(Si)], while this is not the case for $(x_1,c)$ [the
  point corresponding to SN~2000dk is highlighted in both
  Figs.~\ref{fig:rsix1c} and \ref{fig:rsihubdiag}]. This
single SN contributes a large fraction of the residual scatter 
  [${\rm WRMS\ (incl. 00dk)}=0.190\pm0.025$\,mag cf. $0.196\pm0.027$\,mag
  for $(x_1,c)$], and
excluding it from the sample leads to a $\sim10\%$ decrease in the
weighted rms of prediction Hubble residuals, resulting in a
$\sim15$\% improvement over $(x_1,c)$ 
  [${\rm WRMS\ (excl. 00dk)}=0.171\pm0.028$\,mag cf. $0.197\pm0.028$\,mag
  for $(x_1,c)$].

%%% Fig. Hubble diagrams for pEW(SiII 4130) and R(Si)

\subsubsection{Spectroscopic indicators in addition to the SALT2
  fit parameters $(x_1,c)$}

\begin{figure}
\centering
\resizebox{0.475\textwidth}{!}{\includegraphics{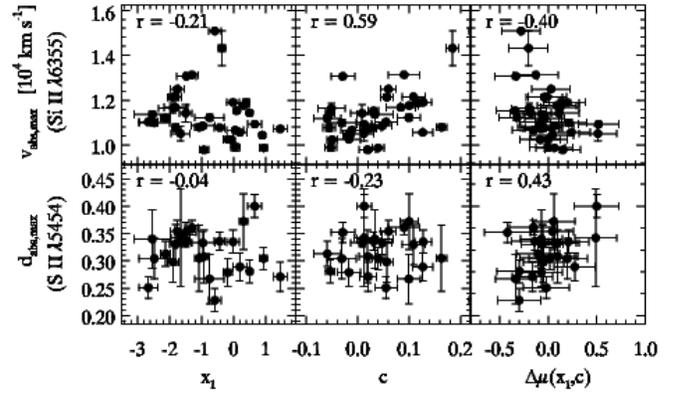}}
\caption{\label{fig:vdabsx1c}
Correlation between $v_{\rm abs}$(\sitwo\,\l6355) and $d_{\rm
  abs}$(\stwo\,\l5454) at maximum light and the SALT2 
fit parameters $(x_1,c)$, and $(x_1,c)$-corrected Hubble residual.
}
\end{figure}

When used in addition to the standard SALT2 fit parameters $(x_1,c)$, both the
absorption velocity of \sitwo\,\l6355 and the relative
absorption depth of \stwo\,\l5454 result in a
$\lesssim10$\% decrease in the weighted rms of prediction residuals 
($\Delta_{x_1,c}=-0.020\pm0.019$\,mag and
$\Delta_{x_1,c}=-0.022\pm0.030$\,mag), although the apparent
  improvement for $d_{\rm abs}$(\stwo\,\l5454) is due to
  the fact that the weighted rms of prediction residuals for the
  $(x_1,c)$ model is somewhat larger for this particular sample (${\rm
  WRMS}=0.221\pm0.031$\,mag).
Both these spectroscopic indicators are uncorrelated with $x_1$ or
color [Fig.~\ref{fig:vdabsx1c}; the apparent correlation of $v_{\rm
 abs}$(\sitwo\,\l6355) with color ($r=0.59$) is destroyed if we ignore
the one point at $c\approx0.2$], and thus provide additional
information independent of light-curve shape or color. The correlation
with $(x_1,c)$-corrected Hubble residuals (Fig.~\ref{fig:vdabsx1c},
{\it right panels}) is only modest ($|r|\approx0.40$ for both
indicators), and should be reviewed as more data become publicly
available. 

%%% Fig. vabs,FWHM,dabs vs. x1,col,hubres

\begin{figure}
\centering
\resizebox{0.475\textwidth}{!}{\includegraphics{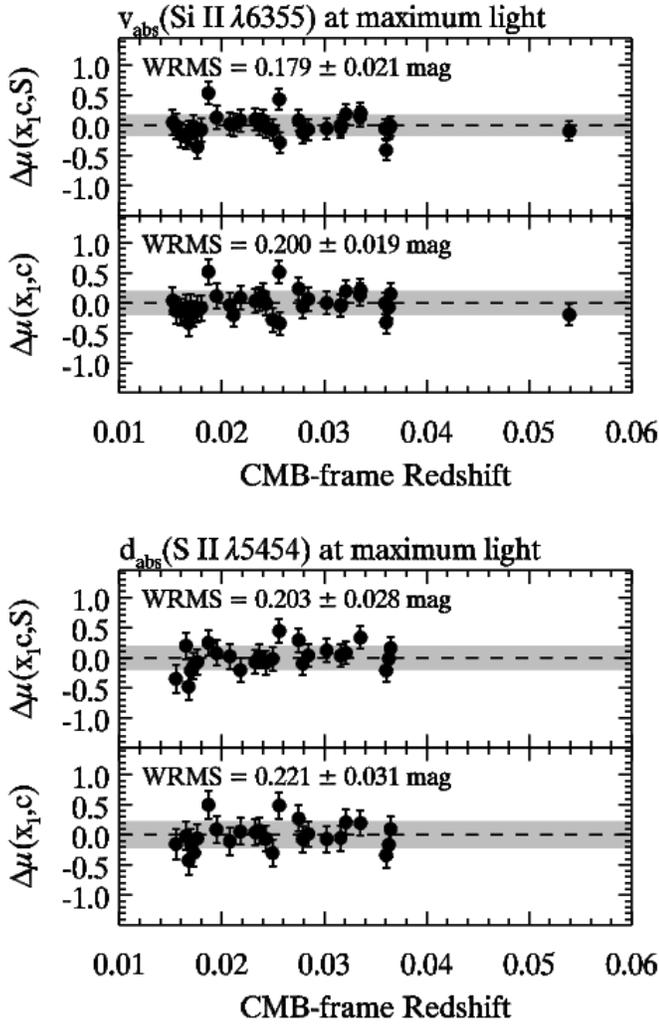}}
\caption{\label{fig:vdabshubdiag}
Hubble diagram residuals for $v_{\rm abs}$(\sitwo\,\l6355) [{\it top}], 
and $d_{\rm abs}$(\stwo\,\l5454) [{\it bottom}] at maximum light. 
In each case we show the Hubble residuals obtained using the
spectroscopic indicator in addition to the SALT2 fit
parameters $(x_1,c)$ ({\it upper panels}), and using $(x_1,c)$ only ({\it lower
 panels}).
We also indicate the weighted rms of Hubble residuals ({\it gray
  highlighted region}).
}
\end{figure}

We show the Hubble residuals obtained when using $v_{\rm
  abs}$(\sitwo\,\l6355) and $d_{\rm abs}$(\stwo\,\l5454) in addition to
the standard $(x_1,c)$ predictors in Fig.~\ref{fig:vdabshubdiag}. One
clearly sees from these diagrams that the impact of the additional
spectroscopic indicator is fairly small, as the sign and magnitude of
the residuals are almost the same for $(x_1,c,\mathcal{S})$ and
$(x_1,c)$. This is further confirmed by looking up the value for the
intrinsic correlation in prediction error for both indicators in
Tables~\ref{tab:vabs} and \ref{tab:dabs}: $\rho_{x_1,c}=0.83\pm0.06$
for $v_{\rm abs}$(\sitwo\,\l6355) and $\rho_{x_1,c}=0.75\pm0.10$ for
$d_{\rm abs}$(\stwo\,\l5454).

%%% Fig. Hubble diagrams for vabs, FWHM, dabs

\subsubsection{Results using multiple indicators}\label{sect:multindic}

We have also considered models involving a linear combination of
two spectroscopic indicators (i.e. $\mu= m_B - M +
\gamma_1 \mathcal{S}_1 + \gamma_2 \mathcal{S}_2$) or a ratio of two
indicators (i.e. $\mu= m_B - M + \gamma
\mathcal{S}_1/\mathcal{S}_2$), also for cases including the SALT2
fit parameters $(x_1,c)$. No combination of two of these spectroscopic
indicators leads to an improvement over the single indicator case,
regardless of the age considered.

%%%%%%%%%%%%%%%%%%%%%%%%%%%%%%%%%%%%%%%%%%%%%%%%%%%%%%%%%%%%%%%%%%
%%
%%   Discussion
%%
%%%%%%%%%%%%%%%%%%%%%%%%%%%%%%%%%%%%%%%%%%%%%%%%%%%%%%%%%%%%%%%%%%

\section{Discussion: do \snia\ spectra really help?}\label{sect:disc}

The central question this paper addresses is whether spectra yield
useful information to predict distances to \sneia\ better than
light-curve width and color alone. The answer to this question can
have a significant impact 
on the way future \snia\ surveys are planned, namely whether or not
they should include spectroscopic (or spectro-photometric)
capabilities. This has been (and remains!) an active area of
discussion for proposals for space-borne missions within the framework
of the Dark Energy Task Force \citep{Albrecht/etal:2009} or the US
Astronomy \& Astrophysics Decadal
Survey\footnote{http://www.nationalacademies.org/astro2010}.

Of all the spectroscopic indicators considered in this paper, the 
concept of flux ratio introduced by \cite{Bailey/etal:2009} appears
to be the most promising, 
yielding up to $\sim30\%$ lower Hubble residual scatter than when
using the standard light-curve parameters. However, given the limited
sizes of the 
SN-Factory (58 \sneia) and CfA (26 \sneia) samples on which the method
has been applied, the results are at best statistically significant at
the $\lesssim2\sigma$ level, and the method should be validated on
much larger samples. It should be noted that the measurement of
flux ratios requires accurate relative flux calibration, as well as
minimal contamination by host-galaxy light. Both requirements impose
strong conditions on future \snia\ surveys that plan to use this
method.

The other spectroscopic indicators we consider in this paper are
intimately linked to line-profile shapes of specific \snia\ spectral
features. One would have hoped that such a physically-motivated
approach would yield interesting results, but this is not the case. At
best, these indicators yield $\lesssim1\sigma$ lower residual
scatter compared with the standard light-curve parameters. This is rather
disappointing, but also points to potential problems with the
measurement method we use. It is largely automated, but requires some
human interaction to ensure the correct local maxima used 
to define the wavelength bounds of each feature are
selected. Moreover, while some indicators (such as the absorption
velocity $v_{\rm abs}$) are largely insensitive to host-galaxy
reddening, others (such as the pseudo-EW) are strongly affected. Recent
unbiased techniques based on wavelet transforms have been proposed
that are largely insensitive to these measurement issues
\citep{Wagers/Wang/Asztalos:2010}, and the present analysis could be
repeated with such techniques.

Last, given our spectroscopic data we have focused exclusively on the
optical region, but there appears to be spectroscopic indicators that
correlate with luminosity in other wavelength regions (UV:
\citealt{Foley/Filippenko/Jha:2008}; NIR: Marion et al., in
preparation). An increased spectroscopic sample at these wavelengths
might reveal spectroscopic quantities that lead to even more precise
distances to \sneia\ than optical flux ratios.

%%%%%%%%%%%%%%%%%%%%%%%%%%%%%%%%%%%%%%%%%%%%%%%%%%%%%%%%%%%%%%%%%%
%%
%%   Conclusions
%%
%%%%%%%%%%%%%%%%%%%%%%%%%%%%%%%%%%%%%%%%%%%%%%%%%%%%%%%%%%%%%%%%%%

\section{Conclusions}\label{sect:ccl}

We have investigated the use of spectroscopic indicators which, when
used alone or in conjunction with light-curve parameters (width and color),
predict distances to \sneia\ better than when using the standard
combination of light-curve width and color. We have carried our a
$K$-fold cross-validation analysis on a large spectroscopic data set
obtained through the CfA Supernova Program. 
We constructed and implemented maximum likelihood estimators for
the rms intrinsic prediction error of a given method, and the
intrinsic covariance of prediction errors of different methods.  We
used these estimates to compare predictive models for SN Ia distances
in a quantitative manner.

We first considered the spectroscopic flux ratios of
\cite{Bailey/etal:2009}, highlighting the importance of an accurate
relative flux calibration and the failure of this method for
highly-reddened objects (SALT2 color $c > 0.5$). At maximum light, our
best single flux ratio $\mathcal{R}(6630/4400)$ from 26 \sneia\ at
$z>0.015$ leads to a $\sim10$\% lower weighted rms of
cross-validated prediction Hubble residuals (${\rm
    WRMS}=0.189\pm0.026$\,mag) than when using the standard SALT2
light-curve width ($x_1$) and color ($c$) parameters (${\rm
    WRMS}=0.204\pm0.029$\,mag), at 0.7$\sigma$ significance. When
used in combination with the SALT2 color parameter, our best
color-corrected flux ratio $\mathcal{R}^c(6420/5290)$ leads to
$\sim15$\% lower weighted rms (${\rm WRMS}=0.175\pm0.025$\,mag),
at 1.4$\sigma$ significance.
We thus confirm the use of flux ratios in improving distance
measurements of \sneia\ magnitudes, although the significance of the
difference with respect to the standard purely photometric approach is
difficult to gauge given our sample size. We also point to differences
between the best ratios found in this paper and those reported by
\cite{Bailey/etal:2009}, in part due to the way these ratios are
selected: \cite{Bailey/etal:2009} select their best ratios based on 
cross-correlation coefficients with uncorrected magnitudes, 
while we directly select them using the rms intrinsic error of
cross-validated distance predictions in the Hubble diagram.

Comparison of our results with synthetic spectra from a 2D survey of
delayed-detonation explosion models of
\cite{Kasen/Roepke/Woosley:2009} shows that a large part of the
variation in our best single flux ratio $\mathcal{R}(6630/4400)$ is
intrinsic and not due to reddening by dust. The correlation of this
ratio with \snia\ magnitudes is due to the luminosity-dependent
spectroscopic variation in the iron-group dominated absorption
features around $\sim4300$\,\AA. While the models confirm the presence
of many flux ratios that correlate strongly with absolute magnitude,
significant deviations exist with respect to the data. Such deviations
can in principle be exploited to impose strong constraints on \snia\
models. 

We extended the analysis of flux ratios to \snia\ spectra at other ages
($-2.5 \le t \le +7.5$\,d from maximum light (see
Table~\ref{tab:kfoldcv_10_othert}). The best set of predictors  
overall in this age range is the color-corrected
$\mathcal{R}^c(4610/4260)$ at $t=-2.5$\,d combined with SALT2 color,
which leads to $\sim30$\% lower weighted rms of prediction Hubble
residuals with respect to $(x_1,c)$ [${\rm WRMS}=0.143\pm0.020$\,mag],
and to $\sim40$\% lower intrinsic prediction error ($\sigma_{\rm
pred}=0.106\pm0.028$\,mag), at $\sim2\sigma$ significance. The
wavelength bins that 
constitute this ratio are part of the two prominent spectral
absorption features predominantly due to iron-group elements, labeled
\fetwo\,\l4300 and \fetwo\,\l4800, and which were found to vary
intrinsically between \sneia\ based on 2D models. Flux ratios at
$t\ge+5$\,d fare worse than at maximum light.

We also considered distance predictions based on two flux
ratios. We find that the improvement over the standard ($x_1,c$)
model is at the $\lesssim2\sigma$ level at best, and tends to be
worse for ages $t\ge+5$\,d. At maximum light, our best secondary
ratios are $\mathcal{R}_2(5160/5290)$ and
$\mathcal{R}^c_2(5690/5550)$, whose wavelength bins are clustered
around the \stwo\,\l\l5454,5640 doublet and the iron-group-dominated
absorption complex \fetwo\,\l4800. Both ratios measure intrinsic
small-scale differences between \sneia\ that are uncorrelated with
light-curve shape or color, and thus provide independent information
on their luminosity.

We also considered spectroscopic indicators associated with spectral
line-profile morphology: the absorption (and peak) velocity, the
full-width at half maximum, the relative absorption depth, the
pseudo-equivalent width, as well as other spectroscopic ratios. 
None of these spectroscopic indicators alone leads to a lower
weighted rms of prediction Hubble residuals. Only when they are combined with
SALT2 color do several indicators compete well with the standard
predictors. Such is the case of the \sitwo\,\l4130 pseudo-EW and
spectroscopic ratio $\mathcal{R}$(Si). Both indicators are correlated
with $x_1$ and act as a replacement to light-curve shape in the
distance prediction. When used in addition to $(x_1,c)$, the
\sitwo\,\l6355 absorption velocity and \stwo\,\l5454 relative
absorption depth lead to a small improvement, albeit statistically
insignificant. Using a linear combination of two such spectroscopic
indicators and ratios thereof leads to no further improvement, whether
at maximum light or at other ages.

Do spectra improve distance measurements of SN Ia?  Yes, but not as
much as we had hoped. The statistical framework developed here should
be applied to an independent and larger sample to find out whether the
effort of obtaining spectra for a cosmological sample will be repaid
with better knowledge of dark energy.

%%%%%%%%%%%%%%%%%%%%%%%%%%%%%%%%%%%%%%%%%%%%%%%%%%%%%%%%%%%%%%%%%%
%%
%%   Acknowledgments
%%
%%%%%%%%%%%%%%%%%%%%%%%%%%%%%%%%%%%%%%%%%%%%%%%%%%%%%%%%%%%%%%%%%%

\begin{acknowledgements}
We acknowledge many useful conversations with members of the RENOIR
group at the CPPM, in particular Florent Marmol and Andr\'e Tilquin.
We thank Stephen Bailey for his patience in explaining the details of
his flux ratio measurements and validation procedure. Alex Conley
shared a non-public custom version of his {\tt simple\_cosfitter} code
and provided invaluable help with cosmology fits. We further thank
Julien Guy and Gautham Narayan for advice on using the SALT2
light-curve fitter, and Dan Kasen for sending us the output of his 2D
radiative transfer calculations.
Support for supernova research at Harvard University, including the
CfA Supernova Archive, is provided in part by NSF grant AST
09-07903.

\bigskip
\noindent
{\it Note added in proof.}
While this paper was in the final stages of the refereeing process, 
\cite{Foley/Kasen:2010} posted a preprint on the arXiv server, in which 
they re-analyze the results of \cite{WangX/etal:2009b} to show that \sneia\ with 
different \sitwo\,\l6355 absorption velocities at maximum light have different 
intrinsic colors. Accounting for these intrinsic color differences reduces the 
scatter of Hubble residuals by $\sim30$\%, while using \sneia\ from a ``normal'' 
subsample reduces the scatter by $\sim40$\%. Although \cite{Foley/Kasen:2010} do 
not cross-validate their results or comment on their statistical significance, 
their analysis suggests that spectroscopy could be used to select a subsample of ``well-behaved'' 
\sneia\ for more precise distance measurements.
\end{acknowledgements}

%%%%%%%%%%%%%%%%%%%%%%%%%%%%%%%%%%%%%%%%%%%%%%%%%%%%%%%%%%%%%%%%%%
%%
%%   Bibliography
%%
%%%%%%%%%%%%%%%%%%%%%%%%%%%%%%%%%%%%%%%%%%%%%%%%%%%%%%%%%%%%%%%%%%

\bibliographystyle{aa} % style aa.bst
\bibliography{specindic} % your references Yourfile.bib

%%%%%%%%%%%%%%%%%%%%%%%%%%%%%%%%%%%%%%%%%%%%%%%%%%%%%%%%%%%%%%%%%%
%%
%%   Appendices
%%
%%%%%%%%%%%%%%%%%%%%%%%%%%%%%%%%%%%%%%%%%%%%%%%%%%%%%%%%%%%%%%%%%%

\appendix

\section{Sampling variance of weighted mean square error}\label{app:wrms}

If we assume the prediction errors are distributed normally with total
variance, Eq. \ref{eqn:magerr}: $\Delta \mu_s \sim N(0, \sigma_s^2)$,
then the sampling variance of the weighted mean square error is 
\begin{equation}
\var[\text{WRMS}^2] =   \frac{2}{\left(\sum_{s=1}^N w_s\right)^{2} }\sum_{s=1}^N w_s^2 \sigma_s^4.
\end{equation}
The standard error on WRMS is then $\sqrt{\var[{\rm WRMS}^2]} / (2
\times {\rm WRMS})$.

\section{Maximum likelihood estimators for intrinsic prediction error
  and covariance}\label{app:mle}

Let the intrinsic variance of predictions ($\sigma_{\rm pred}^2$) from
method $P, Q$ be $\sigma^2_P, \sigma^2_Q$.  Denote the intrinsic
covariance between distance prediction errors from $P$ and $Q$ as
$c_{PQ}$.  The intrinsic correlation is $\rho_{PQ} = c_{PQ} /
(\sigma_P \sigma_Q)$.   Let $\bm{\theta} = (\sigma_P, \sigma_Q,
\rho_{PQ})$ be the vector of this triplet of quantities. These
parameters can be arranged in an intrinsic covariance matrix: 
\begin{equation}
\bm{\Sigma}_\text{int}(\bm{\theta} )= \begin{pmatrix} \sigma^2_P &
  \rho_{PQ} \sigma_P \sigma_Q \\  \rho_{PQ} \sigma_P \sigma_Q &
  \sigma_Q^2 \end{pmatrix}
\end{equation}
We wish to estimate them from the cross-validated predictions.  We
derive estimators for this  intrinsic covariance using maximum
likelihood, and also derive its uncertainty.

We assume that the pair of prediction errors $\bm{\Delta \mu}_s =
(\Delta \mu_s^P, \Delta \mu_s^Q)$ are jointly distributed normally
around zero with total covariance 
\begin{equation}
\bm{\Sigma}_s(\bm{\theta} ) = \bm{\Sigma}_m^s + \bm{\Sigma}_{\rm pec}^s + \bm{\Sigma}_\text{int}(\bm{\theta} ).
\end{equation}
The measurement error covariance matrix, $\bm{\Sigma}_m^s$, contains
the measurement variances, $\sigma^2_{m,s}$, for model $P$ and model
$Q$, on the diagonal, and any covariance due to observational error in
the off-diagonal. Since supernova $s$ is subject to the same random
peculiar velocity under both models $P$ and $Q$, the covariance from
peculiar velocity dispersion is 
\begin{equation}
\bm{\Sigma}_{\rm pec}^s  = \begin{pmatrix}  \sigma^2_{\text{pec},s} &   \sigma^2_{\text{pec},s} \\   \sigma^2_{\text{pec},s} &   \sigma^2_{\text{pec},s} \end{pmatrix}
\end{equation}

The negative log likelihood for the unknown $(\sigma_P, \sigma_Q,
\rho_{PQ})$, given the set of distance predictions is 
\begin{equation}
-L(\bm{\theta} ) =\frac{1}{2} \ \sum_{s=1}^{N} \log \det\left[ 2\pi
  \bm{\Sigma}_s(\bm{\theta} )\right] +  \bm{\Delta \mu}_s^T
\bm{\Sigma}_s(\bm{\theta} ) \bm{\Delta \mu}_s 
\end{equation}
We numerically maximize the likelihood with the constraints $\sigma_P,
\sigma_Q > 0$ and $ | \rho_{PQ} |< 1$.   Once we have found the
maximum likelihood estimate (MLE) $\hat{\bm{\theta}}$, we can compute
its error by numerically evaluating the Hessian of the negative log
likelihood, $\bm{H}(\hat{\bm{\theta}}$). The sampling covariance
(error) of the MLE $\hat{\bm{\theta}}$ is estimated from the Fisher
information: $\bm{V}_{\hat{\bm{\theta}}} =
\bm{H}^{-1}(\hat{\bm{\theta}}) $.  The standard errors in each of
$(\sigma_P, \sigma_Q, \rho_{PQ})$ are the square roots of the diagonal
elements of $\bm{V}_{\hat{\bm{\theta}}}$. The off-diagonal elements
contain the estimation covariance between the three parameters. If the
difference in intrinsic prediction error between the two models is
$\Delta = \sigma_P - \sigma_Q$, the sampling variance of $\Delta$ is  
\begin{equation}
\begin{split}
\var[\Delta] &= \var[\sigma_P] - 2 \cov[\sigma_P, \sigma_Q] + \var[\sigma_Q] \\
&= V_{\hat{\bm{\theta}}}^{(1,1)} -2  V_{\hat{\bm{\theta}}}^{(1,2)} +  V_{\hat{\bm{\theta}}}^{(2,2)}  
\end{split}
\end{equation}
where $\bm{V}_{\hat{\bm{\theta}}}^{(i,j)}$ is the $(i,j)$ element of
the error covariance matrix of the MLE.
This error estimate accounts for covariance from random peculiar
velocities and the intrinsic correlation between two models.
Notably, a large $|\rho_{PQ}|$ will affect the significance of the
difference, $\Delta$.

From the prediction errors of a single method, $\{ \Delta \mu_s\}$, we
can estimate the rms intrinsic prediction error $\sigma_{\rm pred}$.
The negative log likelihood simplifies to
\begin{equation}
-  L(\sigma^2_\text{pred}) = \frac{1}{2} \sum_{s=1}^N \log(  \sigma^2_{m,s} + \sigma_\text{pred}^2 + \sigma^2_{\text{pec},s} ) + \frac{\Delta \mu_s^2}{ \sigma^2_{m,s} + \sigma_\text{pred}^2 + \sigma^2_{\text{pec},s}}
\end{equation}
The maximum likelihood estimate $\hat{\sigma}^2_\text{pred}$ is found
by minimizing this or finding the zero of the score function
$L'(\sigma^2_\text{pred}) = 0$.   If $N$ is large enough, the standard
error on $\hat{\sigma}^2_\text{pred}$ can be estimated using the
Fisher information at the MLE:
\begin{equation}
- L''( \hat{\sigma}^2_\text{pred} ) = \sum_{s=1}^N \frac{\Delta \mu_s^2}{ (\sigma^2_{m,s} + \hat{\sigma}_\text{pred}^2 + \sigma^2_{\text{pec},s})^3} - \frac{1}{2( \sigma^2_{m,s} + \hat{\sigma}_\text{pred}^2 + \sigma^2_{\text{pec},s})^2}
\end{equation}
An estimate of the sampling variance of the maximum likelihood
estimate of the intrinsic variance is the inverse of the Fisher
information $\var(\hat{\sigma}^2_\text{pred}) = [-L''(
  \hat{\sigma}^2_\text{pred} ) ]^{-1}$.   The standard error of
$\hat{\sigma}_{\text{pred}}$ itself is the square root of $\var[
  \hat{\sigma}_{\text{pred}}^2]/(4  \hat{\sigma}_{\text{pred}}^2)$.
This estimate of the intrinsic dispersion ``subtracts'' out the
contribution  of  random peculiar velocities and measurement error to
the total dispersion.

\section{Results for other spectroscopic indicators at maximum light}\label{app:tabs}

We present our results using the absorption velocity ($v_{\rm abs}$),
the full-width at half-maximum (FWHM), the relative absorption depth
($d_{\rm abs}$), the pseudo-equivalent width (pEW), and the various
spectroscopic ratios $\mathcal{R}(X)$
[Eqs.~\ref{eqn:rcasi1}-\ref{eqn:rcasi7}] in
Tables~\ref{tab:vabs}-\ref{tab:rx}.

\begin{table*}
\small
\small
\caption{$v_{\rm abs}$ (units of $10^4$\,\kms) at maximum light from 10-fold CV}\label{tab:vabs}
\begin{tabular}{lrccrrc}
\hline\hline
\multicolumn{1}{c}{Line} & \multicolumn{1}{c}{$\gamma$} & WRMS & $\sigma_{\rm pred}$ & \multicolumn{1}{c}{$\rho_{x_1,c}$} & \multicolumn{1}{c}{$\Delta_{x_1,c}$} & $N_{\rm SNIa}$ \\
\hline
\multicolumn{7}{l}{$v_{\rm abs}$} \\
\hline
\sitwo\,\l4130       & $-0.19 \pm 0.20$ & $0.289 \pm 0.035$ & $0.271 \pm 0.038$ & $ 0.59 \pm 0.12$ & $ 0.073 \pm 0.037\ (2.0\sigma)$ & 33\\
\stwo\,\l5454        & $ 0.33 \pm 0.16$ & $0.261 \pm 0.037$ & $0.241 \pm 0.040$ & $ 0.65 \pm 0.13$ & $ 0.061 \pm 0.038\ (1.6\sigma)$ & 25\\
\sitwo\,\l5972       & $-0.02 \pm 0.19$ & $0.301 \pm 0.040$ & $0.282 \pm 0.043$ & $ 0.47 \pm 0.17$ & $ 0.106 \pm 0.048\ (2.2\sigma)$ & 28\\
\sitwo\,\l6355       & $-0.60 \pm 0.13$ & $0.275 \pm 0.033$ & $0.256 \pm 0.035$ & $ 0.73 \pm 0.09$ & $ 0.090 \pm 0.031\ (2.9\sigma)$ & 35\\
\hline
\multicolumn{7}{l}{$(x_1,v_{\rm abs})$} \\
\hline
\sitwo\,\l4130       & $-0.11 \pm 0.20$ & $0.297 \pm 0.037$ & $0.281 \pm 0.039$ & $ 0.68 \pm 0.10$ & $ 0.081 \pm 0.035\ (2.3\sigma)$ & 33\\
\stwo\,\l5454        & $ 0.20 \pm 0.21$ & $0.257 \pm 0.036$ & $0.234 \pm 0.040$ & $ 0.71 \pm 0.12$ & $ 0.054 \pm 0.034\ (1.6\sigma)$ & 25\\
\sitwo\,\l5972       & $ 0.44 \pm 0.25$ & $0.305 \pm 0.041$ & $0.288 \pm 0.043$ & $ 0.50 \pm 0.16$ & $ 0.107 \pm 0.046\ (2.3\sigma)$ & 28\\
\sitwo\,\l6355       & $-0.41 \pm 0.08$ & $0.261 \pm 0.031$ & $0.242 \pm 0.034$ & $ 0.75 \pm 0.09$ & $ 0.073 \pm 0.028\ (2.6\sigma)$ & 35\\
\hline
\multicolumn{7}{l}{$(c,v_{\rm abs})$} \\
\hline
\sitwo\,\l4130       & $ 0.78 \pm 0.14$ & $0.229 \pm 0.029$ & $0.210 \pm 0.031$ & $ 0.71 \pm 0.10$ & $ 0.008 \pm 0.029\ (0.3\sigma)$ & 33\\
\stwo\,\l5454        & $ 0.75 \pm 0.22$ & $0.240 \pm 0.035$ & $0.222 \pm 0.037$ & $ 0.82 \pm 0.08$ & $ 0.038 \pm 0.025\ (1.5\sigma)$ & 25\\
\sitwo\,\l5972       & $ 0.20 \pm 0.10$ & $0.237 \pm 0.032$ & $0.217 \pm 0.035$ & $ 0.69 \pm 0.11$ & $ 0.035 \pm 0.030\ (1.2\sigma)$ & 28\\
\sitwo\,\l6355       & $ 0.56 \pm 0.13$ & $0.227 \pm 0.028$ & $0.207 \pm 0.030$ & $ 0.70 \pm 0.10$ & $ 0.035 \pm 0.026\ (1.3\sigma)$ & 35\\
\hline
\multicolumn{7}{l}{$(x_1,c,v_{\rm abs})$} \\
\hline
\sitwo\,\l4130       & $ 0.90 \pm 0.11$ & $0.209 \pm 0.026$ & $0.188 \pm 0.029$ & $ 0.90 \pm 0.03$ & $-0.013 \pm 0.017\ (0.8\sigma)$ & 33\\
\stwo\,\l5454        & $ 0.62 \pm 0.24$ & $0.217 \pm 0.031$ & $0.193 \pm 0.034$ & $ 0.96 \pm 0.02$ & $ 0.012 \pm 0.012\ (1.0\sigma)$ & 25\\
\sitwo\,\l5972       & $ 0.72 \pm 0.15$ & $0.204 \pm 0.027$ & $0.179 \pm 0.031$ & $ 0.87 \pm 0.05$ & $-0.003 \pm 0.021\ (0.1\sigma)$ & 28\\
\sitwo\,\l6355       & $ 1.07 \pm 0.12$ & $0.179 \pm 0.021$ & $0.151 \pm 0.025$ & $ 0.83 \pm 0.06$ & $-0.020 \pm 0.019\ (1.1\sigma)$ & 35\\
\hline
\end{tabular}
\end{table*}

\begin{table*}
\small
\caption{FWHM (units of $10^2$\,\AA) at maximum light from 10-fold CV}\label{tab:fwhm}
\begin{tabular}{lrccrrc}
\hline\hline
\multicolumn{1}{c}{Line} & \multicolumn{1}{c}{$\gamma$} & WRMS & $\sigma_{\rm pred}$ & \multicolumn{1}{c}{$\rho_{x_1,c}$} & \multicolumn{1}{c}{$\Delta_{x_1,c}$} & $N_{\rm SNIa}$ \\
\hline
\multicolumn{7}{l}{${\rm FWHM}$} \\
\hline
\sitwo\,\l4130       & $-1.28 \pm 0.09$ & $0.297 \pm 0.036$ & $0.277 \pm 0.040$ & $ 0.57 \pm 0.14$ & $ 0.093 \pm 0.039\ (2.4\sigma)$ & 32\\
\stwo\,\l5454        & $ 0.15 \pm 0.57$ & $0.957 \pm 0.126$ & $0.952 \pm 0.126$ & $-0.01 \pm 0.22$ & $ 0.769 \pm 0.129\ (6.0\sigma)$ & 29\\
\sitwo\,\l5972       & $ 0.14 \pm 0.20$ & $0.434 \pm 0.059$ & $0.424 \pm 0.059$ & $ 0.47 \pm 0.15$ & $ 0.246 \pm 0.063\ (3.9\sigma)$ & 28\\
\sitwo\,\l6355       & $-0.32 \pm 0.07$ & $0.279 \pm 0.033$ & $0.259 \pm 0.036$ & $ 0.72 \pm 0.09$ & $ 0.089 \pm 0.031\ (2.9\sigma)$ & 35\\
\hline
\multicolumn{7}{l}{$(x_1,{\rm FWHM})$} \\
\hline
\sitwo\,\l4130       & $-1.34 \pm 0.09$ & $0.293 \pm 0.035$ & $0.273 \pm 0.039$ & $ 0.49 \pm 0.15$ & $ 0.094 \pm 0.037\ (2.5\sigma)$ & 32\\
\stwo\,\l5454        & $ 0.44 \pm 1.17$ & $0.715 \pm 0.092$ & $0.708 \pm 0.093$ & $ 0.13 \pm 0.21$ & $ 0.524 \pm 0.084\ (6.2\sigma)$ & 29\\
\sitwo\,\l5972       & $ 0.33 \pm 0.24$ & $0.507 \pm 0.068$ & $0.499 \pm 0.069$ & $ 0.40 \pm 0.15$ & $ 0.318 \pm 0.073\ (4.4\sigma)$ & 28\\
\sitwo\,\l6355       & $-0.24 \pm 0.08$ & $0.276 \pm 0.033$ & $0.256 \pm 0.035$ & $ 0.75 \pm 0.09$ & $ 0.085 \pm 0.029\ (2.9\sigma)$ & 35\\
\hline
\multicolumn{7}{l}{$(c,{\rm FWHM})$} \\
\hline
\sitwo\,\l4130       & $-0.89 \pm 0.14$ & $0.258 \pm 0.032$ & $0.236 \pm 0.035$ & $ 0.73 \pm 0.10$ & $ 0.052 \pm 0.031\ (1.7\sigma)$ & 32\\
\stwo\,\l5454        & $ 0.50 \pm 0.44$ & $0.495 \pm 0.065$ & $0.485 \pm 0.066$ & $ 0.22 \pm 0.21$ & $ 0.304 \pm 0.073\ (4.2\sigma)$ & 29\\
\sitwo\,\l5972       & $ 0.28 \pm 0.11$ & $0.359 \pm 0.049$ & $0.348 \pm 0.050$ & $ 0.60 \pm 0.14$ & $ 0.167 \pm 0.050\ (3.3\sigma)$ & 28\\
\sitwo\,\l6355       & $-0.04 \pm 0.09$ & $0.247 \pm 0.030$ & $0.227 \pm 0.032$ & $ 0.79 \pm 0.07$ & $ 0.054 \pm 0.025\ (2.2\sigma)$ & 35\\
\hline
\multicolumn{7}{l}{$(x_1,c,{\rm FWHM})$} \\
\hline
\sitwo\,\l4130       & $ 0.24 \pm 0.51$ & $0.233 \pm 0.030$ & $0.214 \pm 0.031$ & $ 0.93 \pm 0.03$ & $ 0.031 \pm 0.017\ (1.8\sigma)$ & 32\\
\stwo\,\l5454        & $ 0.01 \pm 0.41$ & $0.331 \pm 0.043$ & $0.315 \pm 0.046$ & $ 0.43 \pm 0.17$ & $ 0.135 \pm 0.047\ (2.9\sigma)$ & 29\\
\sitwo\,\l5972       & $ 0.26 \pm 0.08$ & $0.362 \pm 0.050$ & $0.352 \pm 0.050$ & $ 0.66 \pm 0.10$ & $ 0.170 \pm 0.045\ (3.8\sigma)$ & 28\\
\sitwo\,\l6355       & $ 0.28 \pm 0.03$ & $0.193 \pm 0.023$ & $0.168 \pm 0.026$ & $ 0.90 \pm 0.04$ & $-0.006 \pm 0.015\ (0.4\sigma)$ & 35\\
\hline
\end{tabular}
\end{table*}

\begin{table*}
\small
\caption{$d_{\rm abs}$ at maximum light from 10-fold CV}\label{tab:dabs}
\begin{tabular}{lrccrrc}
\hline\hline
\multicolumn{1}{c}{Line} & \multicolumn{1}{c}{$\gamma$} & WRMS & $\sigma_{\rm pred}$ & \multicolumn{1}{c}{$\rho_{x_1,c}$} & \multicolumn{1}{c}{$\Delta_{x_1,c}$} & $N_{\rm SNIa}$ \\
\hline
\multicolumn{7}{l}{$d_{\rm abs}$} \\
\hline
\sitwo\,\l4130       & $-2.51 \pm 0.29$ & $0.283 \pm 0.035$ & $0.267 \pm 0.037$ & $ 0.54 \pm 0.11$ & $ 0.076 \pm 0.036\ (2.1\sigma)$ & 33\\
\stwo\,\l5454        & $-3.07 \pm 0.74$ & $0.280 \pm 0.039$ & $0.259 \pm 0.043$ & $ 0.37 \pm 0.20$ & $ 0.063 \pm 0.050\ (1.3\sigma)$ & 25\\
\sitwo\,\l5972       & $-3.54 \pm 1.31$ & $0.357 \pm 0.046$ & $0.336 \pm 0.051$ & $ 0.34 \pm 0.20$ & $ 0.160 \pm 0.056\ (2.9\sigma)$ & 28\\
\sitwo\,\l6355       & $-1.66 \pm 0.18$ & $0.277 \pm 0.033$ & $0.257 \pm 0.036$ & $ 0.57 \pm 0.13$ & $ 0.088 \pm 0.032\ (2.7\sigma)$ & 35\\
\hline
\multicolumn{7}{l}{$(x_1,d_{\rm abs})$} \\
\hline
\sitwo\,\l4130       & $-3.84 \pm 1.60$ & $0.384 \pm 0.048$ & $0.375 \pm 0.048$ & $ 0.24 \pm 0.18$ & $ 0.179 \pm 0.052\ (3.4\sigma)$ & 33\\
\stwo\,\l5454        & $-2.53 \pm 0.65$ & $0.270 \pm 0.037$ & $0.248 \pm 0.042$ & $ 0.52 \pm 0.17$ & $ 0.052 \pm 0.043\ (1.2\sigma)$ & 25\\
\sitwo\,\l5972       & $-3.13 \pm 1.12$ & $0.260 \pm 0.038$ & $0.239 \pm 0.042$ & $ 0.34 \pm 0.22$ & $ 0.023 \pm 0.022\ (1.0\sigma)$ & 28\\
\sitwo\,\l6355       & $-0.77 \pm 1.14$ & $0.306 \pm 0.037$ & $0.290 \pm 0.039$ & $ 0.62 \pm 0.12$ & $ 0.119 \pm 0.037\ (3.2\sigma)$ & 35\\
\hline
\multicolumn{7}{l}{$(c,d_{\rm abs})$} \\
\hline
\sitwo\,\l4130       & $-2.06 \pm 0.27$ & $0.230 \pm 0.029$ & $0.210 \pm 0.031$ & $ 0.80 \pm 0.08$ & $ 0.019 \pm 0.026\ (0.7\sigma)$ & 33\\
\stwo\,\l5454        & $-3.90 \pm 0.33$ & $0.228 \pm 0.032$ & $0.204 \pm 0.036$ & $ 0.49 \pm 0.17$ & $ 0.008 \pm 0.038\ (0.2\sigma)$ & 25\\
\sitwo\,\l5972       & $-2.65 \pm 0.70$ & $0.243 \pm 0.031$ & $0.214 \pm 0.037$ & $ 0.60 \pm 0.13$ & $ 0.044 \pm 0.037\ (1.2\sigma)$ & 28\\
\sitwo\,\l6355       & $-1.43 \pm 0.14$ & $0.226 \pm 0.027$ & $0.203 \pm 0.030$ & $ 0.78 \pm 0.08$ & $ 0.031 \pm 0.023\ (1.3\sigma)$ & 35\\
\hline
\multicolumn{7}{l}{$(x_1,c,d_{\rm abs})$} \\
\hline
\sitwo\,\l4130       & $-1.87 \pm 1.04$ & $0.248 \pm 0.031$ & $0.231 \pm 0.033$ & $ 0.87 \pm 0.05$ & $ 0.039 \pm 0.020\ (2.0\sigma)$ & 33\\
\stwo\,\l5454        & $-3.26 \pm 0.35$ & $0.203 \pm 0.028$ & $0.175 \pm 0.033$ & $ 0.75 \pm 0.10$ & $-0.022 \pm 0.030\ (0.7\sigma)$ & 25\\
\sitwo\,\l5972       & $-2.24 \pm 0.55$ & $0.217 \pm 0.028$ & $0.188 \pm 0.033$ & $ 0.69 \pm 0.12$ & $ 0.014 \pm 0.030\ (0.5\sigma)$ & 28\\
\sitwo\,\l6355       & $ 0.18 \pm 0.67$ & $0.219 \pm 0.026$ & $0.198 \pm 0.029$ & $ 0.97 \pm 0.01$ & $ 0.028 \pm 0.011\ (2.5\sigma)$ & 35\\
\hline
\end{tabular}
\end{table*}

\begin{table*}
\small
\caption{pEW (units of $10^2$\,\AA) at maximum light from 10-fold CV}\label{tab:pew}
\begin{tabular}{lrccrrc}
\hline\hline
\multicolumn{1}{c}{Line} & \multicolumn{1}{c}{$\gamma$} & WRMS & $\sigma_{\rm pred}$ & \multicolumn{1}{c}{$\rho_{x_1,c}$} & \multicolumn{1}{c}{$\Delta_{x_1,c}$} & $N_{\rm SNIa}$ \\
\hline
\multicolumn{7}{l}{${\rm pEW}$} \\
\hline
\sitwo\,\l4130       & $-2.09 \pm 0.18$ & $0.257 \pm 0.032$ & $0.238 \pm 0.034$ & $ 0.65 \pm 0.12$ & $ 0.041 \pm 0.033\ (1.2\sigma)$ & 33\\
\fetwo\,\l4300       & $-0.20 \pm 0.23$ & $0.298 \pm 0.035$ & $0.282 \pm 0.037$ & $ 0.64 \pm 0.09$ & $ 0.107 \pm 0.035\ (3.1\sigma)$ & 36\\
\fetwo\,\l4800       & $-0.32 \pm 0.05$ & $0.269 \pm 0.032$ & $0.250 \pm 0.034$ & $ 0.67 \pm 0.11$ & $ 0.076 \pm 0.031\ (2.5\sigma)$ & 36\\
\stwo\,\l\l5454,5640 & $-0.65 \pm 0.26$ & $0.269 \pm 0.035$ & $0.249 \pm 0.038$ & $ 0.63 \pm 0.13$ & $ 0.076 \pm 0.037\ (2.1\sigma)$ & 29\\
\sitwo\,\l6355       & $-0.55 \pm 0.05$ & $0.256 \pm 0.030$ & $0.235 \pm 0.033$ & $ 0.73 \pm 0.09$ & $ 0.058 \pm 0.029\ (2.0\sigma)$ & 35\\
\hline
\multicolumn{7}{l}{$(x_1,{\rm pEW})$} \\
\hline
\sitwo\,\l4130       & $-2.53 \pm 0.28$ & $0.262 \pm 0.032$ & $0.244 \pm 0.035$ & $ 0.63 \pm 0.12$ & $ 0.046 \pm 0.033\ (1.4\sigma)$ & 33\\
\fetwo\,\l4300       & $-0.13 \pm 0.26$ & $0.297 \pm 0.035$ & $0.281 \pm 0.037$ & $ 0.69 \pm 0.10$ & $ 0.104 \pm 0.034\ (3.1\sigma)$ & 36\\
\fetwo\,\l4800       & $-0.37 \pm 0.05$ & $0.258 \pm 0.030$ & $0.239 \pm 0.033$ & $ 0.78 \pm 0.08$ & $ 0.064 \pm 0.026\ (2.5\sigma)$ & 36\\
\stwo\,\l\l5454,5640 & $-0.16 \pm 0.62$ & $0.288 \pm 0.038$ & $0.272 \pm 0.040$ & $ 0.68 \pm 0.12$ & $ 0.096 \pm 0.040\ (2.4\sigma)$ & 29\\
\sitwo\,\l6355       & $-0.43 \pm 0.06$ & $0.265 \pm 0.031$ & $0.245 \pm 0.034$ & $ 0.74 \pm 0.09$ & $ 0.067 \pm 0.029\ (2.3\sigma)$ & 35\\
\hline
\multicolumn{7}{l}{$(c,{\rm pEW})$} \\
\hline
\sitwo\,\l4130       & $-1.57 \pm 0.24$ & $0.225 \pm 0.028$ & $0.205 \pm 0.030$ & $ 0.94 \pm 0.03$ & $ 0.006 \pm 0.014\ (0.4\sigma)$ & 33\\
\fetwo\,\l4300       & $ 0.14 \pm 0.17$ & $0.240 \pm 0.029$ & $0.221 \pm 0.031$ & $ 0.84 \pm 0.05$ & $ 0.043 \pm 0.021\ (2.0\sigma)$ & 36\\
\fetwo\,\l4800       & $ 0.23 \pm 0.08$ & $0.230 \pm 0.028$ & $0.212 \pm 0.029$ & $ 0.81 \pm 0.06$ & $ 0.035 \pm 0.022\ (1.6\sigma)$ & 36\\
\stwo\,\l\l5454,5640 & $-0.90 \pm 0.14$ & $0.223 \pm 0.030$ & $0.201 \pm 0.032$ & $ 0.81 \pm 0.08$ & $ 0.026 \pm 0.024\ (1.1\sigma)$ & 29\\
\sitwo\,\l6355       & $-0.31 \pm 0.05$ & $0.231 \pm 0.028$ & $0.210 \pm 0.030$ & $ 0.89 \pm 0.04$ & $ 0.029 \pm 0.018\ (1.6\sigma)$ & 35\\
\hline
\multicolumn{7}{l}{$(x_1,c,{\rm pEW})$} \\
\hline
\sitwo\,\l4130       & $ 0.01 \pm 0.78$ & $0.249 \pm 0.031$ & $0.232 \pm 0.033$ & $ 0.99 \pm 0.00$ & $ 0.022 \pm 0.008\ (2.7\sigma)$ & 33\\
\fetwo\,\l4300       & $ 0.25 \pm 0.16$ & $0.216 \pm 0.026$ & $0.196 \pm 0.028$ & $ 0.99 \pm 0.00$ & $ 0.013 \pm 0.007\ (1.9\sigma)$ & 36\\
\fetwo\,\l4800       & $ 0.17 \pm 0.09$ & $0.212 \pm 0.026$ & $0.193 \pm 0.027$ & $ 0.97 \pm 0.01$ & $ 0.013 \pm 0.010\ (1.3\sigma)$ & 36\\
\stwo\,\l\l5454,5640 & $-0.32 \pm 0.22$ & $0.210 \pm 0.028$ & $0.186 \pm 0.031$ & $ 0.99 \pm 0.01$ & $ 0.013 \pm 0.009\ (1.4\sigma)$ & 29\\
\sitwo\,\l6355       & $ 0.31 \pm 0.13$ & $0.215 \pm 0.026$ & $0.193 \pm 0.028$ & $ 0.95 \pm 0.02$ & $ 0.013 \pm 0.011\ (1.2\sigma)$ & 35\\
\hline
\end{tabular}
\end{table*}

\begin{table*}
\small
\caption{$\mathcal{R}(\rm Ca)$ and $\mathcal{R}(\rm Si)$ at maximum light from 10-fold CV}\label{tab:rx}
\begin{tabular}{lrccrrc}
\hline\hline
\multicolumn{1}{c}{Line} & \multicolumn{1}{c}{$\gamma$} & WRMS & $\sigma_{\rm pred}$ & \multicolumn{1}{c}{$\rho_{x_1,c}$} & \multicolumn{1}{c}{$\Delta_{x_1,c}$} & $N_{\rm SNIa}$ \\
\hline
\multicolumn{7}{l}{$\mathcal{R}_X$} \\
\hline
$\mathcal{R}(\rm CaS)$    & $-0.10 \pm 0.03$ & $0.288 \pm 0.043$ & $0.273 \pm 0.046$ & $ 0.52 \pm 0.15$ & $ 0.085 \pm 0.046\ (1.8\sigma)$ & 22\\
$\mathcal{R}(\rm Si)$     & $-2.30 \pm 0.45$ & $0.282 \pm 0.037$ & $0.259 \pm 0.041$ & $ 0.31 \pm 0.20$ & $ 0.097 \pm 0.048\ (2.0\sigma)$ & 28\\
$\mathcal{R}(\rm SiS)$    & $ 1.52 \pm 0.13$ & $0.286 \pm 0.038$ & $0.270 \pm 0.040$ & $ 0.69 \pm 0.11$ & $ 0.111 \pm 0.040\ (2.8\sigma)$ & 29\\
$\mathcal{R}(\rm SiSS)$   & $ 1.24 \pm 0.18$ & $0.260 \pm 0.031$ & $0.242 \pm 0.033$ & $ 0.73 \pm 0.09$ & $ 0.053 \pm 0.029\ (1.8\sigma)$ & 36\\
$\mathcal{R}(\rm S,Si)^a$ & $ 0.09 \pm 0.06$ & $0.263 \pm 0.036$ & $0.242 \pm 0.040$ & $ 0.59 \pm 0.15$ & $ 0.065 \pm 0.039\ (1.7\sigma)$ & 26\\
$\mathcal{R}(\rm Si,Fe)$  & $-1.69 \pm 0.23$ & $0.275 \pm 0.036$ & $0.257 \pm 0.038$ & $ 0.53 \pm 0.15$ & $ 0.079 \pm 0.041\ (1.9\sigma)$ & 29\\
\hline
\multicolumn{7}{l}{$(x_1,\mathcal{R}_X)$} \\
\hline
$\mathcal{R}(\rm CaS)$    & $-0.15 \pm 0.03$ & $0.273 \pm 0.041$ & $0.259 \pm 0.043$ & $ 0.63 \pm 0.14$ & $ 0.071 \pm 0.042\ (1.7\sigma)$ & 22\\
$\mathcal{R}(\rm Si)$     & $-2.39 \pm 0.20$ & $0.196 \pm 0.033$ & $0.175 \pm 0.035$ & $ 0.27 \pm 0.24$ & $ 0.026 \pm 0.045\ (0.5\sigma)$ & 28\\
$\mathcal{R}(\rm SiS)$    & $ 1.56 \pm 1.52$ & $0.398 \pm 0.052$ & $0.386 \pm 0.053$ & $ 0.35 \pm 0.17$ & $ 0.214 \pm 0.067\ (3.2\sigma)$ & 29\\
$\mathcal{R}(\rm SiSS)$   & $ 1.34 \pm 0.33$ & $0.277 \pm 0.033$ & $0.260 \pm 0.035$ & $ 0.70 \pm 0.09$ & $ 0.071 \pm 0.029\ (2.4\sigma)$ & 36\\
$\mathcal{R}(\rm S,Si)^a$ & $-0.28 \pm 0.13$ & $0.278 \pm 0.038$ & $0.257 \pm 0.041$ & $ 0.64 \pm 0.12$ & $ 0.079 \pm 0.036\ (2.2\sigma)$ & 26\\
$\mathcal{R}(\rm Si,Fe)$  & $ 0.09 \pm 0.85$ & $0.285 \pm 0.038$ & $0.269 \pm 0.040$ & $ 0.64 \pm 0.12$ & $ 0.091 \pm 0.038\ (2.4\sigma)$ & 29\\
\hline
\multicolumn{7}{l}{$(c,\mathcal{R}_X)$} \\
\hline
$\mathcal{R}(\rm CaS)$    & $ 0.08 \pm 0.03$ & $0.248 \pm 0.037$ & $0.234 \pm 0.040$ & $ 0.73 \pm 0.11$ & $ 0.043 \pm 0.035\ (1.2\sigma)$ & 22\\
$\mathcal{R}(\rm Si)$     & $-1.68 \pm 0.19$ & $0.190 \pm 0.025$ & $0.158 \pm 0.030$ & $ 0.59 \pm 0.13$ & $-0.007 \pm 0.030\ (0.2\sigma)$ & 28\\
$\mathcal{R}(\rm SiS)$    & $ 0.92 \pm 0.21$ & $0.240 \pm 0.032$ & $0.220 \pm 0.034$ & $ 0.86 \pm 0.06$ & $ 0.062 \pm 0.022\ (2.8\sigma)$ & 29\\
$\mathcal{R}(\rm SiSS)$   & $ 0.81 \pm 0.19$ & $0.229 \pm 0.027$ & $0.208 \pm 0.030$ & $ 0.95 \pm 0.02$ & $ 0.021 \pm 0.011\ (1.9\sigma)$ & 36\\
$\mathcal{R}(\rm S,Si)^a$ & $ 0.25 \pm 0.04$ & $0.220 \pm 0.032$ & $0.200 \pm 0.033$ & $ 0.88 \pm 0.05$ & $ 0.017 \pm 0.022\ (0.8\sigma)$ & 26\\
$\mathcal{R}(\rm Si,Fe)$  & $-1.99 \pm 0.35$ & $0.216 \pm 0.029$ & $0.196 \pm 0.031$ & $ 0.88 \pm 0.05$ & $ 0.018 \pm 0.021\ (0.9\sigma)$ & 29\\
\hline
\multicolumn{7}{l}{$(x_1,c,\mathcal{R}_X)$} \\
\hline
$\mathcal{R}(\rm CaS)$    & $ 0.04 \pm 0.02$ & $0.215 \pm 0.033$ & $0.202 \pm 0.035$ & $ 0.99 \pm 0.01$ & $ 0.006 \pm 0.012\ (0.5\sigma)$ & 22\\
$\mathcal{R}(\rm Si)$     & $-1.23 \pm 0.87$ & $0.209 \pm 0.027$ & $0.179 \pm 0.032$ & $ 0.79 \pm 0.08$ & $ 0.021 \pm 0.026\ (0.8\sigma)$ & 28\\
$\mathcal{R}(\rm SiS)$    & $-0.98 \pm 0.20$ & $0.205 \pm 0.028$ & $0.184 \pm 0.030$ & $ 0.84 \pm 0.06$ & $ 0.022 \pm 0.017\ (1.3\sigma)$ & 29\\
$\mathcal{R}(\rm SiSS)$   & $-1.06 \pm 0.29$ & $0.225 \pm 0.027$ & $0.205 \pm 0.029$ & $ 0.96 \pm 0.02$ & $ 0.013 \pm 0.013\ (1.0\sigma)$ & 36\\
$\mathcal{R}(\rm S,Si)$\tablefootmark{a} & $-0.11 \pm 0.09$ & $0.216 \pm 0.031$ & $0.195 \pm 0.033$ & $ 0.99 \pm 0.01$ & $ 0.013 \pm 0.011\ (1.2\sigma)$ & 26\\
$\mathcal{R}(\rm Si,Fe)$  & $-0.16 \pm 0.90$ & $0.233 \pm 0.032$ & $0.216 \pm 0.033$ & $ 1.00 \pm 0.00$ & $ 0.035 \pm 0.009\ (3.9\sigma)$ & 29\\
\hline
\end{tabular}
\tablefoot{
\tablefoottext{a}{In fact $\mathcal{R}(\rm S,Si)/10$.}
}
\end{table*}

\end{document}